\documentclass{aa}
\usepackage[varg]{txfonts}
\usepackage{natbib}
\bibpunct{(}{)}{;}{a}{}{,}

\usepackage{dsfont, bm}
\usepackage{scalerel,stackengine}
\usepackage{multirow}
\usepackage{empheq}
\usepackage{geometry}
\usepackage{xcolor, framed}
\usepackage{subcaption}
\usepackage[english]{babel}
\usepackage{graphicx}
\usepackage{hyperref}

\newenvironment{myframedeq}[1][\linewidth]{\FrameSep=4pt\abovedisplayskip=0pt\belowdisplayskip=0pt
\framed\hsize=#1\leftskip=\dimexpr(\textwidth-#1)/2\relax}
{\endframed}

\begin{document}

	\title{Bayesian decomposition of the Galactic multi-frequency sky \\ using probabilistic autoencoders}

	\author{Sara Milosevic\inst{\ref{inst1}, \ref{inst2}}\and Philipp Frank\inst{\ref{inst1} , \ref{inst2}}\and
	Reimar H. Leike\inst{\ref{inst1}, \ref{inst2}}\and Ancla M{\"u}ller\inst{\ref{inst3}}\and Torsten A. En{\ss}lin				\inst{\ref{inst1}, \ref{inst2}}}

	\institute{Ludwig-Maximilians-Universit{\"a}t M{\"u}nchen, Geschwister-Scholl-Platz 1, 80539 M{\"u}nchen, Germany 		\label{inst1} 
	\and  Max-Planck-Institut für Astrophysik, Karl-Schwarzschild-Str. 1, 85748 Garching, Germany \label{inst2}
	\and  Ruhr-Universit{\"a}t Bochum, Universit{\"a}tsstr. 150, 44801 Bochum, Germany \label{inst3}}

\date{\today \, /\, Accepted date }


\abstract{All-sky observations of the Milky Way show both Galactic and non-Galactic diffuse emission, for example from interstellar matter or the cosmic microwave background (CMB). The different emitters are partly superimposed in the measurements, partly they obscure each other, and sometimes they dominate within a certain spectral range. The decomposition of the underlying radiative components from spectral data is a signal reconstruction problem and often associated with detailed physical modeling and substantial computational effort.}{We aim to build an effective and self-instructing algorithm detecting the essential spectral information contained Galactic all-sky data covering spectral bands from $\gamma$-ray to radio waves.}{Utilizing principles from information theory, we develop a state-of-the-art variational autoencoder specialized on the adaption to Gaussian noise statistics. We first derive a generic generative process that leads from a low-dimensional set of emission features to the observed high-dimensional data. We formulate a posterior distribution of these features using Bayesian methods and approximate this posterior with variational inference.}{The algorithm efficiently encodes the information of 35 Galactic emission data sets in ten latent feature maps. These contain the essential information required to reconstruct the initial data with high fidelity and are ranked by the algorithm according to their significance for data regeneration. The three most significant feature maps encode astrophysical components: (1) The dense interstellar medium (ISM), (2) the hot and dilute regions of the ISM and (3) the CMB.}{The machine-assisted and data-driven dimensionality reduction of spectral data is able to uncover the physical features encoding the input data. Our algorithm is able to extract the dense and dilute Galactic regions, as well as the CMB, from the sky brightness values only.}


\maketitle


\section{Introduction}
\label{ch:introduction}
The interstellar medium (ISM) is a key element of the Milky Way and subject to both astrophysical and cosmological studies. It consists of localized components such as molecules and interstellar dust in cold clouds, atomic and ionized hydrogen, and hot plasma in the Galactic halo, as well as components which pervade the entire ISM like cosmic rays and magnetic fields. Our present knowledge about the existence of these components is based on the fact that they all contribute to the interstellar radiation field \citep[e.g.,][]{2011piim.book.....D}. Each component, or the interplay of several components, generates radiation of a specific spectrum and can be reconstructed by component separation algorithms of varying complexity. Some components show very characteristic emission lines such as CO or neutral (HI) and ionized (HII) atomic hydrogen, which permits their Galactic distribution to be determined very precisely. Other components, however, can generate radiation distributed over completely opposite areas of the electromagnetic spectrum, making the component separation task more sophisticated: For example, the interaction of cosmic rays and magnetic fields generates synchrotron radiation and can be observed in the radio regime, as in the $408$~MHz radio map \citep{1982AAS...47....1H}, while the interaction of cosmic rays with interstellar matter imprints in the $\gamma$-ray regime due to hadron-nucleon collisions producing pions \citep[e.g.,][]{1994A&A...286..983M}. Another example is hot ionized plasma, which radiates in the X-ray regime when generated by supernovae, whereas hot molecules ionized by collisions emit in the UV regime \citep[e.g.,][]{ferriere2001interstellar}. In this specific case, both the UV and soft X-ray photons are again absorbed by interstellar dust, which prevents observations in regions of high dust density like the Galactic plane. This interplay also shows the high complexity of radiative extinction, which needs to be taken into account when reconstructing single emission components. 

A famous component that is not subject to dust extinction, but still difficult to measure is the non-Galactic cosmic microwave background (CMB). The reason is that the CMB is superimposed by Galactic foregrounds. Cosmological studies aim to extract the CMB radiation from multiple frequency channels by identifying and systematically removing these Galactic foregrounds. In frequencies below $100$~GHz, Galactic synchrotron and free-free emission contaminate the CMB, while above $100$~GHz, thermal dust emission and the cosmic infrared background dominate \citep{2011ApJS..192...15G, 2016planckcomponent}. To distinguish between different sources of emission, members of the \citet{collaboration2018planck} developed several component separation algorithms, for example the \textsc{Commander} \citep{eriksen2008joint}, \textsc{Sevem} \citep{leach2008component}, or \textsc{Smica} code \citep{10.1111/j.1365-2966.2003.07069.x, cardoso2008component}. The results obtained from those algorithms contain, among others, all-sky maps of the CMB, synchrotron, free-free, thermal and spinning dust, and CO line emission \citep{2016planckcomponent}. These approaches are however based on cosmological, astrophysical and instrumental parameters, and require preprocessed spectral templates or explicit knowledge about physical correlations. To verify these results in a manner independent of such assumptions, approaches that allow automated component identification, like machine learning techniques, have been increasingly pursued in recent years \citep{RN3}. \\

A broad range of machine learning algorithms was applied to cosmological and astrophysical problems \citep{fluke2020surveying}. Here we just name a few: \citet{Beaumont_2011} employed the Support Vector Machine (SVM) algorithm to classify structures in the ISM. The algorithm was able to identify a supernovae remnant behind a molecular cloud based on a sample of manually classified data. \citet{Ucci_2018} examined the composition of the ISM of dwarf galaxies by processing the available spectral information in a machine learning code. The so-called \textsc{Game} algorithm was trained on a large library of synthetic data \citep{10.1093/mnras/sty804} and recovered ISM properties such as metallicity and gas densities and their respective correlations on the basis of spectral emission lines. By training a convolutional neural network on synthetic spectra, \citet{2019AAS...23325209M} were able to decompose the thermal phases of neutral hydrogen HI. This selection of algorithms represents supervised learning approaches, which require labeled or pre-classified data in order to be trained. We want to investigate to what extend unsupervised approaches can be applied to Galactic observations. \\

One unsupervised machine learning approach which automatically identifies relevant features within some input data is called representation learning (RL) \citep{hinton1990connectionist, bengio2013, goodfellow2016deep}. Given an observation, RL methods aim to extract the underlying causes which generated the data, in other words they learn the most informative representation of the observation \citep{bengio2013, goodfellow2016deep}. This can for example be achieved by reducing the dimension of the input data using a neural network to the so-called underlying, explanatory factors of variation \citep{hinton2006, bengio2013, goodfellow2016deep}. 
This concept can be translated to the task of astrophysical component separation by investigating which underlying, data-generating components can be extracted from Galactic all-sky data. These components are constructed to encode mutually independent features of the data and are often used as the input for subsequent analyses. Especially, considering the large quantities of astrophysical data produced every day by surveys such as the Sloan Digital Sky Survey (SDSS), an effective preprocessing strategy needs to be developed to be able to analyze the vast amount of data \citep{Kremer_2017, reis2019effectively}.\\

In the present study, we apply RL to a data set of 35 Galactic all-sky maps recorded in multiple frequencies provided by \citet{muller2018}. On their data set, the authors learned spectral pixel-wise relations using Gaussian Mixture Models (GMMs), which permitted to augment pixels with missing measurement data. They verified the pre-stated hadronic and leptonic component of the $\gamma-$ray sky \citep{selig2015denoised} and presented a higher resolved hadronic component as well as a completion of non-observed information. However, the main component maps of the GMM did not not capture different astrophysical environments. This motivated the approach in the present study to explore other latent variable models, like autoencoders \citep{hinton2006}, which are able to encode useful representations of the data in their latent space.\\

Based on the provided data set, we combine generative modeling with variational inference to learn a lower-dimensional representation of the data, which we call features. Our model approximates the posterior probability on these features, and we efficiently optimize the approximation using a state-of-the-art variational autoencoder (VAE) \citep{kingma2013auto}. Such an approach was, for example, used in the context of the previously mentioned Sloan Digital Sky Survey, as \citet{Portillo_2020} successfully applied variational autoencoders to the SDDS data. The authors efficiently reduced the dimensionality of 1000 input pixels to six latent components, while the VAE was able to outperform principal component analysis considering the spectral dimensionality reduction. Their latent space separates types of galaxies or detects outliers, making it a very useful preprocessing step for large astrophysical data. However, the authors claim that the uncertainty quantification could be improved and suggest to include the pixel-level noise as a separate feature to improve latent variance representations. \\

Our variational autoencoder is based on the principles of information theory, meaning that we follow a Bayesian approach to track all relevant uncertainties and we enable our model to adapt to the introduced model noise. Our mathematical derivation of the loss function is, although based on the specific signal reconstruction problem, a general approach: Starting with a generative data model, we are able to explain all terms occurring in a classical VAE's loss function, but in addition we provide further terms which clearly result from our calculations in Sect.~\ref{ch:datamethods} and, simultaneously, deliver robust results in Sect.~\ref{ch:results}.  

\section{Data and methods}
\label{ch:datamethods}
\subsection{Data}
\label{ch:data}
Observations of the Milky Way can be visualized by all-sky maps showing the sky brightness in a certain frequency range. When we combine data from Galactic all-sky records in multiple frequencies, we can obtain a more complete picture of our Galaxy, but we also gain redundant information. Our goal is to determine a reduced representation of the observed sky, which contains only non-redundant or essential information. For this analysis, we use the aforementioned data set consisting of $39$ Galactic all-sky maps distributed over the entire electromagnetic spectrum compiled by \citet{muller2018}.

To generate the data set, the authors assembled information from all-sky surveys ranging from $\gamma$-ray to radio frequencies \citep[and references therein]{muller2018}. They incorporated all-sky data with at least $80$~$\%$ spatial coverage in HEALPix format \citep{gorski2005} and used the highest resolution map for each frequency. The UV regime is not part of their data set, since the respective GALEX survey \citep{bianchi2017revised} shows too many unobserved regions. To homogenize and unify the data, they (1) converted the sky brightness values in the original maps to flux magnitude values, (2) reduced the noise level in X-ray data by smoothing with a Gaussian kernel, (3) removed extra-Galactic sources and calibration artifacts from all maps except the already cleaned $\gamma$-ray data from \citep{selig2015denoised}, and (4) unified the resolution of all maps to nside~$=128$. \\

Based on the resulting $39$ Galactic all-sky maps, we build a data set $\bm{D}$ (see Table \ref{app:data}) for the present study with three modifications: (1) we discard the hadronic and leptonic component maps by \citet{selig2015denoised}, since these are derived from the Fermi data and thus contain redundant information, (2) we remove the AKARI far infrared map recorded at $65~\mu m$, which exhibits a poor spatial coverage (we can only work with pixels that are covered by all maps), and (3) we neglect the CO line emission map, since this information is partly contained in the Planck $100$, $217$, and $353$~GHz frequency channels \citep{2016planckcomponent}. We then define our data set consisting of $k=35$ all-sky maps as an indexed set $\bm{D}= \left( \bm{d}_1, \ldots, \bm{d}_p \right)$, where $p$ is the number of pixels and the magnitude vectors $\bm{d}_i \in \mathds{R}^{k}$ represent the magnitude flux values of all frequency maps for the $i^{\,\mathrm{th}}$ pixel\footnote{In other words: $\bm{d}_i = \left( d_i^{\,(1)}, d_i^{\,(2)}, \cdot \cdot \cdot, d_i^{\,(k)}\right)^{\mathrm{T}}$, where  $d_i^{\,(\lambda)}$ is the magnitude flux value of the $\lambda^{\,\mathrm{th}}$ sky map at pixel $i$.}.

\begin{figure}
\centering
\includegraphics[width=1\linewidth]{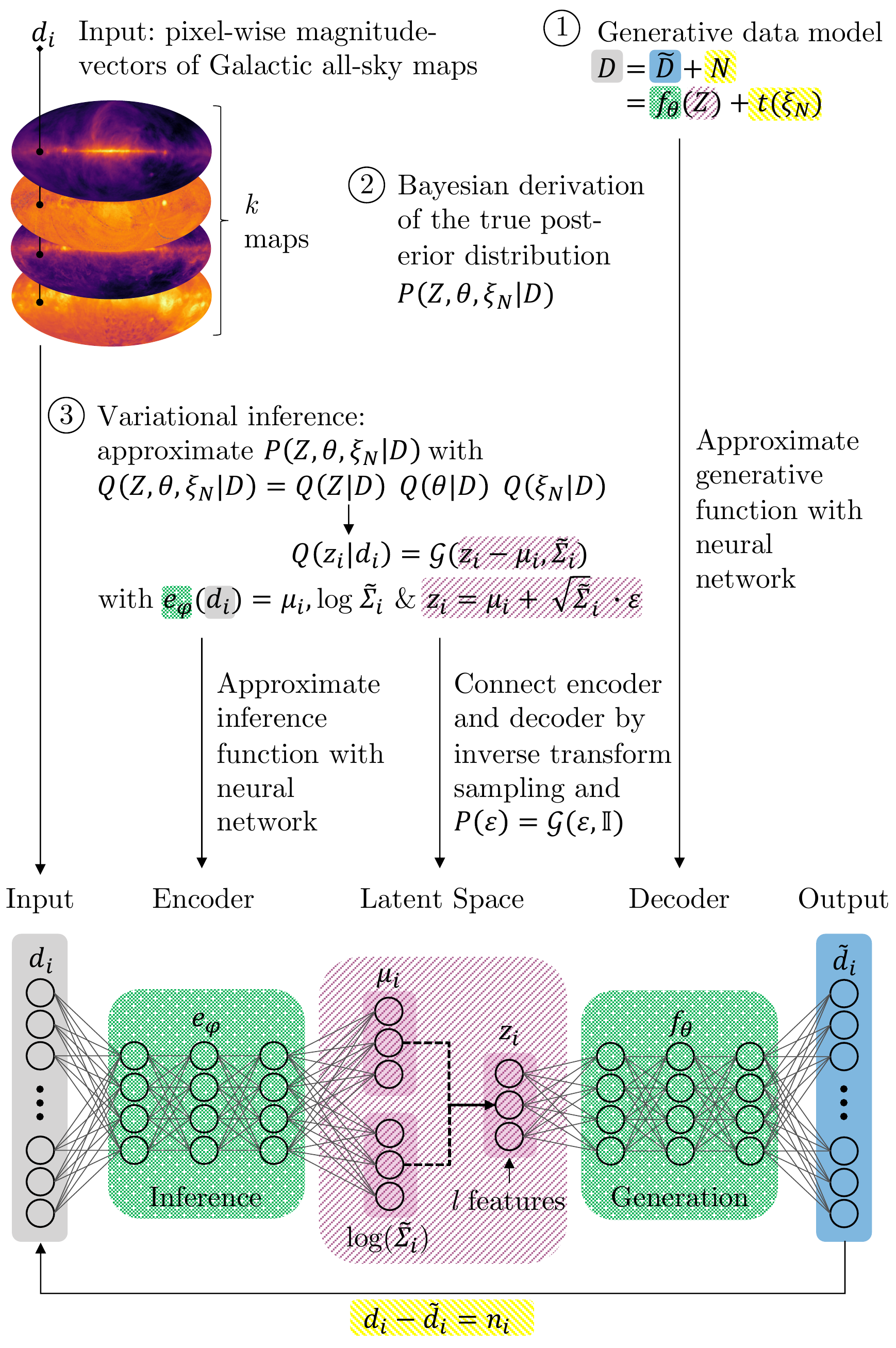}
\caption{Model design of our algorithm (NEAT-VAE). In our mathematical derivation, we first define a generative data model and afterwards build an inference process using information theoretical methods. The combination of both processes allows the resulting algorithm to perform in the opposite direction: It uses its input $\bm{d}_i$ to infer a latent representation $\bm{z}_i$ of the data, which is (among other factors like priors) led by the generative process aiming to regenerate the initial data vector $\bm{d}_i$ from the latent space again. The minimization objective, or loss function, directing the algorithm's learning process is Eq.~\eqref{eq:KLD}.}
\label{fig:NEAT_VAE}
\end{figure}

\subsection{Model design}
\label{ch:model}
The determination of a non-redundant and lower-dimensional representation of the frequency information in our data set is an inverse problem: We are looking for unknown quantities, our so-called features, and procedures applied to them that could have generated the full information in $\bm{D}$. To describe this problem, we use generative modeling to define a data model with corresponding parameters $\bm{\Theta}$. The solution to the inverse problem is then given by the posterior probability distribution $P(\bm{\Theta} \mid \bm{D})$ of the model parameters $\bm{\Theta}$, which we specify in the next paragraph. Using Bayes' theorem, this posterior can be expressed by the data likelihood and the prior distribution up to a $\bm{\Theta}$-independent factor, reading $P(\bm{\Theta} \mid \bm{D}) \propto P( \bm{D}\mid \bm{\Theta}) \times P(\bm{\Theta})$. We perform a pixel-based analysis of image data by assuming magnitude vectors of different pixels to be independent. This allows us to factorize the data likelihood and  prior distributions. In the following, we will calculate the respective distributions per pixel and finally combine all derivations to one solution for the entire data set $\bm{D}$. \\

We start by defining a generative process that leads from an abstract source $\bm{S}$ to the observations in the data set $\bm{D}$. We do not know $\bm{S}$ at this point, but we aim to associate $\bm{S}$ a posteriori with some relevant physical quantities. We assume that each $\bm{d}_i\in \bm{D}$ is generated from a source vector $\bm{s}_i \in \mathds{R}^l$ with $l \leq k$, and the collection of these vectors is building up a source set $\bm{S} = (\bm{s}_1, \ldots, \bm{s}_p)$. This relation is expressed in a data model 
\begin{flalign}
\label{eq:datamodel}
\bm{D} = \bm{\widetilde{D}} + \bm{N}_{\mathrm{model}},
\end{flalign}
where we define the observed data $\bm{D}$ to be composed of $\bm{\widetilde{D}}\coloneqq (\widetilde{\bm{d}}_1, \ldots, \widetilde{\bm{d}}_p)$, which is the output of a generative process $G: \mathds{R}^l \rightarrow \mathds{R}^k$, and some model noise $\bm{N}_{\mathrm{model}}$. The generative process $G(\, \cdot \,)$ mapping the variables of $\bm{S}$ pixel by pixel to $\bm{\widetilde{D}}$, i.e.  $\bm{\widetilde{D}} = G(\bm{S})$, is unknown.

\begin{table}
\caption[Priors on model parameters]{Priors on model parameters. $\bm{z_i}$ are the latent variables generating the data, $\xi_N$ is the transformed expression of the model noise covariance $N$, and $\bm{\theta}$ is the parameter vector of the generative forward model.}
\label{table:priors}
\centering
 \begin{tabular}{|l|l|}
  \hline
  \\[-1em]
  Model parameter & Prior distribution \\
  \\[-1em]
  \hline
  \\[-1em]
  Latent variable &  $P(\bm{z}_i) =  \mathcal{G}(\bm{z}_i, \mathds{1})$ \\
  \\[-1em]
  Latent noise parameter &  $P(\xi_N) =  \mathcal{G}(\xi_N, 1)$\\  
  \\[-1em]
   Generative model parameter &  $P(\bm{\theta}) =$ const.\\
 \hline  
  \end{tabular}
\end{table}

\subsection{Prior distributions}
\label{ch:prior}
We factorize the prior distribution of the set of source vectors $\bm{S}$ as $P(\bm{S}) = \prod^{p}_{i = 1} P(\bm{s}_i)$. The prior on $P(\bm{s}_i)$ can be arbitrarily complex, but without loss of generality we can find a transformation $\bm{s}_i = T(\bm{z}_i)$ of the source vectors to a set of latent vectors $\bm{z}_i \in \mathds{R}^l$ that provides a standardized prior distribution $P(\bm{z}_i) =  \mathcal{G}(\bm{z}_i, \mathds{1})$. This coordinate transformation into the eigenspace of the prior, also known as random variate generation using inverse transform sampling \citep[e.g.,][]{devroye1986non}, or reparametrization trick \citep{ kingma2013auto, rezende2014stochastic, titsias2014doubly}, allows us to absorb all complex and unknown structure of the source space in the transformation $T(\, \cdot \,)$, and provides us with an easy to calculate unit Gaussian prior distribution.
Using this definition, we can rewrite the generative process as a parametrized function of the latent variables $\widetilde{\bm{d}_i} = G(s_i) = G(T(\bm{z}_i)) \eqqcolon f_{\bm{\theta}}(\bm{z}_i)$, with $\bm{\theta}$ denoting the parameters of the transformed generative forward model. \\

We further define the model noise $\bm{N}_{\mathrm{model}} = (\bm{n}_1, \ldots, \bm{n}_p)$ as a set of $p$ noise vectors $\bm{n}_i \in \mathds{R}^k$ and assume the pixel-wise noise is independent and identically distributed with a Gaussian distribution of zero mean and noise covariance $N \in \mathds{R}^{k \times k}$, meaning $P(\bm{N}_{\mathrm{model}}) = \prod^{p}_{i = 1} P(\bm{n}_i) = \prod^{p}_{i = 1}\mathcal{G}(\bm{n}_i, N)$. Later in the discussion we will see that the noise covariance induces a metric in data space, which is between two magnitude vectors $\bm{d}_i$ and $\widetilde{\bm{d}_i}$. $N$ thus indicates how accurately the data set $\bm{d}_i$ can be reconstructed using the latent variables $\bm{z}_i $. Since this accuracy is unknown, we introduce $N$ as an inference parameter. We again perform a transformation to the prior eigenspace of the form $N = t_{\bm{\psi}}(\xi_N)$ with the latent noise parameter $\xi_N \in \mathds{R}$, being distributed as $P(\xi_N) = \mathcal{G}(\xi_N, 1)$ with the transformation parameter $\bm{\psi}$. For our specific application, we choose a log-normal mapping 

\begin{flalign}
\label{eq:noisetrafo}
N = t_{\bm{\psi}}(\xi_N) = \mathds{1}^{k \times k} \, \, \mathrm{exp} \, (\mu_N + \sigma_N \, \xi_N )
\end{flalign}

with mean $\mu_N$ and standard deviation $\sigma_N$. In other words, this is a diagonal noise covariance matrix $N$ with transformation parameters $\bm{\psi}=(\mu_N, \sigma_N)^T$ and a single global noise parameter $\xi_N $, which we aim to learn.  \\

Combining the definitions of the noise and the generative model, we can rewrite the data model from Eq. \eqref{eq:datamodel} as a pixel-wise expression:
\begin{flalign}
\label{eq:pixdatamodel}
\bm{d}_i = f_{\theta}(\bm{z}_i) + \bm{n}_i .
\end{flalign}

At this point, we can define the data model parameter vector $\bm{\Theta}$ described in Sect. \ref{ch:model}. It consists of the latent vectors $\bm{z_i}$,  the latent noise parameter $\xi_N$, which indirectly defines the model noise $\bm{n_i}$, and the parameters of the generative model $\bm{\theta}$. The parameters $\bm{\theta}$ are the parameters of the neural network used to approximate the generative process, and are specified in further detail in Sect. \ref{ch:NEAT_VAE}. Since we do not have prior knowledge on these forward model parameters, we assume a uniform prior distribution on $\bm{\theta}$ (see Table \ref{table:priors}). Thus we have $\bm{\Theta} = (\bm{Z}, \bm{\theta}, \xi_N)$ with the latent space set $\bm{Z}=(\bm{z}_1, \ldots, \bm{z}_p )$.

\subsection{Data likelihood}
We can include our forward model in the pixel-wise likelihood  $P( \bm{D}\mid \bm{\Theta}) = \prod_{i=1}^p  \, P(\bm{d}_i \mid \bm{Z}, \bm{\theta}, \xi_N)$ by marginalizing over the model noise: \\
\begin{flalign}
\label{eq:likelihood}
P(\bm{d}_i \mid \bm{Z}, \bm{\theta}, \xi_N) &= \int d\bm{n}_i  \; P(\bm{d}_i , \bm{n_i}\mid \bm{Z}, \bm{\theta}, \xi_N) \nonumber \\ 
&=\int d\bm{n}_i P(\bm{d}_i \mid \bm{z}_i, \bm{\theta}, \bm{n}_i) \, P(\bm{n}_i \mid \xi_N) \nonumber \\
&= \int d\bm{n}_i \, \delta(\bm{d}_i - f_{\bm{\theta}}(\bm{z}_i) - \bm{n}_i) \, \mathcal{G}(\bm{n}_i, t_{\bm{\psi}}(\xi_N)) \nonumber \\
&= \mathcal{G}(\bm{d}_i - f_{\bm{\theta}}(\bm{z}_i), t_{\bm{\psi}}(\xi_N)) \, ,
\end{flalign}
where we assumed the noise $\bm{n}_i$ to be a priori independent of $\bm{Z}$ and $\bm{\theta}$. 

\subsection{Approximating the posterior distribution}
\label{ch:approxpost}
Combining the prior distributions listed in Table~\ref{table:priors} and the data likelihood in Eq. \eqref{eq:likelihood}, the posterior distribution $P(\bm{\Theta} \mid \bm{D}) $ reads:
\begin{flalign}
P(\bm{Z}, \bm{\theta}, \xi_N\mid\bm{D} ) &\propto \prod_{i=1}^p P(\bm{d}_i \mid \bm{Z}, \bm{\theta}, \xi_N) \, \prod_{i=1}^p P(\bm{z}_i) \,  \,P(\bm{\theta}) \, P(\xi_N) \nonumber \\ 
&\propto \prod_{i=1}^p \left( \mathcal{G}(\bm{d}_i - f_{\bm{\theta}}(\bm{z}_i), t_{\bm{\psi}}(\xi_N)) \, \mathcal{G}(\bm{z}_i, \mathds{1}) \right) \, \times \nonumber \\
\nonumber \\[-1em] 
&\quad \, \, \,  \mathcal{G}(\xi_N, 1)  \, , 
\end{flalign}

where we used $P(\bm{\theta}) =$ const. We are not able to calculate expectation values from this high-dimensional probability distribution, but we can use variational inference \citep[e.g.,][]{Blei_2017} to approximate the posterior distribution $P(\bm{\Theta} \mid \,\bm{D} )$ with an easier to integrate distribution $Q_{\bm{\Phi}}(\bm{\Theta}\mid \bm{D})$ with variational parameters $\bm{\Phi}$. In the following, we define a suitable approximate distribution $Q_{\bm{\Phi}}(\bm{\Theta}\mid \bm{D})$ and use the Kullback-Leibler Divergence\footnote{We consider the spurious amount of artificial information introduced by $Q$ when we calculate $D_{KL} [\, Q(\cdot) \, \mid\mid \, P(\cdot) \,]$, while $D_{KL} [\, P(\cdot) \, \mid\mid \, Q(\cdot) \,]$ expresses the loss of information when using $Q$ instead of $P$ \citep{leike2017optimal}. The latter quantity is what one ideally would like to base approximate (conservative) inference on, but unfortunately it cannot be calculated with an intractable posterior $P$. For this reason, we use the variational inference approach and minimize the former KL-Divergence $D_{KL} [\, Q(\cdot) \, \mid\mid \, P(\cdot) \,]$, that is minimizing the amount of spurious information added when going from $P$ to $Q$.} as a measure to evaluate the dissimilarity of $P(\bm{\Theta} \mid \,\bm{D} )$ and $Q_{\bm{\Phi}}(\bm{\Theta}\mid \bm{D})$. \\

Assuming that $\bm{Z}$, $\bm{\theta}$ and $\xi_N$ are a posteriori independent, the approximate distribution $Q_{\bm{\Phi}}$ can be written as
\begin{flalign}
Q_{\bm{\Phi}}(\bm{Z}, \bm{\theta}, \xi_N \mid \bm{D}) &= Q_{\bm{\Phi}}(\bm{Z} \mid \bm{D}) \, Q_{\bm{\Phi}}(  \bm{\theta}\mid \bm{D}) \, Q_{\bm{\Phi}}(\xi_N\mid \bm{D}) \, .
\end{flalign}
For  $Q_{\bm{\Phi}}(\bm{Z} \mid \bm{D}) \,  = \, \prod_{i=1}^{p} \, Q_{\bm{\Phi}}(\bm{z}_i\mid \bm{d}_i)$  we choose a pixel-wise independent Gaussian distribution \mbox{$\mathcal{G}(\bm{z}_i - \bm{\mu}_i, \Sigma_i)$} based on the maximum entropy principle \citep{jaynes1982rationale}. The principle states that if only the mean $\bm{\mu}$ and covariance $\Sigma$ of some data are known, then the knowledge contained in that data set can best be expressed by a Gaussian distribution with exactly this mean $\bm{\mu}$ and covariance $\Sigma$ \citep[e.g.,][]{ensslin2019information}. Later we will see that by construction $\bm{\mu}_i$ and $\Sigma_i$ are functions of the input data, making the choice of a Gaussian distribution for $Q_{\bm{\Phi}}(\bm{Z} \mid \bm{D})$ valid. For $Q_{\bm{\Phi}}(\bm{\theta} \mid \bm{D}) $ and $Q_{\bm{\Phi}}(\xi_N \mid \bm{D})$ we choose maximum a posteriori solutions. In practice we evaluate the approximation at the variational parameter values $\{\widehat{\bm{\theta}}, \widehat{\xi_N} \} \in \bm{\Phi}$, expressed by $Q_{\bm{\Phi}}(\bm{\theta} \mid \bm{D}) = \delta(\bm{\theta} - \widehat{\bm{\theta}})$ and $Q_{\bm{\Phi}}(\xi_N \mid \bm{D}) = \delta(\xi_N - \widehat{\xi_N})$. \\

To approximate $P(\bm{\Theta} \mid \,\bm{D} )$ with $Q_{\bm{\Phi}}(\bm{\Theta}\mid \bm{D})$, we use the Kullback-Leibler Divergence 
\begin{flalign}
\label{eq:FullKLD}
&D_{KL} [\, Q_{\bm{\Phi}}(\bm{Z}, \bm{\theta},  \xi_N \mid \bm{D}) \, \mid\mid \, P(\bm{Z}, \bm{\theta},  \xi_N \mid \bm{D}) \,] =  \nonumber \\
&= \int d\bm{Z} \, d\bm{\theta} \, d \xi_N \, Q_{\bm{\Phi}}(\bm{Z}, \bm{\theta},  \xi_N \mid \bm{D}) \, \ln \Bigg( \frac{Q_{\bm{\Phi}}(\bm{Z}, \bm{\theta},  \xi_N \mid \bm{D})}{P(\bm{Z}, \bm{\theta},  \xi_N \mid \bm{D})}\Bigg).   & 
\end{flalign}
Inserting the derived expressions from the previous sections and using Monte Carlo Methods to approximate integrals with finite sums, we arrive at 
\begin{myframedeq}
\begin{flalign}
\label{eq:KLD}
 D_{KL}[Q_{\bm{\Phi}}(\cdot) \mid\mid P(\cdot)] &= \frac{1}{2} \sum_{i=1}^p \Bigg[ - \mathrm{tr} \,(\ln \Sigma_i)  - l \left( 1+ \mathrm{ln}\,(2\pi) \right) +  \frac{1}{p} \, \widehat{ \xi_N}^2  \nonumber \\
 & \quad +\, \mathrm{tr}\, \left( \frac{1}{ t_{\bm{\psi}}(\widehat{ \xi_N})} \left(\bm{d}_i - f_{\widehat{\bm{\theta}}}(\bm{z}_i)\right) \left(\bm{d}_i - f_{\widehat{\bm{\theta}}}(\bm{z}_i)\right)^T \right) \nonumber \\
 & \quad  + \, \mathrm{tr} \, \left( \Sigma_i + \bm{\mu}_i \bm{\mu}_i^T \right)  + \mathrm{tr}\,\left(\ln \, t_{\bm{\psi}}(\widehat{ \xi_N})\right) \Bigg] + \mathcal{H}_0,  
\end{flalign}
\end{myframedeq}
where we absorb all constant terms into $\mathcal{H}_0$. The full calculations with all intermediate steps are carried out in Appendix~\ref{app:calculations}. 

\subsection{NEAT-VAE}
\label{ch:NEAT_VAE}
Our goal is to infer a lower-dimensional representation of the data, which in our case is expressed by the latent source space $\bm{Z}$. We can achieve this by minimizing Eq. \eqref{eq:KLD}, which describes the spurious amount of artificial information introduced by the approximation of $P$ with $Q_{\bm{\Phi}}$. In the following, we will explain how this objective function can be translated into the framework of a latent variable model called variational autoencoder \citep{kingma2013auto}. \\

A basic autoencoder learns a low-dimensional, latent representation of higher-dimensional input data. This is achieved by training the autoencoder to reconstruct the original input as accurately as possible from the reduced, latent representation \citep[e.g.,][]{rumelhart1985learning, hinton2006, goodfellow2016deep}. In this context, training describes the minimization of an objective or loss function, for example the mean squared error between input data and the reconstructed output data. The dimensionality reduction of the input to the latent space occurs in the so-called encoder, the latent space itself is called bottleneck layer, and in the decoder, the latent space gets translated back to data space. Variational autoencoders (VAEs) offer a probabilistic framework to jointly optimize latent variable (or generative) models and inference models \citep{kingma2019introduction}. Eq. \eqref{eq:KLD} can be transformed to such a VAE framework (see Fig.~\ref{fig:NEAT_VAE}) as follows:
\begin{itemize}
\item \textbf{Generative Process}: Neural networks are generalized function approximators. Since the exact form of the data generating process $f_{\bm{\theta}}(\bm{z_i}) = \widetilde{\bm{d}}_i$ is unknown, we can use a neural network with input $\bm{z_i}$ and output $\widetilde{\bm{d}}_i$ to approximate $f_{\bm{\theta}}$ and use back propagation to optimize the parameters $\bm{\theta}$ of the network, which corresponds to the generative decoder. 
\item \textbf{Variational Inference}: The inference of the latent space variables is approximated by $Q_{\bm{\Phi}}(\bm{z}_i\, \mid \bm{d}_i)= \mathcal{G}(\bm{z}_i - \bm{\mu}_i, \Sigma_i)$. We can build a function delivering posterior samples following this variational distribution, expressed by a neural network with input $\bm{d}_i$ and output $\bm{z}_i$. Let the pixel-wise mean $\bm{\mu}_i \in \mathds{R}^l$ and covariance $\Sigma_i \in \mathds{R}^{l \times l}$ be determined by a parametrized function of the input data $e_{\bm{\phi}}(\bm{d}_i) = \left(\bm{\mu}_i, \mathrm{log}\,\Sigma_i\right)$, where $\bm{\phi}$ contains the parameters of the function $e_{\bm{\phi}}$ and the matrix logarithm of $\Sigma_i$ is calculated. This parametrization ensures the variance to be positive and the calculation to be numerically stable, since the logarithmic function maps the small values of the variance to a larger space. By inverse transform sampling, we can then define the posterior latent space variables as $\bm{z}_i = \bm{\mu}_i + \mathrm{exp}\left(\frac{1}{2} \mathrm{log}\,\left(\Sigma_i \right)\right) \cdot \bm{\epsilon}_i = \bm{\mu}_i + \sqrt{\Sigma_i} \, \bm{\epsilon}_i$ with an auxiliary variable $\bm{\epsilon}_i$ and $P(\bm{\epsilon}_i) = \mathcal{G}(\bm{\epsilon}_i, \mathds{1})$. In practice we approximate the function $e_{\bm{\phi}}(\bm{d}_i)$ by the variational inference encoder. We take $\sqrt{\Sigma_i}$ to be diagonal and describe it by its diagonal vector $\mathrm{diag}\left(\sqrt{\Sigma_i}\right)$, allowing us to calculate $\bm{\mu}_i$ and $\mathrm{diag}\left(\sqrt{\Sigma_i}\right)$ as two distinct outputs of the encoder network.

\item \textbf{Independent representation}: Based on the input data, the encoder network delivers latent space variables $\bm{z_i}$ with mean $\bm{\mu}_i$ and covariance vector $\widetilde{\Sigma}_i = \mathrm{diag}\left(\sqrt{\Sigma_i} \, \sqrt{\Sigma_i}^{\mathrm{\, T}}\right)$. By using this definition we find the optimal independent approximation to the posterior $P(\bm{\Theta} \mid \bm{D}) $. This leads to a disentanglement of the input information, meaning each dimension of $\bm{z_i}$, which is equivalent to the hidden neurons in the bottleneck layer, encodes (approximately) mutually independent features of the data. 

\end{itemize}
The minimization objective in Eq. \eqref{eq:KLD} contains the loss function of a classic VAE with three modifications: (1) We include the size of the latent space $l$ and its corresponding weight (see Appendix~\ref{app:calculations} for the derivation) to be able to compare different latent spaces with each other, (2) besides the network weights, we aim to optimize the noise covariance and thus the latent noise variable $\xi_N$. The prior on this latent noise adds an extra term proportional to $\widehat{\xi_N}^{ 2}$ to the objective function, and (3) by including noise in the data model, the likelihood contains a factor $1/ t(\widehat{\xi_N})$ and contributes an additional term $\mathrm{tr}\left(\ln \, t(\widehat{\xi_N})\right)$ from the normalization. Since we expand the VAE framework by the adaption to noise, we name our algorithm NEAT-VAE (NoisE AdapTing Variational AutoEncoder). \\
Using the transformations illustrated before, the final objective function depends only on the generative decoder parameters $\widehat{\bm{\theta}}$, the variational encoder parameters $\bm{\phi}$ and the latent noise parameters $\widehat{\xi_N}$. We implement an autoencoder architecture (encoder network, bottleneck layer, decoder network) with Eq. \eqref{eq:KLD} as the respective loss function in the PyTorch framework\footnote{https://pytorch.org/}, which is publicly available at \url{https://gitlab.mpcdf.mpg.de/msara/neat_vae}. The framework allows us to calculate derivatives of Eq. \eqref{eq:KLD} with respect to $\widehat{\bm{\theta}}$, $\bm{\phi}$ and $\widehat{\xi_N}$ using automated back propagation, and to minimize the loss with a build-in optimizer. The conducted experiments are described in Appendix~\ref{app:hyperparameters}.

\section{Results and discussion}
\label{ch:results}
Applied to our set of Galactic full-sky observations, the NEAT-VAE framework yields a posterior probability distribution of the latent space variables that capture the essential information in our data. The derived loss function forces the latent variables to be statistically independent of each other, and thereby to represent individual physical components. We obtained the posterior by simultaneously optimizing two processes: an inference mechanism reducing the observed data $\bm{D}$ to a lower-dimensional latent space $\bm{Z}$, and a generative process mapping these latent space variables back to the higher-dimensional data space $\widetilde{\bm{D}}$. These two processes, described by artificial encoder and decoder networks, support each other during training: the decoder reconstructs the data space based on the latent variables the encoder delivers. This reconstruction is constantly compared to the input data by the likelihood term in Eq. \eqref{eq:KLD}, ensuring the encoder to adapt the inference function and thus to provide improved latent space variables to the decoder. In addition, we aim to find latent space variables that encode mutually independent features of the input data. In the following, we analyze the resulting latent space variables $\bm{z}_i \in \bm{Z}$  and their full-sky representations. The dimension of the latent space $\bm{Z}$ corresponds to the number of hidden neurons in the bottleneck layer. From here on, we call the subsequent hidden neurons `features' and the resulting full-sky representations `feature maps'. 

\begin{figure}[t!]
\centering
\includegraphics[width=0.5\textwidth]{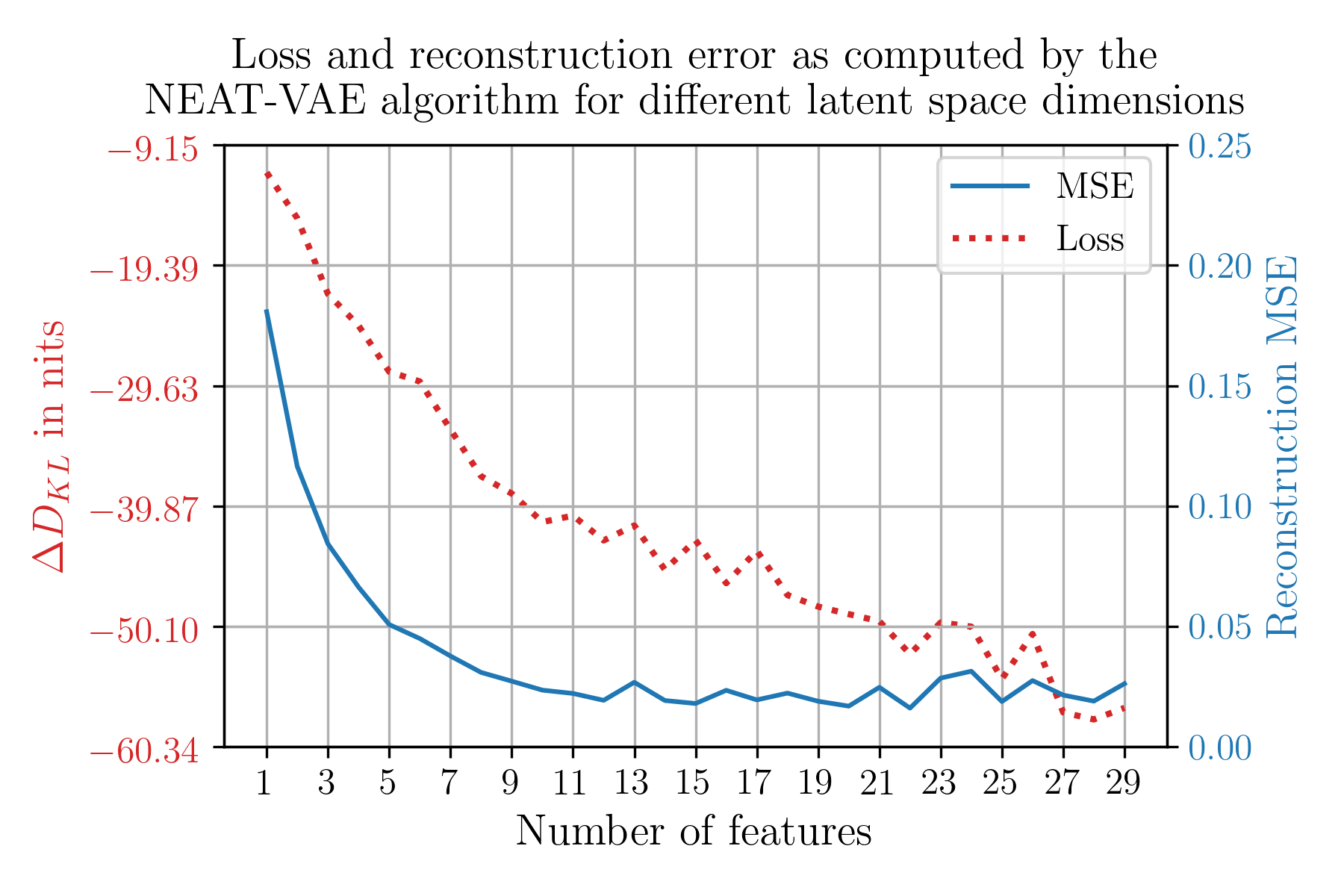}
\vspace*{-0.5em}
\caption[Reconstruction mean squared error (MSE) and loss for different latent space dimensions]{Reconstruction mean squared error (MSE) and values of the minimized Kullback-Leibler Divergence (Loss) $\Delta D_{KL}$ ($D_{KL}$ in Eq. \eqref{eq:KLD} except constant terms) depending on the dimension of the latent space (x-axis). The values are determined by the NEAT-VAE with a configuration of six layers, $30$ hidden neurons in the encoder and decoder layers, noise transformation parameters $\mu_N = -7$ and $\sigma_N = 1$, learning rates of $0.005$ for the network weights and $0.001$ for $\xi_N$, and a batchsize of $128$. We do not track all normalization constants through the calculations, which leads to negative values for the loss.}
\label{fig:RECLoss}
\end{figure}

\subsection{Dimensionality reduction}
We first analyze how the dimension of the latent space in the NEAT-VAE correlates with the reconstruction quality of the generative process. We quantify the reconstruction by the mean squared error of input maps and reconstructed maps  
\begin{equation}
\mathrm{MSE}\,(\bm{d}, \widetilde{\bm{d}} ) = \frac{1}{p} \, \sum_{i=1}^p \, (\bm{d}_i - \widetilde{\bm{d}}_i )^2,
\end{equation} 
with the number of pixels $p$, the $35$-dimensional data vectors $\bm{d}_i \in \bm{D}$ and the corresponding reconstruction vectors $\widetilde{\bm{d}}_i \in \widetilde{\bm{D}}$. We observe that only three features are required to achieve an MSE of below $0.1$, which describes an average deviation of the reconstructed values compared to the input values of a natural logarithm-based flux magnitude. For small values, the absolute uncertainty on logarithmic scale equals the relative uncertainty on linear scale, meaning MSE $=0.1$ corresponds to a relative uncertainty of $\approx10 \%$. The MSE decreases for a further growing number of features, and stagnates around a value of $0.02$, that is  a relative uncertainty of $\approx2 \%$ at approximately ten features. We interpret this as an indication for a high redundancy of the information contained in the $35$ Galactic all-sky maps, since increasing the number of features that are able to encode the input data beyond this point do not increase the quality of the reconstruction any further. \\

\subsection{Morphology of features}
\label{ch:morphology}
Based on the experiments with different latent space dimensions, we examine the spatial structures in the feature maps in more detail. We recognize some spatially correlated structures of the input data in the feature maps. We assume this to be a meaningful result, since the autoencoder only learns correlations among the magnitude flux values $\bm{d}_i$ within one data pixel and is not informed about spatial structures. We also observe feature maps with the same morphology to occur in latent spaces of different dimensions. To investigate whether there is a pattern or an order among the feature maps with respect to information content, we calculate the significance of each feature for the reconstruction of the data by using the following measure
\begin{equation}
\label{eq:Sign}
S_{\mathrm{feature}} = \frac{1}{p} \, \sum_{i=1}^p \,  \frac{ \left( z_{\mathrm{feature \, map}, \, i} - \overline{z}_{\mathrm{feature \, map}} \right)^2}{ \sigma_{\mathrm{feature}, \, i}^2},
\end{equation}
where $z_{\mathrm{feature \, map}, \, i}$ is the intensity value at the $i^{\mathrm{th}}$ pixel, $\overline{z}_{\mathrm{feature \, map}}$ is the mean intensity value of a single feature map, and $\sigma_{\mathrm{feature}, \, i}^2$ denotes the posterior variance as calculated by the algorithm at the $i^{\mathrm{th}}$ pixel. This means we calculate the feature map variance, which describes the fluctuations within the posterior mean map, weighted by the posterior feature variance averaged over all pixels $p$. The significance thus expresses the ratio of the magnitude of fluctuations within a map compared to the uncertainties of the map. In this context, a high significance marks the features that the autoencoder is most certain about to be required for the reconstruction, while a feature significance below $1$ corresponds to an insignificant feature, as the posterior uncertainty is larger compared to the posterior mean values of the map. Averaged over all the experiments we conducted (see Appendix \ref{app:hyperparameters}), we call the three most significant features A, B and C. \\

\begin{figure*}[t!]\captionsetup[subfigure]{singlelinecheck=false}
\begin{subfigure}[t]{\textwidth}
  \centering
  \caption{\hspace{3.25cm} Posterior mean \hspace{2.75cm}  Feature A \hspace{2.6cm} Posterior variance}
   \vspace{-1.3\baselineskip}
  \includegraphics[width=0.49\linewidth]{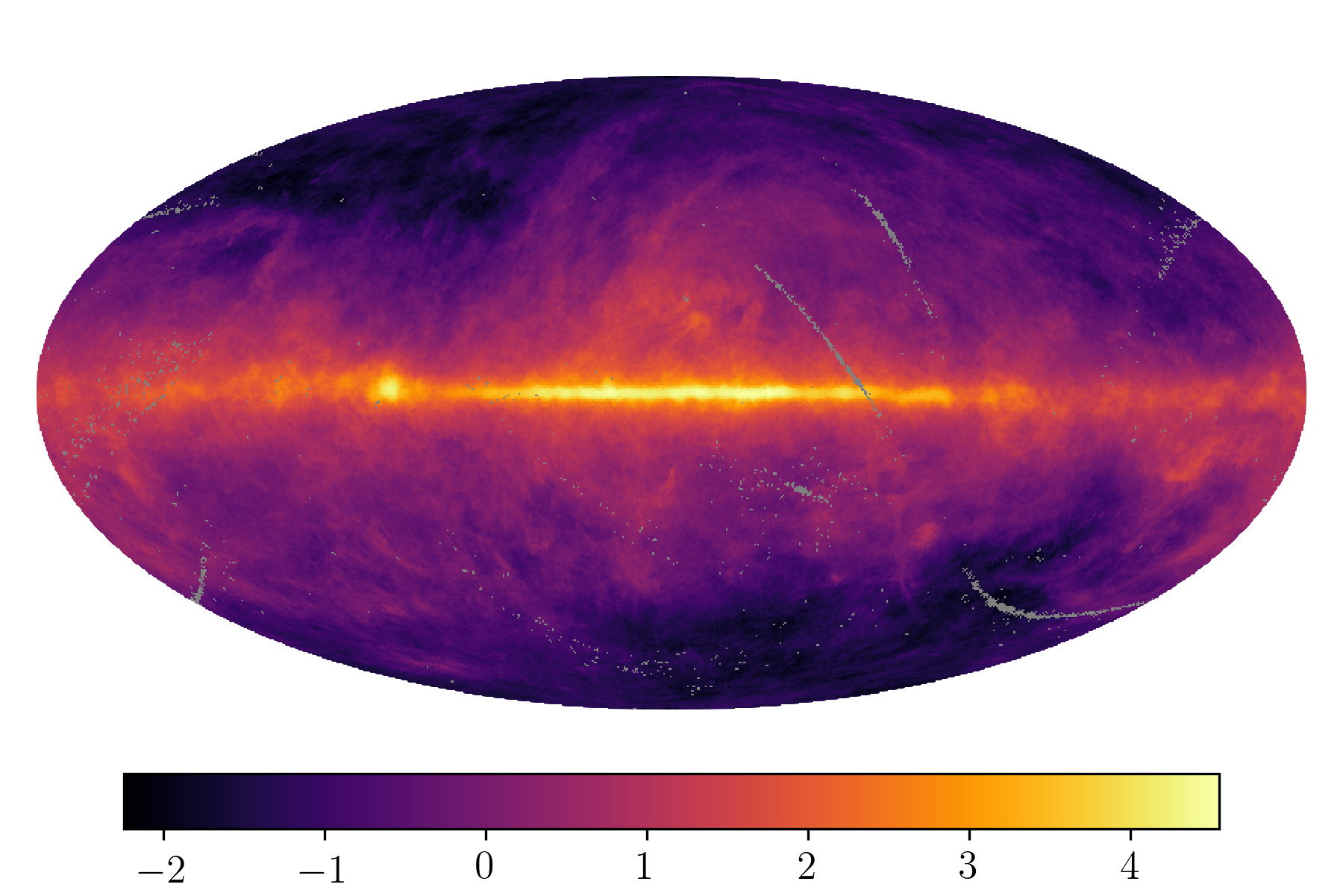}  
  \includegraphics[width=0.49\linewidth]{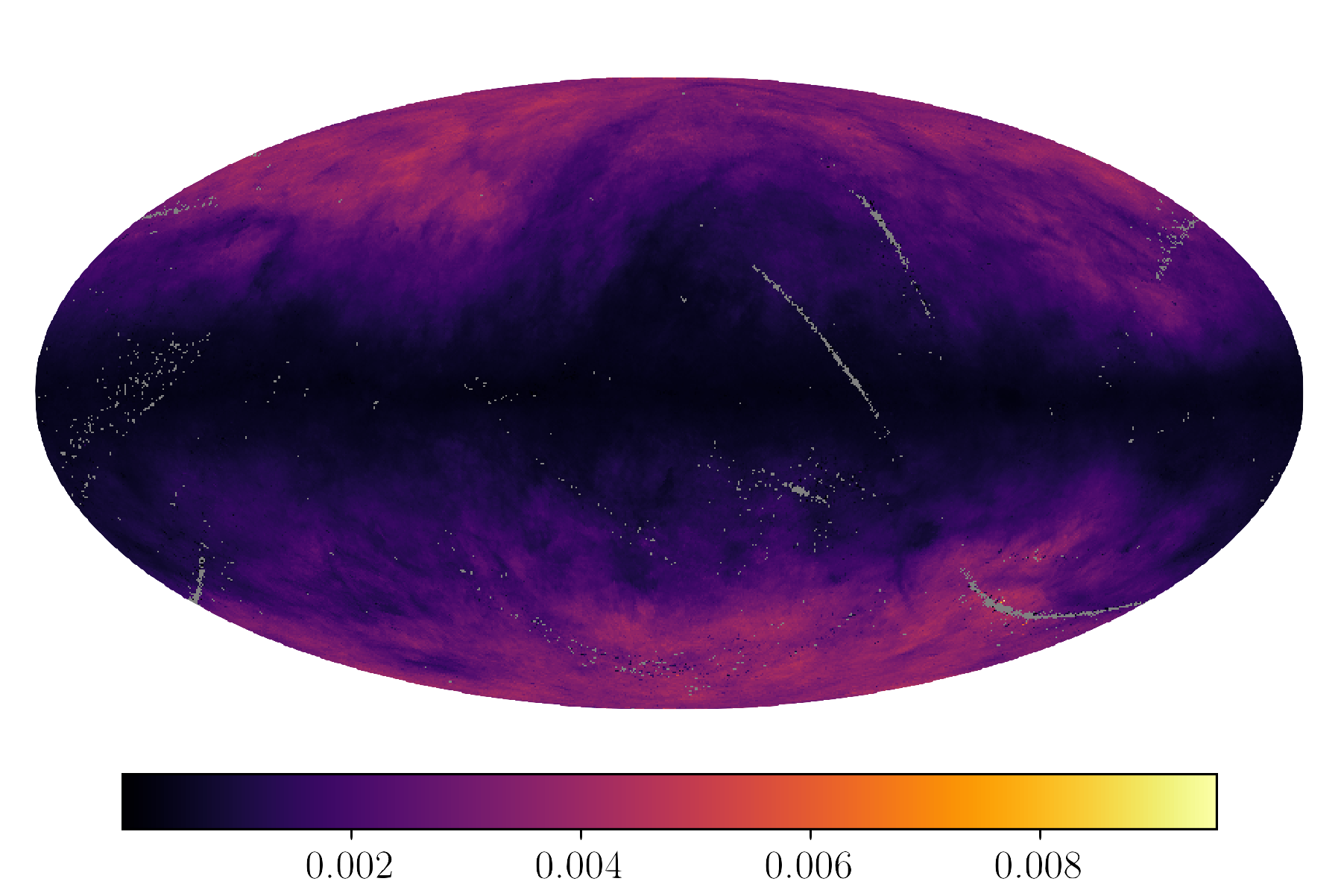}  
  \label{fig:FeatA}
\end{subfigure}
\par\bigskip 
\begin{subfigure}[t]{\textwidth}
  \centering
   \caption{\hspace{3.25cm} Posterior mean \hspace{2.75cm}  Feature B \hspace{2.6cm} Posterior variance} 
   \vspace{-1.3\baselineskip}
   \includegraphics[width=0.49\linewidth]{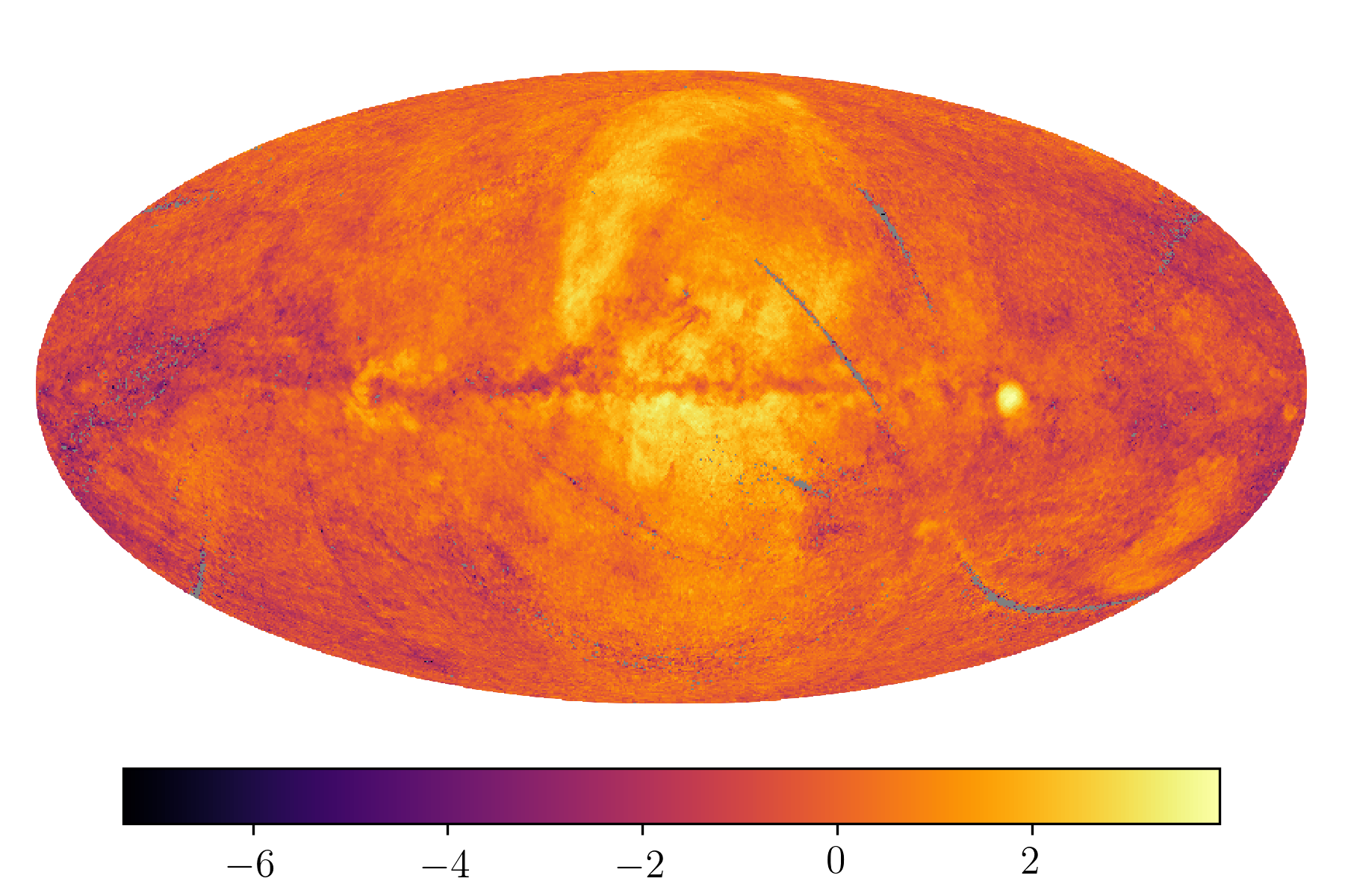}  
  \includegraphics[width=0.49\linewidth]{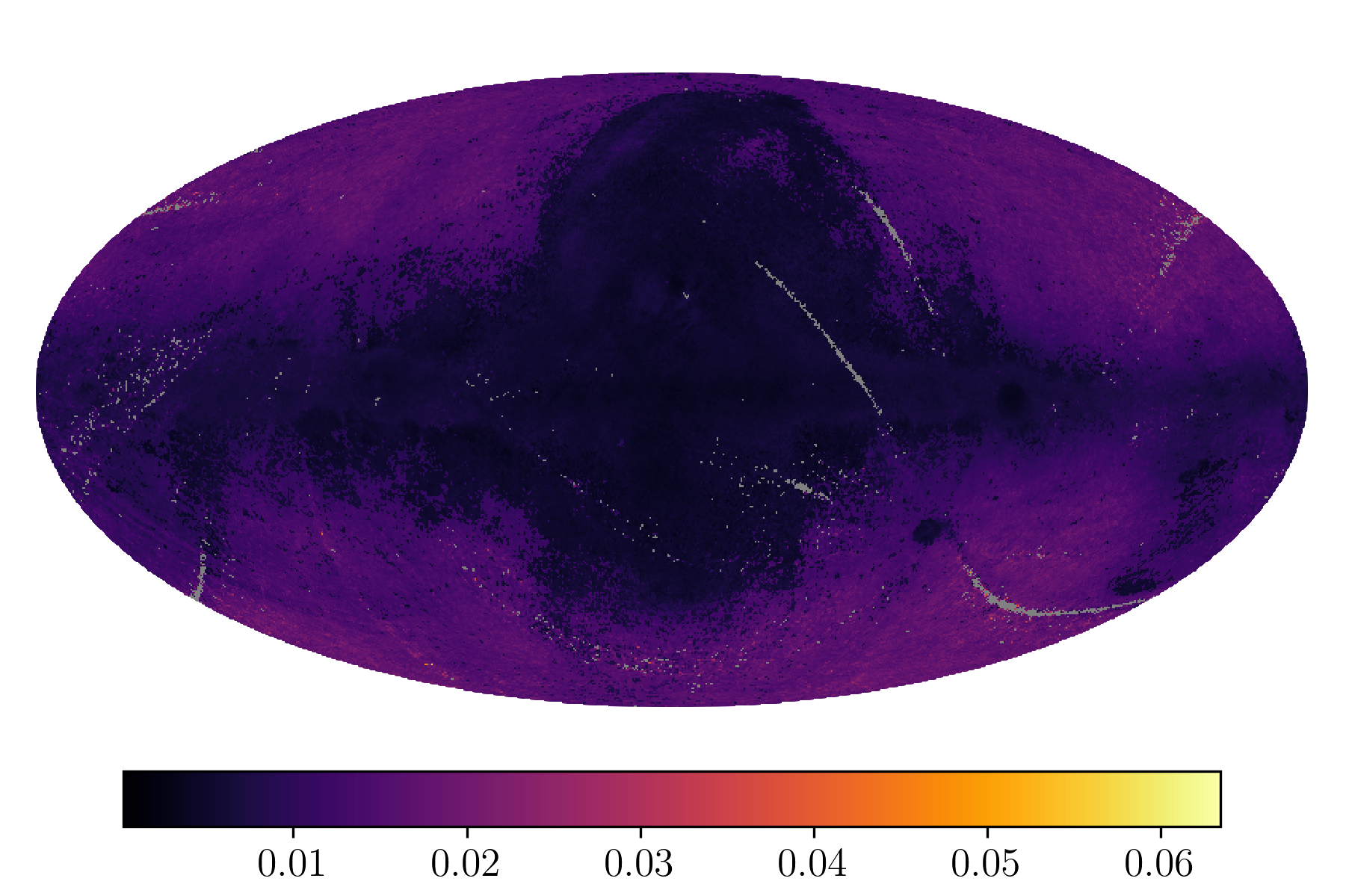}  
  \label{fig:FeatB}
\end{subfigure}
\par\bigskip 
\begin{subfigure}[t]{\textwidth}
  \centering
   \caption{\hspace{3.25cm} Posterior mean \hspace{2.75cm}  Feature C \hspace{2.6cm} Posterior variance }
    \vspace{-1.3\baselineskip}
   \includegraphics[width=0.49\linewidth]{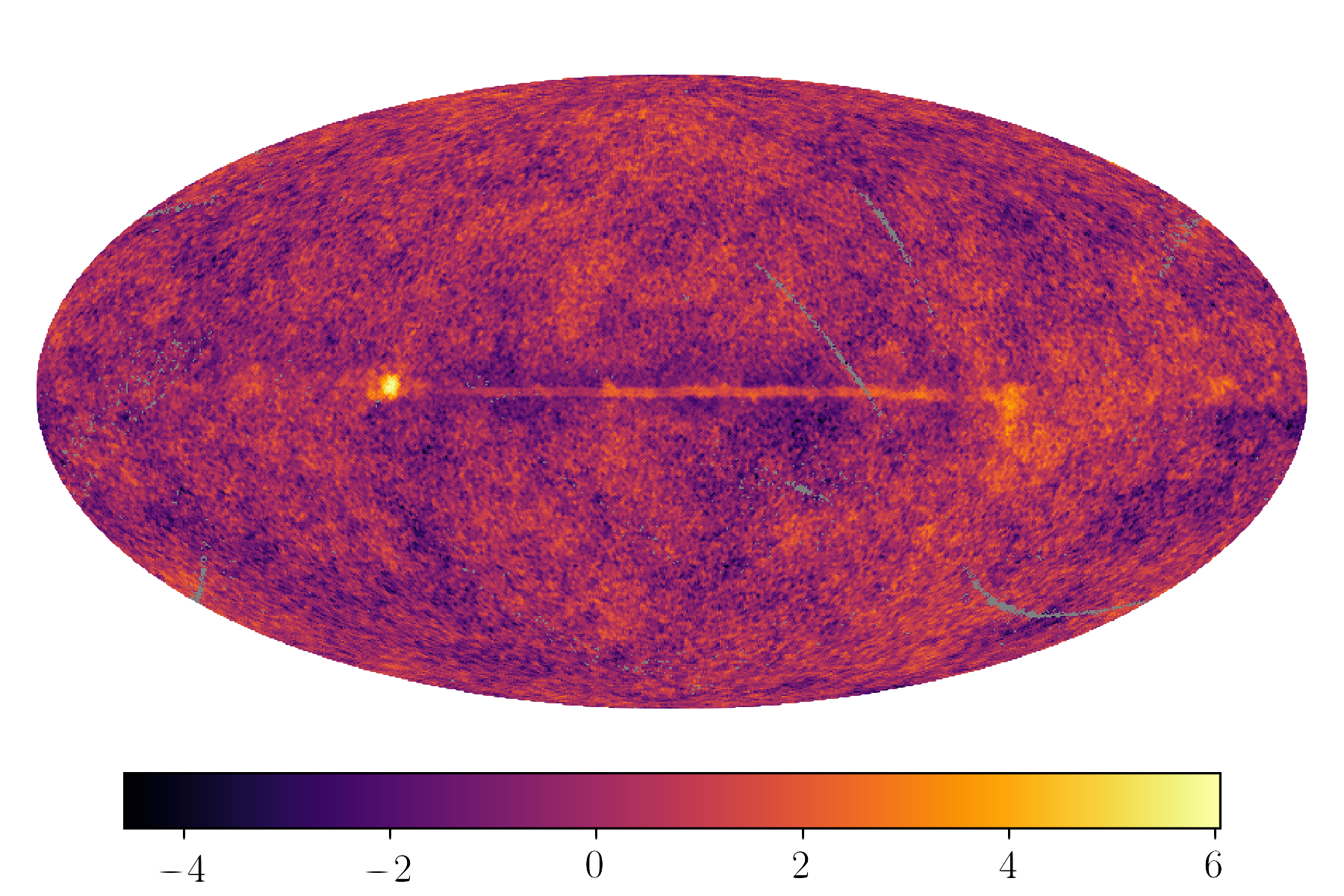}  
  \includegraphics[width=0.49\linewidth]{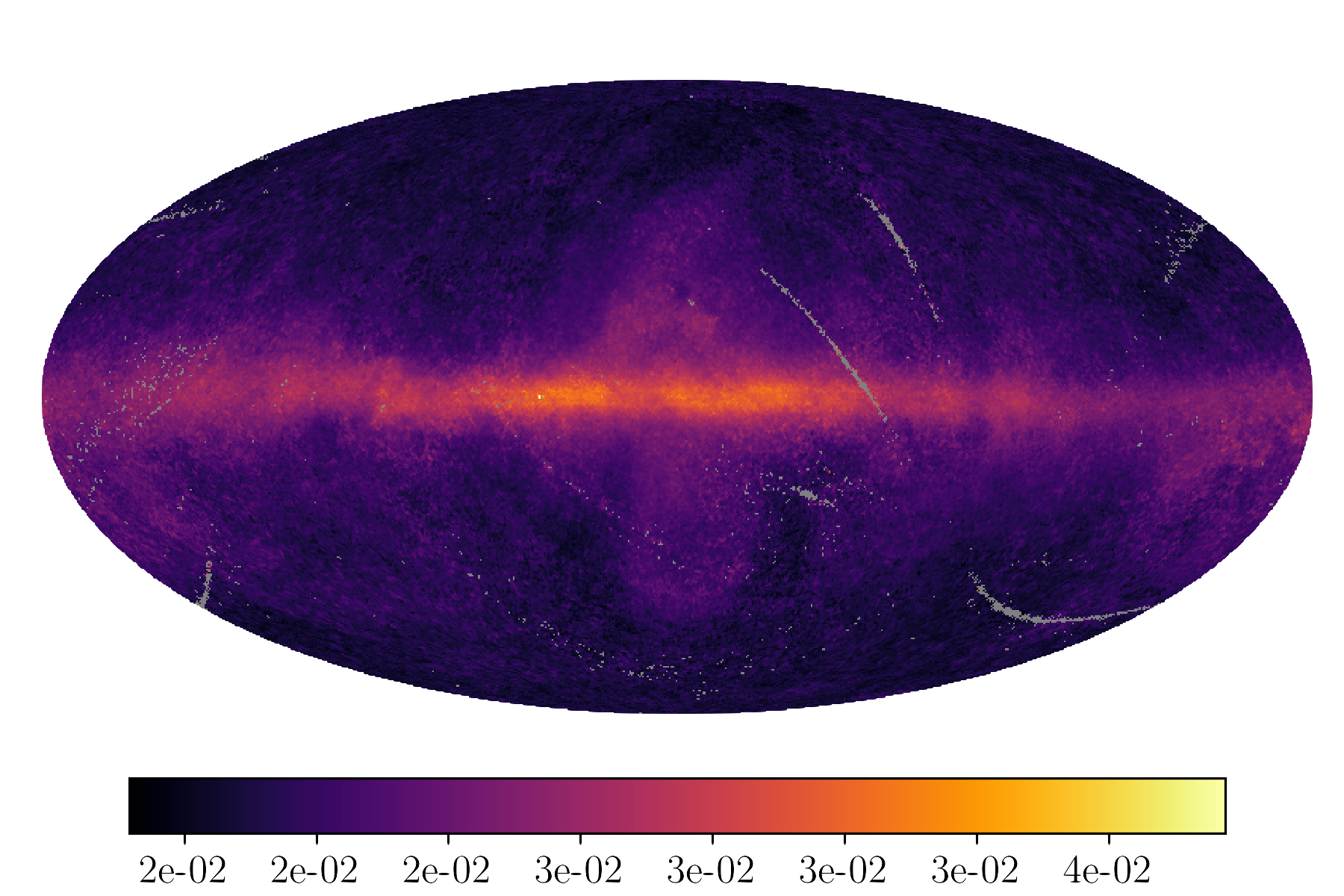}  
  \label{fig:FeatC}
\end{subfigure}
\caption[All-sky projection of main features]{Most essential components of the Galactic emission data in HEALPix pixelization. Left panels show the posterior mean and right panels show the posterior variance of the three most significant hidden neurons (also called features) in the latent space of the NEAT-VAE. In this case, the algorithm was trained to reduce its input of 35 Galactic all-sky maps to ten features in the latent space with the configuration described in Fig.~\ref{fig:RECLoss}. The mean feature maps (left panels) are displayed in order of significance according to Eq. \eqref{eq:Sign}, meaning these features show the highest ratio of feature fluctuations to feature uncertainties within the latent space (see Sect. \ref{ch:morphology} for details).  The colors in the posterior mean of feature C (left panel in (c)) are inverted for illustration purposes. The grey pixels correspond to missing values in the input data.}
\label{fig:SignFeat}
\end{figure*}

For the configuration shown in Fig.~\ref{fig:SignFeat}, the features have significance values $S_{\mathrm{featureA}}=1.67 \times 10^3$, $S_{\mathrm{featureB}}=1.41 \times 10^2$, and $S_{\mathrm{featureC}}=4.68 \times 10^1$. The remaining features have significance values ranging from $8.75$ to $3.82 \times 10^1$ and encode artifacts and other morphologies of the input data, as displayed in Appendix \ref{app:insignificantfeat}. A similar behavior was also observed by \citet{muller2018}: The authors build a Gaussian mixture model (GMM) to reconstruct observations from a certain frequency range based on data of complementary frequencies. The GMM components used in their study also encoded artifacts of the input data when the number of components was increased. 
However, posterior samples of the latent space of our algorithm show that the NEAT-VAE does not always assign information to each and every feature: Starting from twelve features and adding further neurons to the latent space, the significance of the added features drops below one. This means that the posterior variance of a feature map is greater than the fluctuations within the map, and the resulting posterior samples show white noise statistics. On average, ten to eleven features are significant throughout our experiments with varying latent space size. We assume that from this point on, our algorithm has identified all mutually independent features of the input data. When further dimensions are added to the latent space, the algorithm `tunes out' those degrees of freedom by making them insignificant. 

In the next sections, we focus on the three most significant features of the configuration with in total ten latent space features (since from this point on the reconstruction does not change significantly, see Fig.~\ref{fig:RECLoss}), and their physical interpretation. Visual inspections of the other features (Appendix \ref{app:insignificantfeat}) indicate that the separation into independent components is not fully reached, possibly due to the finite amount of optimization. Morphological structures similar to the North Galactic Spur or the Fermi bubbles imprint onto several features simultaneously, as well as measurement artifacts. \\

The posterior mean values in Fig.~\ref{fig:SignFeat} reflect the internal representation of the three most significant latent space features by our algorithm. We observe that the overall sign of the maps changes for varying hyperparameter configurations. We did not observe the change of sign to follow a specific rule or to depend on the sign of other features from the same hyperparameter configuration. Hence we assume the overall sign to not have any physical interpretation and only to depend on the initialization of the network parameters. We however observe that the relation of positive and negative values within single maps is constant throughout different choices of hyperparameters, that is the overall feature map structure. Since these signs do not change the morphology of the map, we chose the sign of the color code in the map display of Fig.~\ref{fig:SignFeat} according to that of the input map which most closely resembles it. This is positive for stronger emitting regions and negative for weaker emitting regions.  \\

We can investigate which data information a feature map encodes by considering the generative process of the NEAT-VAE and using the back propagation algorithm: Since the amount of parameters in the latent space is much smaller compared to the data space, the autoencoder is forced to extract the essential information of the input data in order to generate it again. The generative process therefore offers, at least approximately, an explanation for the information flow within the autoencoder, and we can visualize to which extend specific features are generating the individual Galactic all-sky maps using sensitivity analysis \citep[e.g.,][]{zurada1994sensitivity}. Here, we compute the gradients of the reconstructed output maps with respect to the feature maps using the back propagation algorithm. The values of the resulting decoder Jacobian are displayed in HEALPix pixelization in Appendix \ref{app:gradientmaps}.\\

Another measure for quantifying which input information is encoded by the features is the mutual information. The mutual information
\begin{equation}
\label{eq:MI}
I(X;Y) = \sum_{x,y} \mathcal{P}(x,y) \, \mathrm{ln}\left(\frac{\mathcal{P}(x,y)}{\mathcal{P}(x) \, \mathcal{P}(y)}\right)
\end{equation}
can be calculated from two-dimensional histograms of feature maps ($x$) and output maps ($y$). For a given number of bins, the joint probability distribution $\mathcal{P}(x,y)$ is represented in an $(x, y)$-shaped matrix by counts per bin, while the marginalized distributions $\mathcal{P}(x)$ and $\mathcal{P}(y)$ are obtained by summing over the respective $y$ and $x$-axes of this matrix. In our interpretations, we will use both measures to evaluate the encoded information in the latent space (mutual information of feature and input maps), as well as the generative process from the latent space (decoder Jacobian maps), in order to determine the physical content of the features A, B and C. \\

\begin{figure*}[t!]
\begin{subfigure}[t]{0.5\textwidth}
\centering
\caption{Thermal dust emission}
\includegraphics[width=0.95\linewidth]{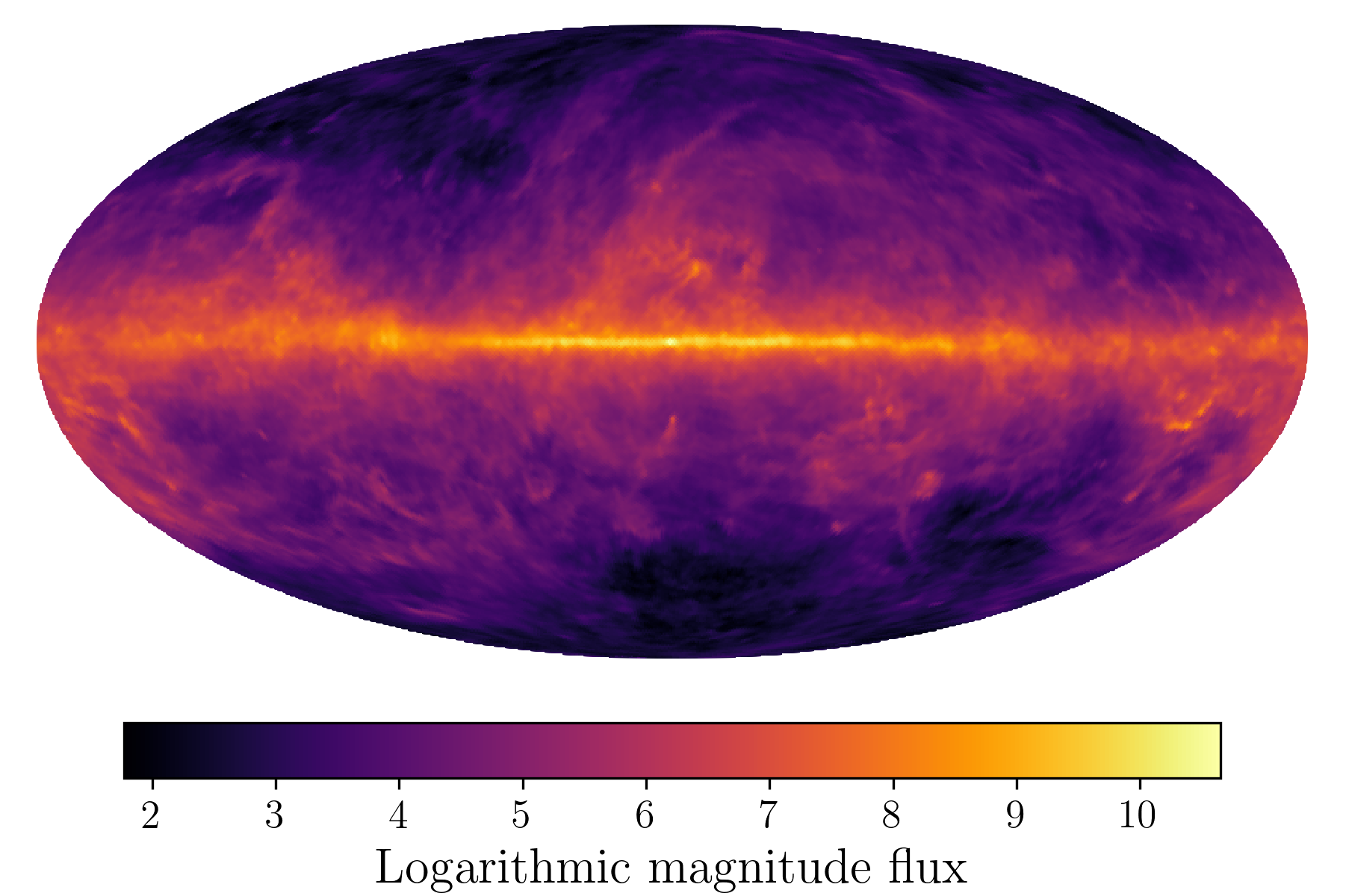}
\label{fig:dust}
\end{subfigure}
\begin{subfigure}[t]{0.5\textwidth}
\centering
\caption{Correlation feature A - dust}
\vspace{-1.5\baselineskip}
\includegraphics[width=1.05\linewidth]{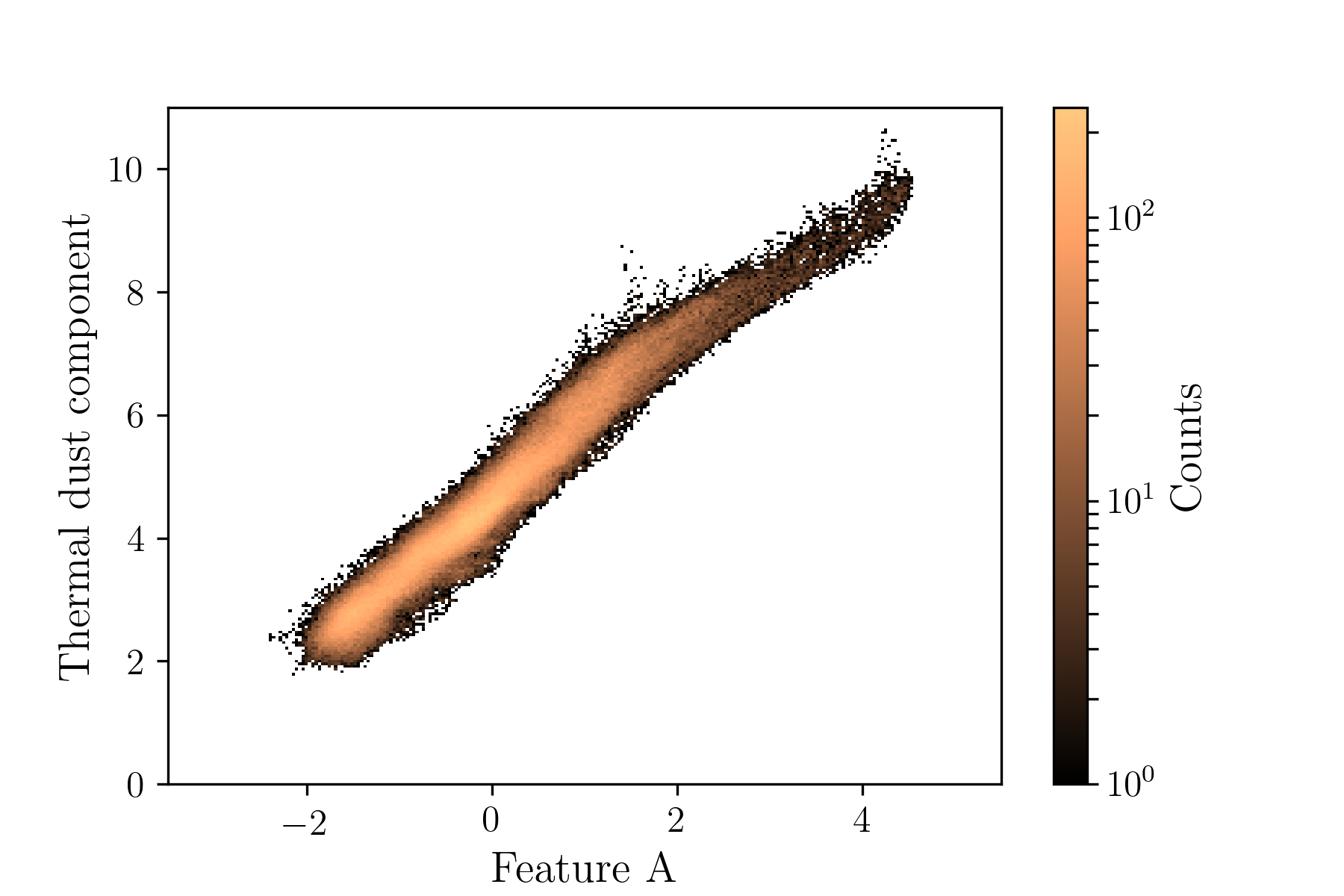}
\label{fig:Dustcorr}
\end{subfigure}
\begin{subfigure}[t]{0.5\textwidth}
\centering
\caption{Hadronic $\gamma$-ray component}
\includegraphics[width=0.95\linewidth]{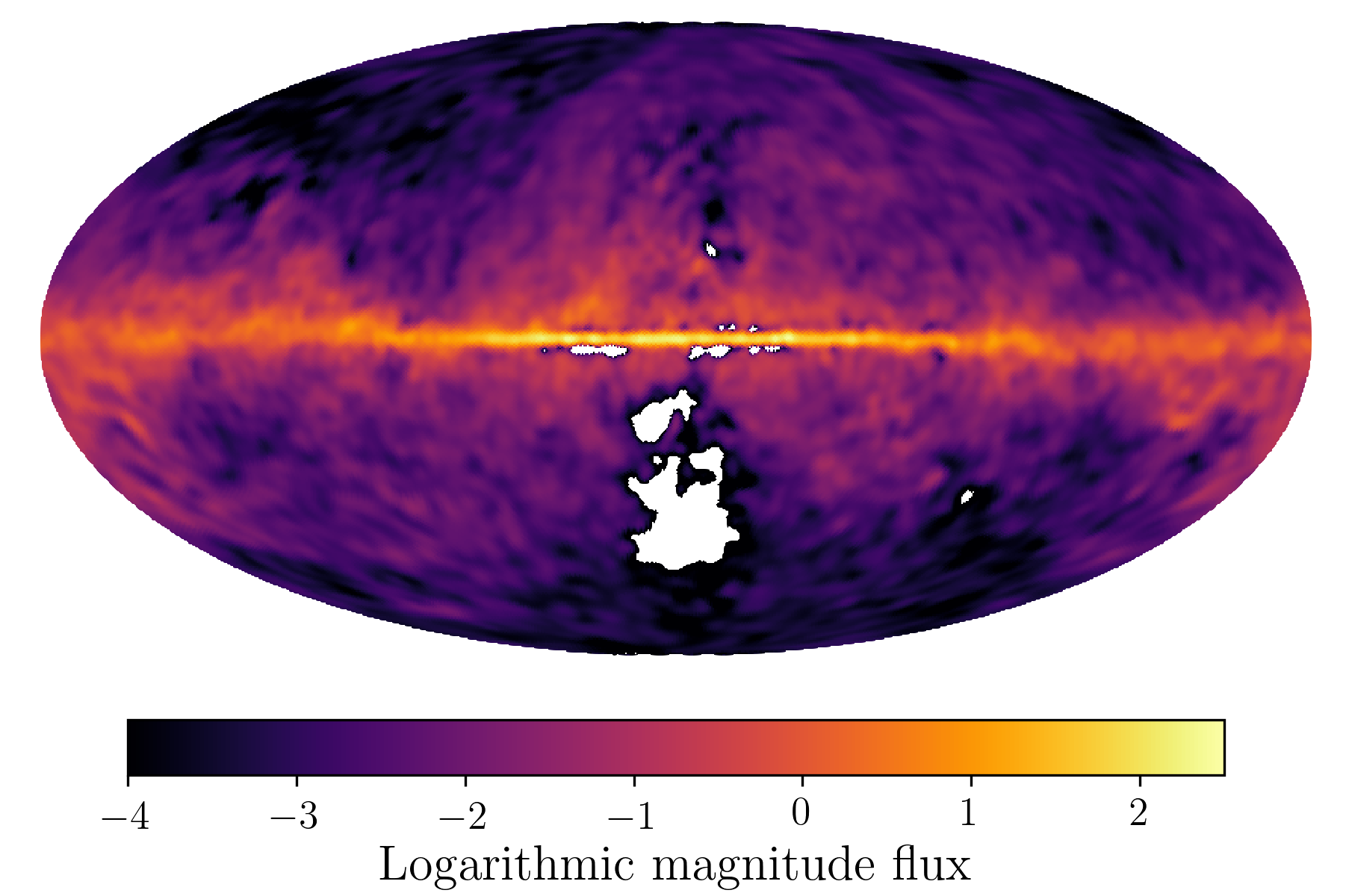}
\label{fig:hadroniccomp}
\end{subfigure}
\begin{subfigure}[t]{0.5\textwidth}
\centering
\caption{Correlation feature A - hadronic component}
\vspace{-1.5\baselineskip}
\includegraphics[width=1.05\linewidth]{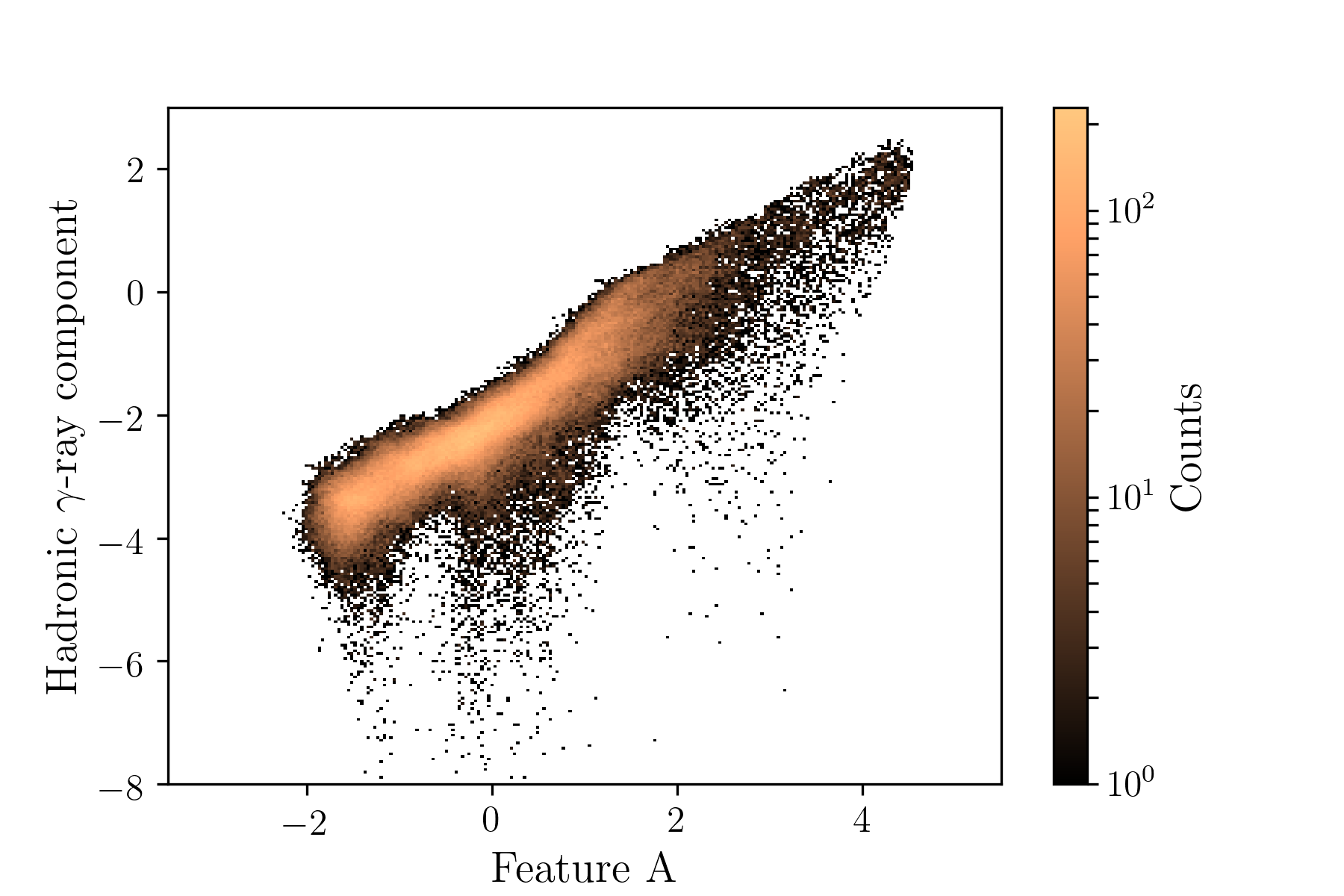}
\label{fig:hadroniccorr}
\end{subfigure}
\vspace*{-0.5em}
\caption[Correlations of feature A]{Correlation of feature A with astrophysical components tracing interstellar matter. From top left to bottom right: (a) thermal dust component in logarithmic scaling \citep{2016planckcomponent}, (b) 2D histogram of the posterior mean intensity of feature A and the thermal dust component (mutual information $I(X;Y) = 1.72$), (c) hadronic component of the $\gamma$-ray spectrum in logarithmic scaling \citep{selig2015denoised} with white pixels denoting missing values, (d) 2D histogram of the posterior mean intensity of feature A and the hadronic component ($I(X;Y) = 1.07$). For the 2D histograms, the intensity ranges of the maps are divided into $256$ equal bins and the number of intensity pairs per bin is displayed as counts on logarithmic scaling. Bright colors denote a high number of counts. Mutual information is calculated as described in Eq.~\eqref{eq:MI} with $512$ bins.}
\label{fig:featAcorr}
\end{figure*}

\subsection{Identifying the information encoded by feature A}
Feature A, which is the most significant feature in $98\%$ of the examined hyperparameter configurations (see Appendix~\ref{app:hyperparameters}), is displayed in Fig. \ref{fig:FeatA}. From a visual analysis, we recognize a positive color-coded Galactic plane and negative color-coded Galactic poles in the posterior mean. In the eastern part of the Galactic plane, we see a bright, circular structure in the Cygnus region. Further in the east, south of the Galactic plane, structures similar to the Perseus region occur. The circular shaped structure north of the Galactic center resembles the Ophiuchus region and southwestern of the Galactic center structures similar to the Small and Large Magellanic clouds can be recognized. In the western part of the Galactic plane, the bright structures look like the Orion region. The posterior variance shows a high certainty in the region of the Galactic plane and the structures of the surrounding latitudes, while at the southern and northern Galactic poles, the uncertainty increases. \\

\textit{Interpretation.} We find compelling evidence that feature A traces the dense and dusty parts of the ISM: Based on calculations of the mutual information with $512$ bins, the top three data sets contributing to feature A are the AKARI far-infrared $140~\mu$m input data with mutual information $I(X;Y) = 1.91$,  the IRIS infrared $100~\mu$m input data with $I(X;Y) = 1.84$, and the AKARI far-infrared $160~\mu$m input data, also with $I(X;Y) = 1.84$. This frequency band of the interstellar radiation field is dominated by the infrared emission of dust \citep[e.g.,][p.121]{2011piim.book.....D}. Feature A shares the least mutual information with the ROSAT X-ray $1.545~$keV input data with $I(X;Y) = 0.20$. Next, we analyze the decoder Jacobian maps (see Appendix \ref{app:gradientmaps}), which display the gradients of the reconstructed Galactic all-sky maps with respect to latent space features in HEALPix format. We observe the following:
The lower and mid-latitudes of the $\gamma$-ray regime strongly depend on feature A, and going from higher to lower energies, the overall-dependence on feature A grows. Especially, the low-energy regime of the Fermi data is dominated by hadronic interactions of cosmic rays with the interstellar medium (ISM) and thus shows the gas distribution \citep{selig2015denoised}. In the X-ray regime, we have small to no dependence; however, the mid- and high latitudes of the soft X-ray data ($0.212~$keV and $0.197~$keV) show negative gradients toward feature A. Here, negative gradients state that when increasing values in feature A, the corresponding values in the reconstructed X-ray maps will decrease. Such an inverse-proportional behavior is very plausible by the physical context: Radiation from X-rays is extincted by cold interstellar gas  \citep{ferriere2001interstellar}, thus the absence of X-ray emission reveals regions where interstellar matter is present. The $\mathrm{H}_{\alpha}$ map at $656.3$ nm, which displays emission due to hydrogen transitions occurring from the second excited state $n=3$ to the first excited state $n=2$ \citep[e.g.,][]{2003ApJS..146..407F}, positively depends on feature A. The physical interpretation is that neutral hydrogen preferably resides in the dense ISM, as traced by feature A. We observe a very strong dependence of the Galactic planes of the $12$ and $25~\mu$m infrared data on feature A, while with further decreasing energies, the infrared data down to $545$ GHz show an overall, positive dependence. In this regime, emission from thermal dust is observed \citep{klessen2014physical, akrami2018planck}. With further decreasing energy, the Planck microwave data depends strongly on feature A, in particular the Galactic plane. The $21$ cm line emission shows an overall dependence on feature A and describes the total neutral atomic hydrogen column density, since the displayed emission occurs due to the transition between two levels of the ground state of atomic hydrogen \citep{RN7}. In the synchrotron radio regime at $1420~$MHz and $408~$MHz, both feature A and feature B play an important role in determining the emission structures, which we will address in Sect.~\ref{ch:internalcomputations}.

The positive dependencies of sky maps on feature A describe areas of our Galaxy with high density of interstellar matter, while the negative gradients correlate with extinction of emissions by interstellar matter. We assume that positive gradients mark the pixels in latent space that are used to generate the corresponding pixels in data space, while negative gradients mark an anti-correlation of feature and data pixels. On the basis of this specific combination of gradients and the mutual information, we infer that feature A encodes dense regions of the ISM. \\

\textit{Discussion.}
We can test our hypothesis by investigating the correlation of feature A with dust as a tracer for the ISM \citep{doi:10.1146/annurev-astro-081811-125610}. Fig.~\ref{fig:Dustcorr} shows the relationship of feature A and the thermal dust emission (Fig.~\ref{fig:dust}) as calculated by the Planck \textsc{Commander} code \citep{eriksen2008joint, 2016planckcomponent}. The posterior mean of feature A has a strong, positive, and linear correlation with the thermal dust component. Both the \textsc{Commander}  and the NEAT-VAE algorithm perform a Bayesian, pixel-wise analysis based on a data model containing the linear sum of a signal function and noise, but with two main differences: First, we only employ statistical information in our prior knowledge, while the \textsc{Commander} priors include detailed physical models for the various emission processes contributing to the radio to far-infrared sky, as well as calibration and correction factors, and a prior for the CMB. Second, the \textsc{Commander} algorithm has $11$ million free parameters to tune \citep{2016planckcomponent}, while our model has just about $5,000$. The algorithms are not directly comparable, since the \textsc{Commander} code was especially developed to separate the Galactic foregrounds to reconstruct the CMB, while the NEAT-VAE only seeks to find an essential representation of the input data. But in this context, the NEAT-VAE summarized the input data into categories, one of which already contains dust emission. With this result, we assume it is possible to derive the dust component based on feature A with little computational effort. 

We investigate another tracer for interstellar matter, namely the hadronic $\gamma$-ray component derived by \citet{selig2015denoised}, shown in Fig.~\ref{fig:hadroniccomp}. It represents the $\gamma$-ray emission due to the interaction of cosmic ray protons with interstellar matter, and is positively correlated with feature A, see Fig.~\ref{fig:hadroniccorr}. This correlation is reasonable, since \citet{selig2015denoised} composed the hadronic component of the low-energy $\gamma$-ray maps, on which feature A also depends on. These results meet our initial aim of finding a reduced representation of the input data that combines redundant information in one feature, in this case tracers for dense regions in the ISM. 

The high significance of feature A is likely to result from the choice of input data, since most of the maps in our data set $\bm{D}$ represent emission from the dense interstellar matter. We did not include the CO line emission data, but from the positive gradients of the Planck $100$ - $353$ GHz channels with respect to feature A, which include the information of the CO line emission \citep{2016planckcomponent}, we assume that the Galactic plane of feature A most likely would generate the CO data. This would support our interpretation of feature A, since CO is a tracer for molecular interstellar gas \citep{10.1007/978-94-009-3861-8_2}, and thus, for regions of high density.

\subsection{Identifying the information encoded by feature B}
Feature B, the second most significant feature in $98 \%$ of our experiments, is displayed in Fig.~\ref{fig:FeatB}. We see a negative color-coded equator which resembles the Galactic plane, with positive color-coded bulges north and south of the plane. Especially the northern bulge structure looks similar to the morphology of the North Polar Spur. The positive color-coded, circular structure in the western part of the Galactic plane lies in the Vela region, whereas in the east, the location of the Cygnus region appears negative color-coded with a positive, ring-like structure surrounding it. The uncertainty of this map appears to be low in all latitudes, but is an order of magnitude higher compared to the variance map of feature A. \\

\begin{figure}[t]
\centering
\includegraphics[width=1.05\linewidth]{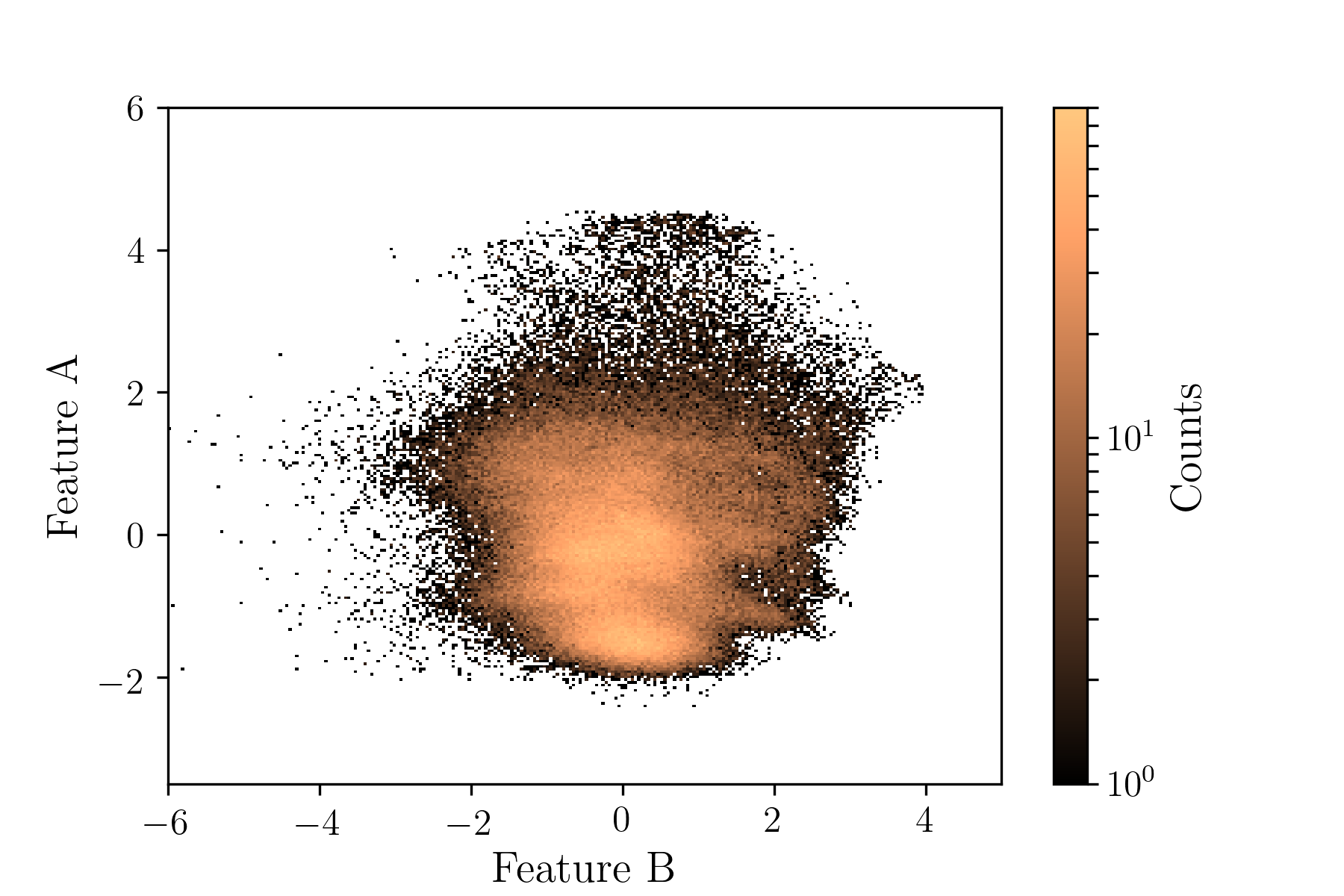}
\vspace*{-0.5em}
\caption[Correlations of feature B]{Correlation of feature B and feature A with mutual information $I(X;Y) = 0.33$. The 2D histogram is computed as described in Fig.~\ref{fig:featAcorr}. \textcolor{blue}{}}
\label{fig:CorrBA}
\end{figure}

\begin{figure*}[t!]
\begin{subfigure}[t]{0.5\textwidth}
\centering
\caption{CMB}
\includegraphics[width=0.9\linewidth]{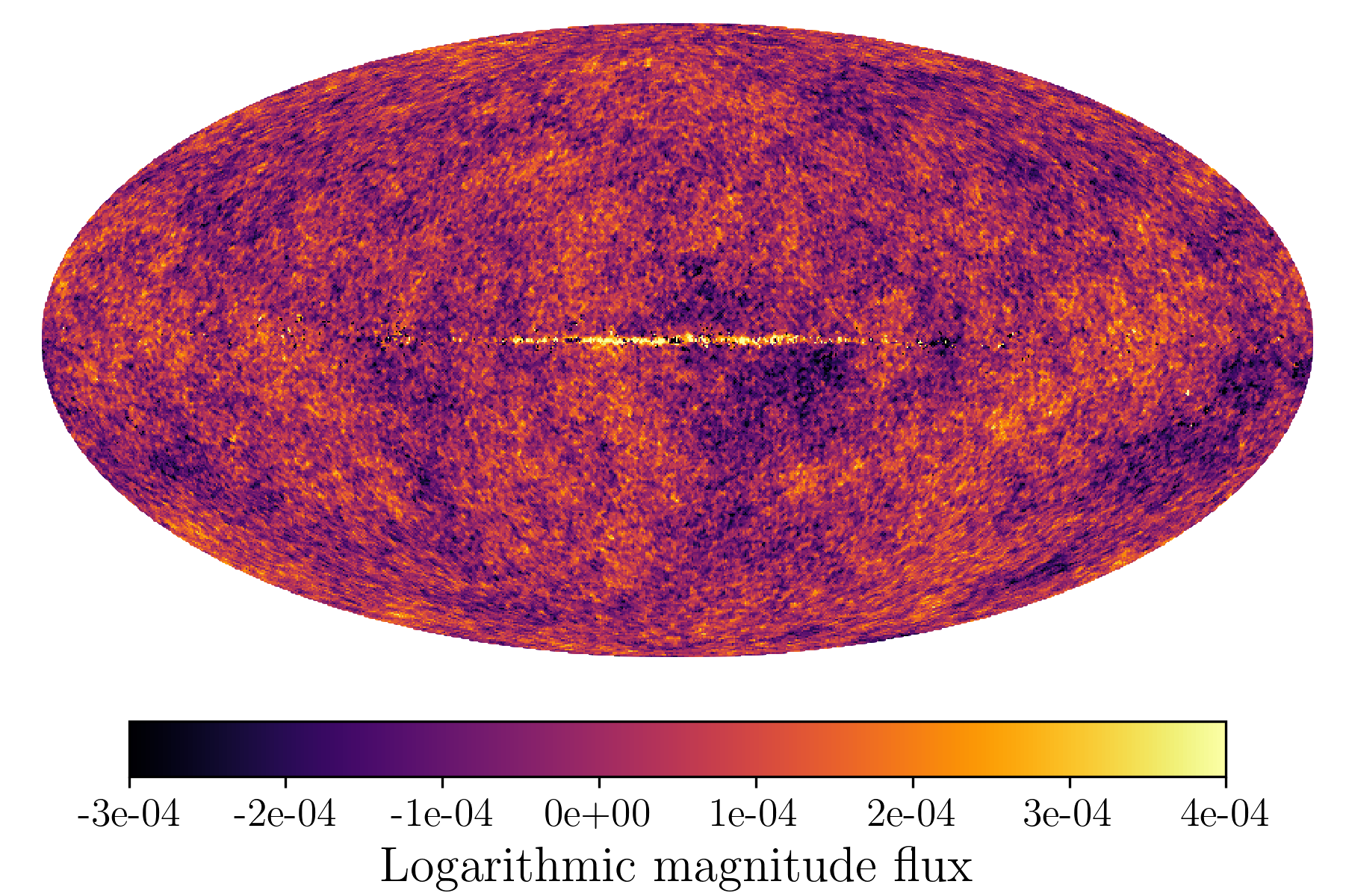}
\label{fig:CMB}
\end{subfigure}
\begin{subfigure}[t]{0.5\textwidth}
\centering
\caption{Correlation feature C - CMB}
\vspace{-1.5\baselineskip}
\includegraphics[width=1\linewidth]{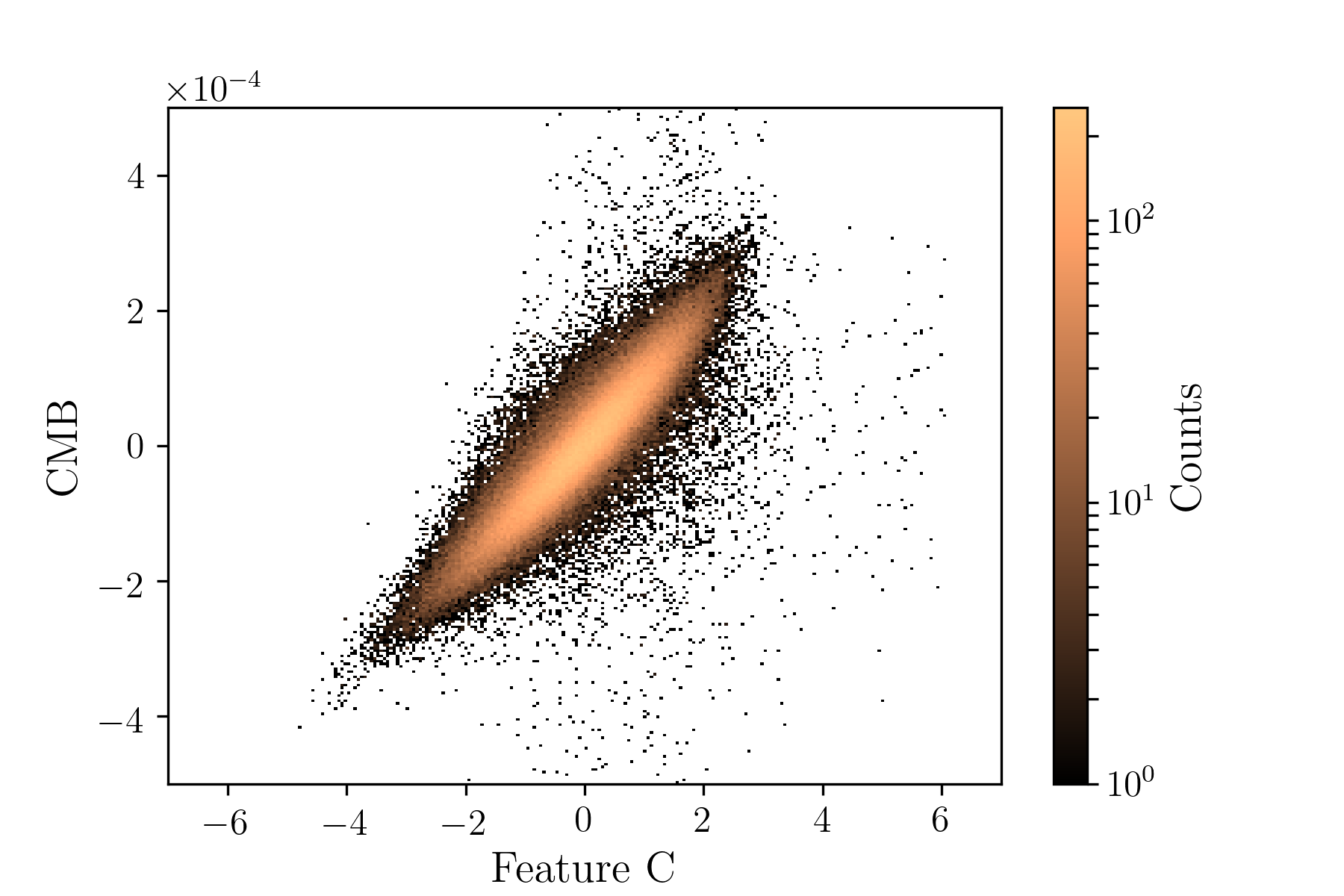}
\label{fig:CorrCCMB}
\end{subfigure} \\
\begin{subfigure}[t]{0.5\textwidth}
\centering
\caption{Correlation feature C - feature A}
\vspace{-1.5\baselineskip}
\includegraphics[width=1\linewidth]{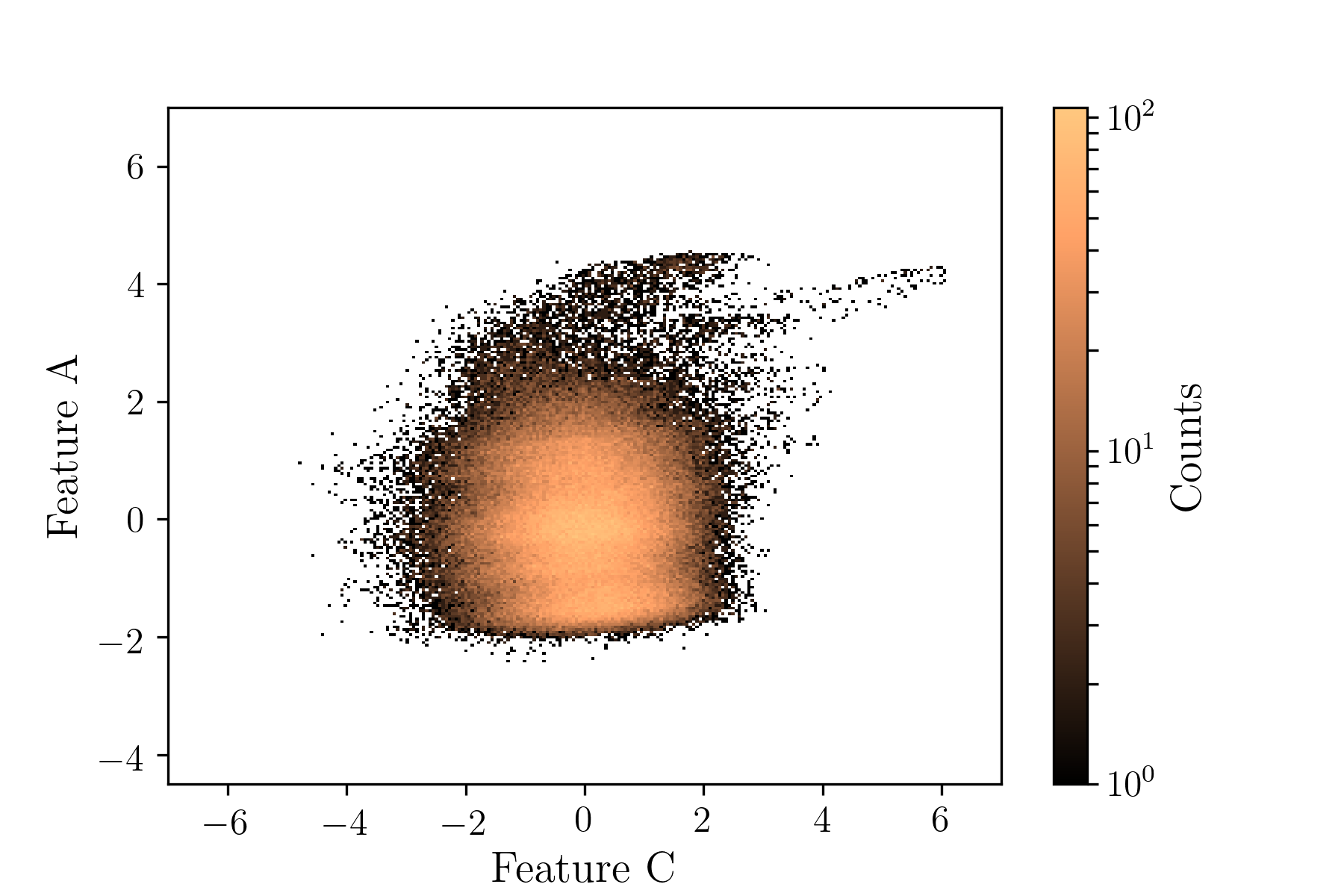}
\label{fig:CorrCA}
\end{subfigure}
\begin{subfigure}[t]{0.5\textwidth}
\centering
\caption{Correlation feature C - feature B}
\vspace{-1.5\baselineskip}
\includegraphics[width=1\linewidth]{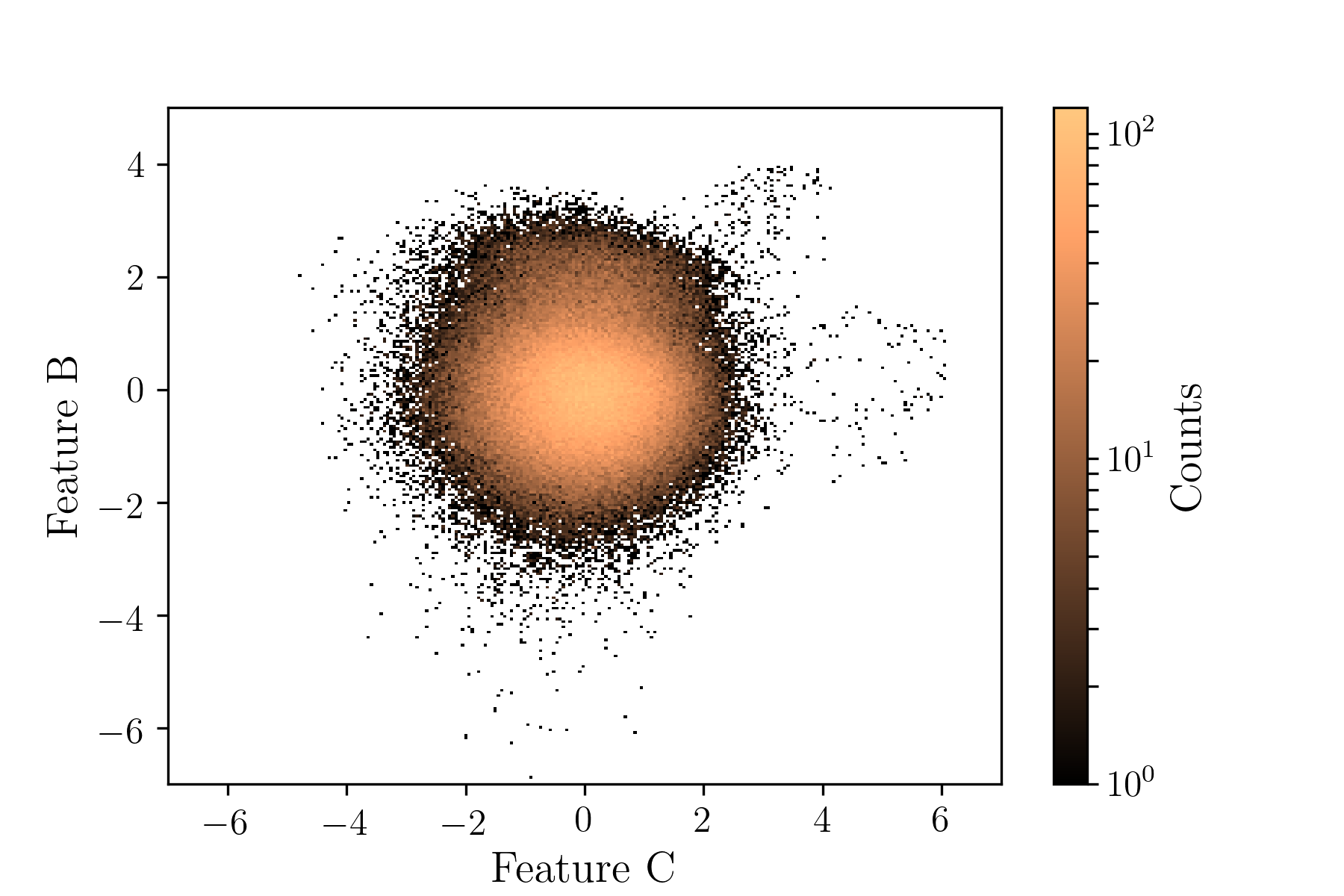}
\label{fig:CorrBC}
\end{subfigure}\\
\vspace*{-0.5em}
\caption[Correlations of feature C]{Panel (a) shows the CMB as derived by the \citet{akrami2018planckIV}, panels (b)-(d) display correlations of the posterior mean of feature C with (b) the CMB, mutual information $I(X;Y)=0.88$, (c)  feature A, $I(X;Y)=0.29$, and (d) feature B, $I(X;Y)=0.20$. The histograms and mutual information are compiled as described in Fig.~\ref{fig:featAcorr}. Bright colors in the histograms denote a high number of counts.}
\label{fig:featCcorr}
\end{figure*}

\textit{Interpretation.} Feature B shows highest mutual information with the X-ray input data of ROSAT at $0.885~$keV ($I(X;Y)=0.82$), $0.725~$keV ($I(X;Y)=0.78$) and $1.145~$keV ($I(X;Y)=0.67$), and lowest mutual information with the Planck $70~$GHz data ($I(X;Y)=0.18$). In general, X-ray emission is assumed to arise from hot, ionized gas in regions of the ISM with low density \citep[e.g.,][]{ferriere2001interstellar}. Considering the gradient maps in Appendix \ref{app:gradientmaps}, there is little to no dependence of the reconstructed $\gamma$-ray data on feature B. Starting from X-ray data at $1.545~$keV, the positive gradients get stronger with decreasing energy, and especially the bulge-like area of the reconstructed X-ray data strongly depends on feature B. The soft X-ray regime around $0.25$ keV, which coincides with hydrogen cavities \citep[e.g.,]{1977ApJ...217L..87S,snowden1990model}, also shows a weak, positive dependence on feature B. The infrared and microwave regime again show little to no dependence, we however observe small negative gradients to be present where in the corresponding gradients maps of feature A, strong, positive gradients occur. The reconstructed $1420~$MHz and $408~$MHz radio maps show positive dependence on feature B, especially in the bulge-like areas. This data set detects the synchrotron radiation generated by the interaction of cosmic ray electrons with magnetic fields in the ISM  \citep{ginzburg1965cosmic}, and the intensity of this radiation depends on the density of relativistic particles and the magnetic field strength. The former one is predominantly found in the hot areas of the ISM, if only for the much larger volume occupation of this phase within the Milky Way \citep{doi:10.1146/annurev.astro.43.072103.150615}. Thus we assume that feature B encodes the enhanced presence of such cosmic ray electrons. With exception of the low energy X-ray and radio synchrotron regime, we observe the data predominantly generated by feature A to have little to no dependence on feature B, which we interpret as feature B being complementary to dense regions of the ISM. In combination with the indications from the mutual information with X-ray input data and positive gradients, we assume that feature B encodes tracers for dilute and hot regions of the ISM.\\

\textit{Discussion.} Fig.~\ref{fig:CorrBA}, showing the correlation between Feature A and B, supports our interpretation: The features, in our case the dense and dilute regions of the ISM, are basically uncorrelated. This relationship is reasonable in the sense that the features describe two distinct categories of the ISM: The data generated by feature A are based on emissions from cold and warm ionized gas (as observed in the $21$ cm and $\mathrm{H}_{\alpha}$ line emissions), dust, and interactions with interstellar matter (cosmic rays). The ionization of the warm gas occurs mainly due to photo-ionization by O and B stars, the hottest and most massive stars of the Milky Way. Due to their their ratio of mass and luminosity, these stars have a short main sequence life cycle and are thus found near their initial birthplaces, the dense ISM \citep[e.g.,][p.60]{blome1997bergmann}. Feature B, however, encodes data which represent radiation of an even hotter medium, the hot ionized plasma, which is generated by supernovae explosions \citep[e.g.,][]{1980HiA.....5..365K}. 

The radiative processes encoded by features A and B can thus be traced back to two fundamentally different origins, namely emission from interstellar matter and from stellar explosions. One regime of the electromagnetic spectrum missing in our input data is the ultraviolet (UV) frequency band. In this regime, the emission of very hot gas which gets ionized by collisions can be observed, but UV radiation does not penetrate dense regions of the ISM and is thus mostly absorbed in low and mid-latitudes \citep{ferriere2001interstellar}. Due to this extinction, it is likely that UV radiation would be interpreted by the NEAT-VAE as redundant with the soft X-ray data, which shows similar dust absorption patterns. This would support  our interpretation of feature B to encode the hot and dilute ISM.

\subsection{Identifying the information encoded by feature C}
The third most significant feature in $74 \%$ of the investigated configurations, Feature C, is displayed in Fig \ref{fig:FeatC}. Here, we inverted the colors of the posterior mean such that the fluctuations resemble the color-coding of the CMB, see Fig.~\ref{fig:CMB}. The regions of the Galactic plane and Cygnus have a strong, positive color-coding, while most of the other structures fluctuate around zero. The uncertainty is highest in the Galactic plane, especially in the Galactic center. There are bulge-like shapes in the variance map north and south of the Galactic center denoting a medium level of uncertainty. Toward higher latitudes, the uncertainty is very low. \\

\textit{Interpretation.} The mutual information of feature C is highest with the Planck data of $70$, $100$ and $143$ GHz with values of $I(X;Y)=\{ 1.01, 0.95, 0.82\}$, respectively, and the least mutual information is observed with the ROSAT X-ray $1.545~$keV data with $I(X;Y)=0.12$. According to the decoder Jacobian maps in Appendix \ref{app:gradientmaps}, feature C mostly generates the Planck $30 - 217~$GHz channels, especially the $70$ and $100~$GHz data, in which the CMB can be observed \citep{akrami2018planck}. All other reconstructed maps show little to no dependence on feature C, with exception of a weak, positive dependence of the soft ROSAT X-ray data on feature C that we will address in Sect.~\ref{ch:internalcomputations}. Based on the mutual information and the strong positive gradients in the microwave regime around $100~$GHz, we assume feature C to encode the CMB.\\

\textit{Discussion.} Considering our physical interpretation of features A and B, it is reasonable that the CMB emission is encoded in a separate feature, since the underlying physical processes in the Early Universe shaping the CMB fluctuations are independent of the processes in the ISM. To test our hypothesis we compare feature C to the CMB all-sky map derived by the Planck \textsc{Commander} code in Fig.~\ref{fig:CMB}. Feature C shows a positive, linear correlation ($I(X;Y) = 0.88$) with the CMB (Fig.~\ref{fig:CorrCCMB}). A main difference between feature C and the CMB can be seen in the Galactic plane, which feature C encodes (in addition to the Cygnus region) with high intensities, but also with high uncertainty (see Fig.~\ref{fig:FeatC}). We assume the high intensities in the Galactic plane of feature C to originate from the model's architecture: Since we do not take spatial correlations into account, the algorithm learns spectral relations of each pixel independently. The Galactic planes of the input maps are dominated by high intensity values, while for other latitudes, more low-intensity values are present. We assume that this generates two distinct clusters of intensity ranges, leading to two different spectral relations learned by the algorithm. We can also observe this distinction in intensity by analyzing the contributions of pixels for data generation (see Appendix~\ref{app:gradientmaps}): The structure of the Galactic plane is recognizable in most of the decoder Jacobian maps, even though the algorithm has no information about spatial correlations. This again indicates that the relations learned in different intensity regimes represent different processes in the NEAT-VAE structure, which however are rather to be associated with internal computations than with physical properties.

\subsection{Non-physical interpretation of the NEAT-VAE}
\label{ch:internalcomputations}
We finally examine the decoder Jacobian maps which do not have a clear physical interpretation. For example, we observe feature A (the dense ISM) contributing to the $1420$ and $408~$MHz radio data, as well as feature C (the CMB) to supposedly generate the soft X-ray data. 

The NEAT-VAE algorithm recombines all features of the latent space in a highly non-linear way to generate the output maps, meaning that the gradients in Appendix \ref{app:gradientmaps} cannot always be considered independently or in a linear way. Especially when the absolute gradient values of the reconstructed maps are large for more than one feature, a holistic analysis is required. In most of the displayed cases in Appendix \ref{app:gradientmaps}, there is one feature dominantly generating the reconstruction of input data. One exception, however, is the radio data, where we observe positive gradients with respect to both feature A and B. A speculative physical explanation is that feature A marks regions of enhanced magnetic field strength and feature B of enhanced relativistic electron densities, both quantities that in combination determine the synchrotron emission. However, we rather assume that these two features have no physical meaning for synchrotron emission, but the internal computations of our algorithm run most efficiently when latent values of features A and B are used for radio data generation. One always has to consider the fact that our algorithm has no information about physical relations and primarily optimizes a statistical function. Another example for such a behavior can be found in the soft X-ray data, where we observe strong negative gradients with respect to feature A and weak positive gradients toward both features B and C. This mixing of features in a partially contrasting manner indicates the non-linearity of the function mapping from the latent space values to the reconstructed output data: We hypothesize that some values might be used by the algorithm just to tune some others out, since the decoder has to regenerate the data based on all available features. This discussion shows that the decoder Jacobians give a valid first interpretation of the strongest correlations, but have to be analyzed with care. Not each contribution to the gradient represents a meaningful relationship, especially since all gradients are entangled and cannot be considered independently.

\section{Conclusions}
In summary, we derived a probabilistically motivated machine learning framework, the NEAT-VAE, which successfully identifies the most significant sky emission components according to pixel-wise sky brightness values across the full electromagnetic spectrum. At this task, the algorithm performs computationally efficient and in a fully unsupervised manner. The three most significant resulting sky components express physical relationships in the considered data set and can be assigned to emission processes of the dense ISM, the hot and dilute ISM, and the CMB. \\

We achieved this performance by developing a Bayesian formulation of the component separation problem for astrophysical data, which serves as the minimization objective, or loss function, for a state-of-the-art variational autoencoder. Using Bayes' theorem, we combined a generative data model with a variational inference process, incorporated a very limited amount of prior knowledge of generic nature (e.g., independence of features, Gaussian statistics of typical noise processes), and approximated unknown functions by neural networks. The resulting algorithm is able to group essential information contained in a set of Galactic all-sky observations into mutually independent features. This property results from the algorithm's architecture and the diagonal covariance approximation we chose: First, an autoencoder maps its input to a lower dimensional latent space, from which it aims to reconstruct its input again. By reducing the dimension of the data in this so-called information bottleneck, the autoencoder is forced to learn a useful representation of the input data. By additionally stating a diagonal covariance matrix in the variational approximation of the latent space posterior distribution (see Sect.~\ref{ch:approxpost}), we find the latent spaces' optimal independent approximation. We observe that there is an order among the latent space features based on their significance to encode the data set. This computational significance correlates with physical significance, with the most significant features having a clear physical interpretation and representing emissions from the dense ISM, the hot and dilute ISM, and the non-Galactic CMB, respectively. Our interpretations are based on the analysis of (1) the mutual information of the features and the input data, (2) the generative properties of the features to reconstruct the input data, and (3) the features' correlations with other astrophysical quantities. Thus we were able to verify that the NEAT-VAE algorithm detects the most informative components of emission into which the Galactic input data can be decomposed. However, as discussed in Sect.~\ref{ch:internalcomputations}, the analysis of our generative process is not always straightforward and has to be performed with care, since the examined gradients might represent internal computation strategies of the algorithm rather than physically meaningful generative properties. \\

The relevance of our work lies in the insights we obtained in the context of representation learning (RL). In RL, the reduced representation of high-dimensional data is often used as a preprocessing step in order to perform the actual analysis more efficient. For example, the task of deriving the thermal dust emission is computationally very expensive. Our feature A, however, which encodes the dense ISM, is highly correlated with the thermal dust component. We hypothesize that a component separation algorithm only analyzing the information contained in feature A, or using feature A as a starting point, is likely to perform faster and more efficient compared to analyzing a larger data set with redundant and entangled information. \\

The loss function of the NEAT-VAE and the associated hyperparameter tuning mainly determine the performance of the algorithm. Although the loss function used to direct the learning is based on very general assumptions, the Bayesian framework in which it was derived describes the initial problem probabilistically. This allows us to track uncertainties and evaluate the significance of each feature. Hyperparameter tuning can be reduced by including those parameters in the Bayesian inference process, as in our case the noise covariance of the model noise. In general, the joint optimization of an inference and a generative process is a state-of-the-art approach to disentangle data, which we showed to be very successful for unsupervised learning applied on astrophysical observations. Especially the fact that astrophysical data are a superposition of several radiative processes, and thus are rich in relationships between emitting as well as absorbing components, makes them a very suitable target for decomposing algorithms. \\

The fact that we neither incorporated spatial correlations nor physical priors in the learning process shows how much information can be retrieved from data-driven approaches: Alone the sky brightness values of single pixels in Galactic all-sky maps are sufficient to detect emission from the dense ISM, the dilute ISM and the CMB. By additionally combining Bayesian methods with neural networks, we were able to track uncertainties and efficiently approximate unknown functions. 
This work shows that machine-assisted, Bayesian signal inference can be performed with neural networks, and that the minimization of a loss function based on information theoretic principles reveals meaningful relations in astrophysical data. 

\begin{acknowledgements}
We acknowledge Jakob Knollm{\"u}ller for helpful discussions.
\end{acknowledgements}

\bibliographystyle{aa}
\bibliography{Bibliography}

\onecolumn
\begin{appendix}

\section{Astrophysical data set}
\begin{table*}[h]
\caption[Overview of data sets]{Overview of data sets used in this study. Data was preprocessed by \citet{muller2018} as described in Sect. \ref{ch:data}. Our data compilation $\bm{D}$ consists of 35 data sets including a certain number of data sets (declared in the column `Maps') from each mission.}
\label{app:data}
\centering
 \begin{tabular}{|l|l|l|l|}
  \hline
  \\[-1em]
  Mission/ survey & Frequency range & Maps & References \\
  \\[-1em]
  \hline
  \\[-1em]
  Fermi $\gamma$-ray & $0.85$--$217.22$~GeV & $9$ & \citet{Atwood_2009}, \citet{2012ApJS..203....4A}, \\  & & &\citet{selig2015denoised}\\
  \\[-1em]
  ROSAT X-ray &  $0.197$--$1.545$~keV & $6$ &\citet{1995ApJ...454..643S, 1997ApJ...485..125S}, \citet{1998AN....319...93F}, \\  & & &  \citet{freyberg1999rosat}\\  
  \\[-1em]
    $\mathrm{H}_{\alpha}$ line emission& $656.3$~nm& $1$ &\citet{vtss}, \citet{2001PASP..113.1326G}, \\ & & & \citet{madsen2001wisconsin}, \citet{2003ApJS..146..407F}\\
    \\[-1em]
    IRIS infrared & $12$--$100$~$\mu$m & $4$&\citet{1984ApJ...278L...1N} , \\ & &  & \citet{2005ApJS..157..302M}\\
   \\[-1em]
     AKARI far-infrared & $90$--$160$~$\mu$m & $3$& \citet{2015PASJ...67...50D}\\
    \\[-1em]
    Planck microwave & $30$--$857$~GHz & $9$&\citet{2016AA...594A...1P}\\
    \\[-1em]
    HI line emission radio & $21$~cm & $1$ &\citet{2016AA...594A.116H}\\
    \\[-1em]
     Reich radio & $1\,420$~MHz & $1$ &\citet{1982AAS...48..219R}, \citet{1986AAS...63..205R}, \\ & & & \citet{2001AA...376..861R}\\
    \\[-1em]
    Haslam radio & $408$~MHz & $1$ &\citet{1982AAS...47....1H, 2015MNRAS.451.4311R}\\
 \hline  
  \end{tabular}
\end{table*}

\section{Calculations}
\label{app:calculations}
Here, we provide the analytic steps between Eqs. \eqref{eq:FullKLD} and \eqref{eq:KLD}. For completeness, we start our calculations with the full posterior $P(\bm{\Theta} \mid \bm{D}) = P( \bm{D}\mid \bm{\Theta}) \times P(\bm{\Theta}) / P(\bm{D}) $:

\begin{flalign}
\label{teq:KLD}
D_{KL}[Q_{\bm{\Phi}}(\cdot) \mid\mid P(\cdot)] 
&= \int d\bm{Z} \, d\bm{\theta} \, d \xi_N \, Q_{\bm{\Phi}}(\bm{Z}, \bm{\theta},  \xi_N \mid \bm{D}) \, \ln \left( \frac{Q_{\bm{\Phi}}(\bm{Z}, \bm{\theta},  \xi_N \mid \bm{D})}{P(\bm{Z}, \bm{\theta},  \xi_N \mid \bm{D})}\right)   \nonumber \\
&= \int \left( \prod_{i=1}^{p} d\bm{z}_i \, Q_{\bm{\Phi}}(\bm{z}_i \mid \bm{d}_i) \right) \,\int d\bm{\theta} \,  Q_{\bm{\Phi}}(  \bm{\theta}\mid \bm{D}) \, \int d \xi_N  \, Q_{\bm{\Phi}}( \xi_N\mid \bm{D}) \, \times   \nonumber  \\
& \quad \ln \left( \frac{\prod_{i=1}^{p} Q_{\bm{\Phi}}(\bm{z}_i \mid \bm{d}_i) \, Q_{\bm{\Phi}}(  \bm{\theta}\mid \bm{D}) \, Q_{\bm{\Phi}}( \xi_N\mid \bm{D}) }{\frac{P(\bm{\theta})}{P(\bm{D})} \, P(\xi_N) \, \prod_{i=1}^p \, \mathcal{G}\left(\bm{d}_i - f_{\bm{\theta}}(z_i), t_{\bm{\psi}}( \xi_N)\right)\, \mathcal{G}(\bm{z}_i, \mathds{1})}  \right)    \nonumber  \\
&= \int \left( \prod_{i=1}^{p} d\bm{z}_i \, Q_{\bm{\Phi}}(\bm{z}_i \mid \bm{d}_i) \right) \,\int d\bm{\theta} \,  Q_{\bm{\Phi}}(  \bm{\bm{\theta}}\mid \bm{D}) \, \int d \xi_N  \, Q_{\bm{\Phi}}( \xi_N\mid \bm{D}) \, \times  \nonumber  \\
& \quad \Bigg[ \left(\sum_{i=1}^{p} \ln Q_{\bm{\Phi}}(\bm{z}_i \mid \bm{d}_i) \right) + \ln Q_{\bm{\Phi}}(  \bm{\bm{\theta}}\mid \bm{D}) + \ln Q_{\bm{\Phi}}( \xi_N\mid \bm{D}) - \ln P(\bm{\bm{\bm{\theta}}})   \nonumber  \\
& \quad - \ln P( \xi_N) + \ln P(\bm{D}) -  \sum_{i=1}^{p} \ln \left(\mathcal{G}(\bm{z}_i, \mathds{1}) \, \mathcal{G}\left(\bm{d}_i - f_{\bm{\bm{\theta}}}(z_i), t_{\bm{\psi}}( \xi_N)\right)\right) \Bigg]   \nonumber \\ 
&=  \underbrace{\int \left( \prod_{i=1}^{p} d\bm{z}_i Q_{\bm{\Phi}}(\bm{z}_i \mid \bm{d}_i) \right) \,\left( \sum_{i=1}^{p} \ln Q_{\bm{\Phi}}(\bm{z}_i \mid \bm{d}_i)\right)}_{=\sum_{i=1}^p \, \left\langle \ln Q_{\bm{\Phi}}(\bm{z}_i \mid \bm{d}_i) \right\rangle_{Q_{\bm{\Phi}}(\bm{z}_i \mid \bm{d}_i)}} \underbrace{\int d\bm{\bm{\theta}} \,  Q_{\bm{\Phi}}(  \bm{\bm{\theta}}\mid \bm{D})}_{= \, 1} \, \underbrace{\int d \xi_N  \, Q_{\bm{\Phi}}( \xi_N\mid \bm{D})}_{= \, 1}   \nonumber \\
& \quad + \underbrace{\int \left( \prod_{i=1}^{p} d\bm{z}_i Q_{\bm{\Phi}}(\bm{z}_i \mid \bm{d}_i)\right)}_{= \, 1} \,\underbrace{\int d\bm{\bm{\theta}} \,  Q_{\bm{\Phi}}(  \bm{\bm{\theta}}\mid \bm{D}) \, \ln Q_{\bm{\Phi}}(  \bm{\bm{\theta}}\mid \bm{D})}_{= \, \int d\bm{\bm{\theta}} \, \delta(\bm{\bm{\bm{\theta}}} - \widehat{\bm{\bm{\theta}}}) \, \ln\delta(\bm{\bm{\bm{\theta}}} - \widehat{\bm{\bm{\theta}}})\, =\, \mathrm{const.}} \underbrace{\int d \xi_N  \, Q_{\bm{\Phi}}( \xi_N\mid \bm{D})}_{= \, 1}   \nonumber \\
& \quad + \underbrace{\int \left(\prod_{i=1}^{p} d\bm{z}_i Q_{\bm{\Phi}}(\bm{z}_i \mid \bm{d}_i)\right)}_{= \, 1} \,\underbrace{\int d\bm{\bm{\theta}} \,  Q_{\bm{\Phi}}(  \bm{\bm{\theta}}\mid \bm{D})}_{= \, 1} \, \underbrace{\int d \xi_N  \, Q_{\bm{\Phi}}( \xi_N\mid \bm{D}) \ln Q_{\bm{\Phi}}( \xi_N\mid\bm{D})}_{= \, \int d \xi_N \, \delta( \xi_N - \widehat{ \xi}_N) \, \ln\delta( \xi_N - \widehat{ \xi}_N)\, =\, \mathrm{const.}}    \nonumber \\ \displaybreak
& \quad - \underbrace{\int \left(\prod_{i=1}^{p} d\bm{z}_i Q_{\bm{\Phi}}(\bm{z}_i \mid \bm{d}_i)\right)}_{= \, 1} \,\underbrace{\int d\bm{\bm{\theta}} \,  Q_{\bm{\Phi}}(  \bm{\bm{\theta}}\mid \bm{D}) \, \ln P(\bm{\bm{\bm{\theta}}})}_{= \, \mathrm{const.}, \, \mathrm{since} \,P(\bm{\bm{\bm{\theta}}})\,=\,\mathrm{const}.} \underbrace{\int d \xi_N  \, Q_{\bm{\Phi}}( \xi_N\mid \bm{D})}_{= \, 1}   \nonumber\\ 
& \quad - \underbrace{\int \left(\prod_{i=1}^{p} d\bm{z}_i Q_{\bm{\Phi}}(\bm{z}_i \mid \bm{d}_i)\right)}_{= \, 1} \,\underbrace{\int d\bm{\bm{\theta}} \,  Q_{\bm{\Phi}}(  \bm{\bm{\theta}}\mid \bm{D})}_{= \, 1} \, \underbrace{\int d \xi_N  \, Q_{\bm{\Phi}}( \xi_N\mid \bm{D}) \ln P( \xi_N)}_{= \, \left\langle \ln \mathcal{G}( \xi_N, 1) \right\rangle_{Q_{\bm{\Phi}}( \xi_N\mid \bm{D})}} \nonumber  \\
& \quad + \underbrace{\int \left(\prod_{i=1}^{p} d\bm{z}_i Q_{\bm{\Phi}}(\bm{z}_i \mid \bm{d}_i)\right)}_{= \, 1} \,\underbrace{\int d\bm{\bm{\theta}} \,  Q_{\bm{\Phi}}(  \bm{\bm{\theta}}\mid \bm{D})}_{= \, 1} \, \underbrace{\int d \xi_N  \, Q_{\bm{\Phi}}( \xi_N\mid \bm{D}) \ln P(\bm{D})}_{= \, \mathrm{const}., \, \mathrm{since} \,P(\bm{D})\,=\,\mathrm{const}.}  \nonumber \\
& \quad - \underbrace{\int \left(\prod_{i=1}^{p} d\bm{z}_i Q_{\bm{\Phi}}(\bm{z}_i \mid \bm{d}_i) \right)\sum_{i=1}^{p} \ln \mathcal{G}(\bm{z}_i, \mathds{1})}_{= \sum_{i=1}^{p} \left\langle\, \ln \mathcal{G}(\bm{z}_i, \mathds{1})  \right\rangle_{Q_{\bm{\Phi}}(\bm{z}_i \mid \bm{d}_i)}}\,\underbrace{\int d\bm{\bm{\theta}} \,  Q_{\bm{\Phi}}(  \bm{\bm{\theta}}\mid \bm{D})}_{= \, 1} \, \underbrace{\int d \xi_N  \, Q_{\bm{\Phi}}( \xi_N\mid \bm{D})}_{= \, 1}   \nonumber \\
& \quad - \int \left(\prod_{i=1}^{p} d\bm{z}_i Q_{\bm{\Phi}}(\bm{z}_i \mid \bm{d}_i)\right) \int d\bm{\bm{\theta}} \,  Q_{\bm{\Phi}}(  \bm{\bm{\theta}}\mid \bm{D}) \int d \xi_N  \, Q_{\bm{\Phi}}( \xi_N\mid \bm{D}) \times   \nonumber \\
& \quad \quad \sum_{i=1}^{p} \ln\mathcal{G}\left(\bm{d}_i - f_{\bm{\bm{\theta}}}(z_i), t_{\bm{\psi}}( \xi_N)\right).  \nonumber
\end{flalign}
Inserting the definitions for $Q_{\bm{\Phi}}( \bm{\theta}\mid \bm{D})= \delta(\bm{\theta} - \widehat{\bm\theta})$, $ Q_{\bm{\Phi}}( \xi_N\mid \bm{D})= \delta( \xi_N - \widehat{ \xi}_N)$ and $Q_{\bm{\Phi}}(\bm{z}_i \mid \bm{d}_i) = \mathcal{G}\left(\bm{z}_i - \bm{\mu}_i, \Sigma_i \right)$, we get \\
\begin{empheq}[box=\fbox]{align}
D_{KL}[Q_{\bm{\Phi}}(\cdot) \mid\mid P(\cdot)]  &= \sum_{i=1}^p \, \left\langle \ln \mathcal{G}\left(\bm{z}_i - \bm{\mu}_i, \Sigma_i \right) \right\rangle_{\mathcal{G}\left(\bm{z}_i - \bm{\mu}_i, \Sigma_i \right)} - \left\langle \ln \mathcal{G}( \xi_N, 1) \right\rangle_{\delta( \xi_N - \widehat{ \xi}_N)}  - \sum_{i=1}^{p} \left\langle\, \ln \mathcal{G}(\bm{z}_i, \mathds{1})  \right\rangle_{\mathcal{G}(\bm{z}_i - \bm{\mu}_i, \Sigma_i)}\nonumber \\
&\quad  - \sum_{i=1}^{p} \left\langle \ln\mathcal{G}\left(\bm{d}_i - f_{\widehat{\bm{\bm{\theta}}}}(z_i), t_{\bm{\psi}}(\widehat{\xi}_N) \right) \right\rangle_{\mathcal{G}(\bm{z}_i - \bm{\mu}_i, \Sigma_i)}  
 + \mathcal{H}_0,  
\end{empheq}

where $\mathcal{H}_0$ collects all terms independent of the parameters. We rename the subsequent terms of the Kullback-Leibler Divergence as follows and calculate each expression respectively.
\begin{itemize}
\item $\bm{\mathrm{T}1} = \left\langle \ln \mathcal{G}(\bm{z}_i - \bm{\mu}_i, \Sigma_i) \right\rangle_{\mathcal{G}(\bm{z}_i - \bm{\mu}_i, \Sigma_i)} $ 
\vspace{1em}
\item $\bm{\mathrm{T}2} = \left\langle \ln \mathcal{G}( \xi_N, 1) \right\rangle_{\delta( \xi_N - \widehat{ \xi}_N)}$
\vspace{1em}
\item $\bm{\mathrm{T}3} = \left\langle\, \ln \mathcal{G}(\bm{z}_i, \mathds{1})  \right\rangle_{\mathcal{G}(\bm{z}_i - \bm{\mu}_i, \Sigma_i)}$
\vspace{1em}
\item $\bm{\mathrm{T}4} =\left\langle \ln\mathcal{G}\left(\bm{d}_i - f_{\widehat{\bm{\bm{\theta}}}}(z_i), t_{\bm{\psi}}(\bm{\widehat{\xi}}_N)\right) \right\rangle_{\mathcal{G}(\bm{z}_i - \bm{\mu}_i, \Sigma_i)} $
\end{itemize}
\subsection*{First energy term T1:}
\begin{flalign}
\label{teq:HQQ}
\bm{\mathrm{T}1}&= \int d\bm{z}_i \, \mathcal{G}(\bm{z}_i - \bm{\mu}_i \, , \, \Sigma_i) \, \ln \, \left( \frac{\exp \left(-\frac{1}{2}(\bm{z}_i - \bm{\mu}_i)^T \, \Sigma_i^{-1} \, (\bm{z}_i - \bm{\mu}_i)\right)}{\sqrt{| 2 \pi \Sigma_i|}} \right)  \nonumber \\
&=  \int d\bm{z}_i \, \mathcal{G}(\bm{z}_i - \bm{\mu}_i \, , \, \Sigma_i) \, \left[ -\frac{1}{2}(\bm{z}_i - \bm{\mu}_i)^T \, \Sigma_i^{-1} \, (\bm{z}_i - \bm{\mu}_i) \, -\frac{1}{2}\ln \big(\underbrace{|2 \pi \Sigma_i|}_{(2\pi)^l \, \mid\Sigma_i|}\big)\right]   \nonumber 
\intertext{The trace of a scalar is the scalar itself, resulting in $(\bm{z}_i - \bm{\mu}_i)^T \, \Sigma_i^{-1} \, (\bm{z}_i - \bm{\mu}_i) =  \mathrm{tr}\left((\bm{z}_i - \bm{\mu}_i)^T \, \Sigma_i^{-1} \, (\bm{z}_i - \bm{\mu}_i)\right)$. Further, the trace is invariant under cyclic permutations: $\mathrm{tr}\left((\bm{z}_i - \bm{\mu}_i)^T \, \Sigma_i^{-1} \, (\bm{z}_i - \bm{\mu}_i)\right) = \mathrm{tr}\left( \, \Sigma_i^{-1} \, (\bm{z}_i - \bm{\mu}_i) (\bm{z}_i - \bm{\mu}_i)^T\right)$. Finally, the trace is a linear operator and therefore commutes with the expectation:}
 &=  - \, \frac{1}{2} \, \mathrm{tr} \, \left(  \int d\bm{z}_i \, \mathcal{G}(\bm{z}_i - \bm{\mu}_i \, , \, \Sigma_i) \,  \underbrace{\Sigma_i^{-1} \, (\bm{z}_i - \bm{\mu}_i) (\bm{z}_i - \bm{\mu}_i)^T}_{= \, \Sigma^{-1}_i \, \Sigma_i \, = \, \mathds{1}} \right) \nonumber \\ 
 & \quad  -  \frac{1}{2} \int d\bm{z}_i \, \mathcal{G}(\bm{z}_i - \bm{\mu}_i \, , \, \Sigma_i) \Big[ \underbrace{\ln(|\Sigma_i|)}_{= \, \mathrm{tr}(\ln \Sigma_i)} +\, l \, \ln(2 \pi)\Big]   \nonumber \\
 &= - \, \frac{1}{2} \Big[ \underbrace{\mathrm{tr}({\mathds{1}})}_{= \, l} + \mathrm{tr}(\ln \Sigma_i) + \, l \, \ln(2 \pi) \Bigr]   \nonumber \\
 &=  - \,  \frac{1}{2} \, \Big[\mathrm{tr}(\ln \Sigma_i) + l \left( 1+ \mathrm{ln}\,(2\pi) \right) \Big]  &&
\end{flalign}
In the upper calculation, we carry the term $l \left( 1+ \mathrm{ln}\,(2\pi) \right)$ through our computations, since we change the number of latent space features $l$ in our different experiments. 
\subsection*{Second energy term T2:}
\begin{flalign}
\label{teq:HNN}
\bm{\mathrm{T}2} &= \int d \xi_N \, \ln \Biggl( \frac{1}{\sqrt{2 \pi}} \, \exp(- \frac{1}{2} \,  \xi_N^2 ) \Biggr) \, \delta( \xi_N - \widehat{ \xi_N})   \nonumber\\
 &= -\frac{1}{2} \, \widehat{ \xi_N}^2 - \underbrace{\frac{1}{2}  \ln(2 \pi)}_{= \, \mathrm{const.}}  \nonumber \\
 &= - \frac{1}{2} \, \widehat{ \xi_N}^2 + C_2  &&
 \end{flalign}

\subsection*{Third energy term T3:}
 \begin{flalign}
\label{teq:HpriorQ}
 \bm{\mathrm{T}3} &= \int d\bm{z}_i \, \mathcal{G}(\bm{z}_i - \bm{\mu}_i, \Sigma_i) \, \ln \Biggl( \frac{\exp(- \frac{1}{2} \, \bm{z}_i^T \, \mathds{1}^{-1} \, \bm{z}_i)}{\sqrt{|2 \pi \mathds{1}|}} \Biggr)   \nonumber \\
 &= -  \Bigg[ \int d\bm{z}_i \, \mathcal{G}(\bm{z}_i - \bm{\mu}_i \, , \, \Sigma_i) \, \frac{1}{2} \, \underbrace{\bm{z}_i^T \, \mathds{1}^{-1} \, \bm{z}_i}_{= \mathrm{tr}( \bm{z}_i \, \bm{z}_i^T )} + \underbrace{\int d\bm{z}_i \, \mathcal{G}(\bm{z}_i - \bm{\mu}_i \, , \, \Sigma_i) \,  \frac{1}{2} \ln(|2 \pi \mathds{1}|)}_{= \, \mathrm{const.}} \Bigg]  \nonumber \\
&=  -  \frac{1}{2} \, \mathrm{tr} \, \Biggl( \int d\bm{z}_i \, \mathcal{G}(\bm{z}_i - \bm{\mu}_i \, , \, \Sigma_i) \,  \bm{z}_i \, \bm{z}_i^T \Biggr) + C_3    \nonumber \\
\intertext{The expression $\bm{z}_i \, \bm{z}_i^T$ can be rewritten as $\bm{z}_i \, \bm{z}_i^T = (\bm{z}_i - \bm{\mu})(\bm{z}_i - \bm{\mu})^T - \bm{\mu} \bm{\mu}^T + \bm{z}_i\bm{\mu}^T + \bm{\mu} \bm{z}_i^T$. We insert this in the line above:}
&=  -  \frac{1}{2} \, \mathrm{tr} \, \Biggl(  \int d\bm{z}_i \, \mathcal{G}(\bm{z}_i - \bm{\mu}_i \, , \, \Sigma_i) \Bigl[ \underbrace{(\bm{z}_i - \bm{\mu}_i)(\bm{z}_i - \bm{\mu}_i)^T}_{= \Sigma_i} - \bm{\mu}_i \bm{\mu}_i^T + \bm{z}_i\bm{\mu}_i^T + \bm{\mu}_i \bm{z}_i^T \Bigr] \Biggr) + C_3  \nonumber \\
&= -  \frac{1}{2} \, \mathrm{tr} \, \Bigl(\underbrace{ \Sigma_i - \bm{\mu}_i \bm{\mu}_i^T +\bm{\mu}_i \bm{\mu}_i^T +\bm{\mu}_i \bm{\mu}_i^T}_{= \Sigma_i + \bm{\mu}_i \bm{\mu}_i^T} \Bigr) + C_3  \nonumber \\
&=   - \frac{1}{2} \, \mathrm{tr} \, \Bigl( \Sigma_i + \bm{\mu}_i \bm{\mu}_i^T \Bigr) + C_3  &&
\end{flalign}

\subsection*{Fourth energy term T4:}
\begin{flalign}
\label{teq:HPQ}
\bm{\mathrm{T}4} &=  \int d\bm{z}_i \,\mathcal{G}(\bm{z}_i - \bm{\mu}_i, \Sigma_i) \, \ln\mathcal{G}(\bm{d}_i - f_{\widehat{\bm{\bm{\theta}}}}(\bm{z}_i),t_{\bm{\psi}}( \widehat{ \xi_N}))   \nonumber \\
&= \int d\bm{z}_i \, \mathcal{G}(\bm{z}_i - \bm{\mu}_i, \Sigma_i) \, \ln\Biggl( \frac{\exp(-\frac{1}{2}(\bm{d}_i - f_{\widehat{\bm{\bm{\theta}}}}(\bm{z}_i))^T \, t_{\bm{\psi}}(\widehat{ \xi_N})^{-1} \, (\bm{d}_i - f_{\widehat{\bm{\bm{\theta}}}}(\bm{z}_i)))}{\sqrt{| 2 \pi \, t_{\bm{\psi}}(\widehat{ \xi_N})|}} \Biggr)   \nonumber \\
&= -\int d\bm{z}_i \, \mathcal{G}(\bm{z}_i - \bm{\mu}_i \, , \, \Sigma_i) \, \Big[ \frac{1}{2} (\bm{d}_i - f_{\widehat{\bm{\bm{\theta}}}}(\bm{z}_i))^T \, t_{\bm{\psi}}(\widehat{ \xi_N})^{-1} \, (\bm{d}_i - f_{\widehat{\bm{\bm{\theta}}}}(\bm{z}_i)) \,   \nonumber \\
&\quad + \frac{1}{2}\ln(\underbrace{|2 \pi \, t_{\bm{\psi}}(\widehat{ \xi_N})|}_{= (2\pi)^k \, |t_{\bm{\psi}}(\widehat{ \xi_N})|})\Big]  \nonumber \\
&= - \, \frac{1}{2}\int d\bm{z}_i \, \mathcal{G}(\bm{z}_i - \bm{\mu}_i \, , \, \Sigma_i) \, \Big[(\bm{d}_i - f_{\widehat{\bm{\bm{\bm{\theta}}}}}(\bm{z}_i))^T \, t_{\bm{\psi}}(\widehat{ \xi_N})^{-1} \, (\bm{d}_i - f_{\widehat{\bm{\bm{\bm{\theta}}}}}(\bm{z}_i)) \Big]\nonumber \\ &\quad - \frac{1}{2} \int d\bm{z}_i \, \mathcal{G}(\bm{z}_i - \bm{\mu}_i \, , \, \Sigma_i)\, \big[\underbrace{k \, \ln (2\pi)}_{= \mathrm{const.}} + \mathrm{tr}(\ln \, t_{\bm{\psi}}(\widehat{ \xi_N}))\big]   \nonumber \\
&= - \, \frac{1}{2}\int d\bm{z}_i \, \mathcal{G}(\bm{z}_i - \bm{\mu}_i \, , \, \Sigma_i) \,\mathrm{tr}\Big[  t_{\bm{\psi}}(\widehat{ \xi_N})^{-1} \, (\bm{d}_i - f_{\widehat{\bm{\bm{\bm{\theta}}}}}(\bm{z}_i))\,(\bm{d}_i - f_{\widehat{\bm{\bm{\bm{\theta}}}}}(\bm{z}_i))^T \Big] \nonumber \\ 
&\quad \, - \,\frac{1}{2}\,\mathrm{tr}\,(\ln \, t_{\bm{\psi}}(\widehat{ \xi_N})) + C_4  &&
\end{flalign}

We cannot exactly evaluate the integral in this expression using analytic techniques. Fortunately, we can make use of Monte Carlo methods to approximate the expectation with a finite sum:
\begin{flalign}
\label{teq:MCestimator}
\int d\bm{z}_i \, \mathcal{G}(\bm{z}_i - \bm{\mu}_i \, , \, \Sigma_i) \, \,\mathrm{tr}\Big[  t_{\bm{\psi}}(\widehat{ \xi_N})^{-1} \, (\bm{d}_i - f_{\widehat{\bm{\bm{\bm{\theta}}}}}(\bm{z}_i))\,(\bm{d}_i - f_{\widehat{\bm{\bm{\bm{\theta}}}}}(\bm{z}_i))^T \Big]  
 \approx \frac{1}{M} \, \sum_{m=1}^M  \, \mathrm{tr} \,\Big( t_{\bm{\psi}}(\widehat{ \xi_N})^{-1} \, (\bm{d}_i - f_{\widehat{\bm{\bm{\bm{\theta}}}}}(\bm{z}_i^{(m)}))\,(\bm{d}_i - f_{\widehat{\bm{\bm{\bm{\theta}}}}}(\bm{z}_i^{(m)}))^T \Big).  
\end{flalign}
With this method we can achieve a high accuracy estimator (even with a small number of samples) for the expectation value, as long as we draw samples $\bm{z}_i^{(m)}$ from the distribution $\mathcal{G}(\bm{z}_i - \bm{\mu}_i \, , \, \Sigma_i)$ \citep{bishop2006pattern}. This means the samples $\bm{z}_i$ are a function of $\bm{\mu}_i $ and $\Sigma_i$ by $\bm{z}_i = \bm{\mu}_i + \sqrt{\Sigma_i} \, \bm{\epsilon}_i$ with an auxiliary variable $\bm{\epsilon}_i$ and $P(\bm{\epsilon}_i) = \mathcal{G}(\bm{\epsilon}_i, \mathds{1})$ (see Sect.~\ref{ch:NEAT_VAE} for details). We sample once for each pixel, i.e. $M=1$. The full Kullback-Leibler Divergence \eqref{teq:KLD} then reads

\begin{empheq}[box=\fbox]{align}
\label{appeq:KLD}
 D_{KL}[Q_{\bm{\Phi}}(\cdot) \mid\mid P(\cdot)] &= \frac{1}{2} \sum_{i=1}^p \Bigg[ - \mathrm{tr} \,(\ln \Sigma_i) - l \left( 1+ \mathrm{ln}\,(2\pi) \right) +  \frac{1}{p} \, \widehat{ \xi_N}^2+ \, \mathrm{tr} \, \Bigl( \Sigma_i + \bm{\mu}_i \bm{\mu}_i^T \Bigr)    +\, \mathrm{tr}\, \left( \frac{1}{ t_{\bm{\psi}}(\widehat{ \xi_N})} (\bm{d}_i - f_{\widehat{\bm{\bm{\bm{\theta}}}}}(\bm{z}_i))(\bm{d}_i - f_{\widehat{\bm{\bm{\bm{\theta}}}}}(\bm{z}_i))^T \right)  \\ \nonumber
 & \quad \quad \quad \quad  + \mathrm{tr}\,(\ln \, t_{\bm{\psi}}(\widehat{ \xi_N}) ) \Bigg] +  \mathcal{H}_0,  
\end{empheq}
where we absorbed all constant terms in a redefined $\mathcal{H}_0$.

\section{Hyperparameters}
\label{app:hyperparameters}
Hyperparameters for the NEAT-VAE are (1) the number of network layers, (2) the number of hidden neurons per layer, (3) the number of neurons in the bottleneck layer, (4) the mean $\mu_N$ and (5) the standard deviation $\sigma_N$ of the log-normal model for the noise covariance transformation, the learning rates of (6) network weights and (7) noise parameter weights, and (8) the batch size. We examined 50 configurations of the NEAT-VAE, which consist of different arrangements of hyperparameters. The values for the hyperparameters were randomly chosen from limited intervals, which we specified for each hyperparameter based on prior experiments (see Table \ref{table:hyperparameters}). We implemented rectified linear unit (ReLU) functions as activation functions in each layer except the output layer and used the Adam optimizer \citep{kingma2014adam} to update the network weights ${\bm{\phi}}$ and $\widehat{\bm{\theta}}$, and the latent noise value $\widehat{\xi_N}$. We trained each configuration for $(10^5 \times \mathrm{batchsize})/p$ epochs with $p$ denoting the number of pixels. We build our NEAT-VAE as a descriptive rather than a predictive model, since we aim to learn the underlying, independent features generating one certain data set. For this reason, we do not split the data into training, validation and test sets, nor do we analyze overfitting. For reproducibility, we fixed the random seed to $123$.

\begin{table}[h!]
\caption[Hyperparameters of NEAT-VAE]{Hyperparameters of NEAT-VAE. The number of layers, neurons per layer and bottleneck neurons determine the network architecture. $\mu_N$ and $\sigma_N$ are transformation parameters of the noise covariance matrix $N$. The optimization of learnable parameters is determined by the learning rates (LR) for network weights ${\bm{\phi}}$, $\widehat{\bm{\theta}}$ and the latent noise $\widehat{\xi_N}$, which are tuned to minimize the objective function in Eq. \eqref{appeq:KLD}. When using mini-batching, the batch size determines how many data samples are used to compute the loss function before back propagation and model updating is performed.}
\label{table:hyperparameters}
\centering
 \begin{tabular}{|l|l|}
  \hline
  \\[-1em]
  Hyperparameter  & Sampling sets \\
  \\[-1em]
  \hline
  \\[-1em]
  Layers &  $\{6, 8, 10, 16, 24\}$\\
  \\[-1em]
  Hidden neurons &  $\{30, \ldots, 37\}$\\
  \\[-1em]
  Bottleneck neurons &  $\{10, \ldots, 35\}$\\
  \\[-1em]
  $\mu_N$ &  $\{-11, -9.21, -8, -5, -1\}$\\ 
  \\[-1em]
  $\sigma_N$ &  $\mathrm{Unif}[1, 2]$\\
  \\[-1em]
  LR network weights &  $\{0.005, 0.001, 0.0005, 0.01\}$\\
  \\[-1em]
  LR $\widehat{\xi}_N$ &  $\{0.0025, 0.0005, 0.00025, 0.005\}$\\
  \\[-1em]
  Batch size &  $\{16, 64, 128, 256, 512\}$\\
 \hline  
  \end{tabular}
\end{table}

\clearpage
\newpage
\section{Other features}
\label{app:insignificantfeat}
In the experiment discussed in Sect.~\ref{ch:results}, our NEAT-VAE algorithm mapped $35$ Galactic all-sky maps to ten latent space features that encode the essential information required to reconstruct the input again. The latent space exhibits an order in significance of the features, which we measured by the ratio of mean fluctuations compared to feature posterior variance (see Eq. \eqref{eq:Sign}). Features A, B and C, which have the highest significance, are shown in Fig.~\ref{fig:SignFeat}. The remaining seven features (with their mean and variance in HEALPix representation) are displayed in the following in order of descending significance. By a visual analysis, we can recognize artifacts, for example, of the IRIS scanning scheme, in features D, F, G and H. Feature H also seems to encode structures near the Galactic plane of the $\mathrm{H}_{\alpha}$ map, the mean of feature J encodes structures similar to the Fermi Bubbles in high energy $\gamma$-ray data. All-sky images of the $35$ Galactic input data are displayed in the leftmost columns of Appendix \ref{app:gradientmaps}. 

\begin{figure}[h!]
\begin{subfigure}{\textwidth}
\centering
  \includegraphics[width=0.45\linewidth]{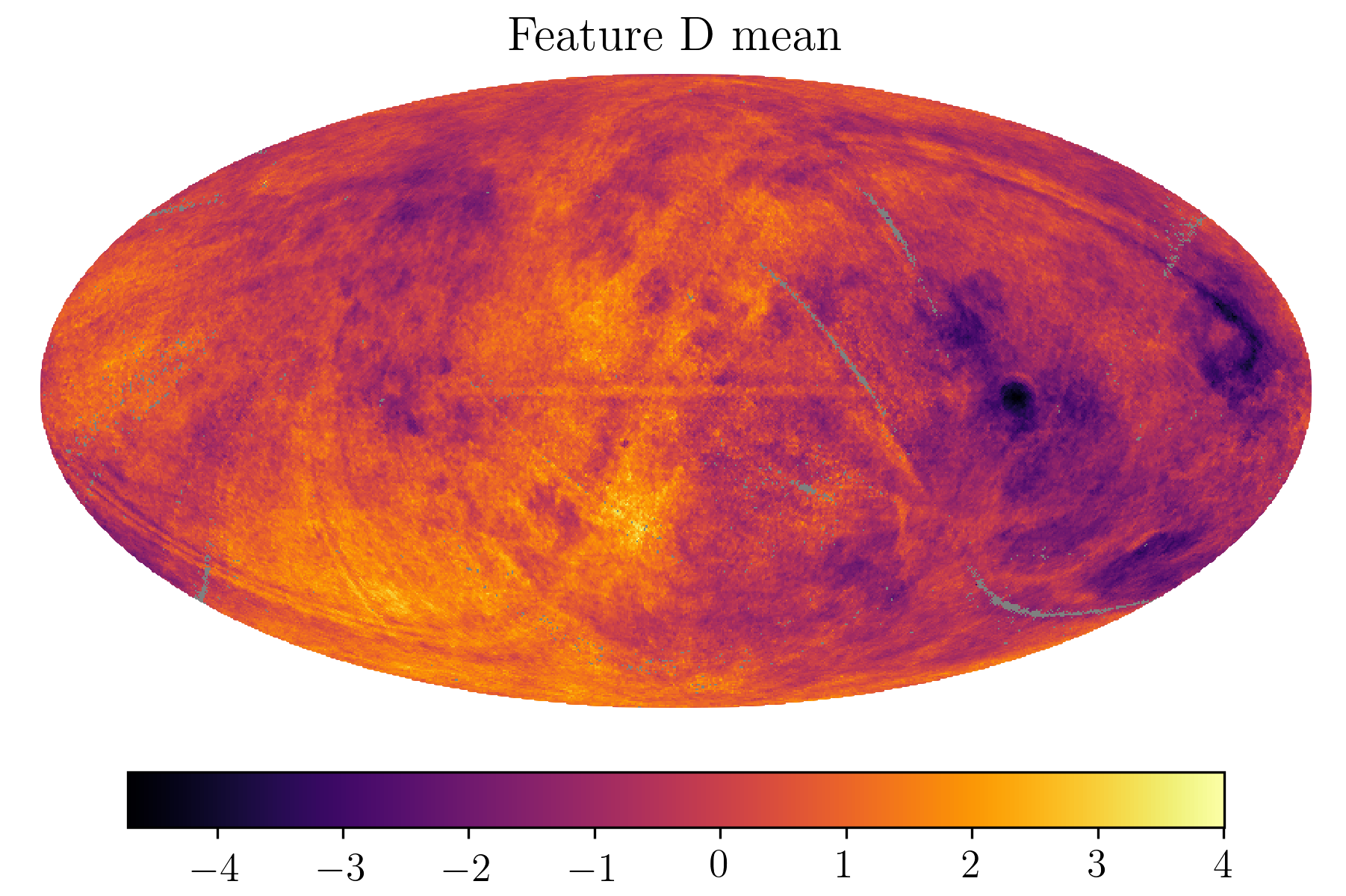}  
  \includegraphics[width=0.45\linewidth]{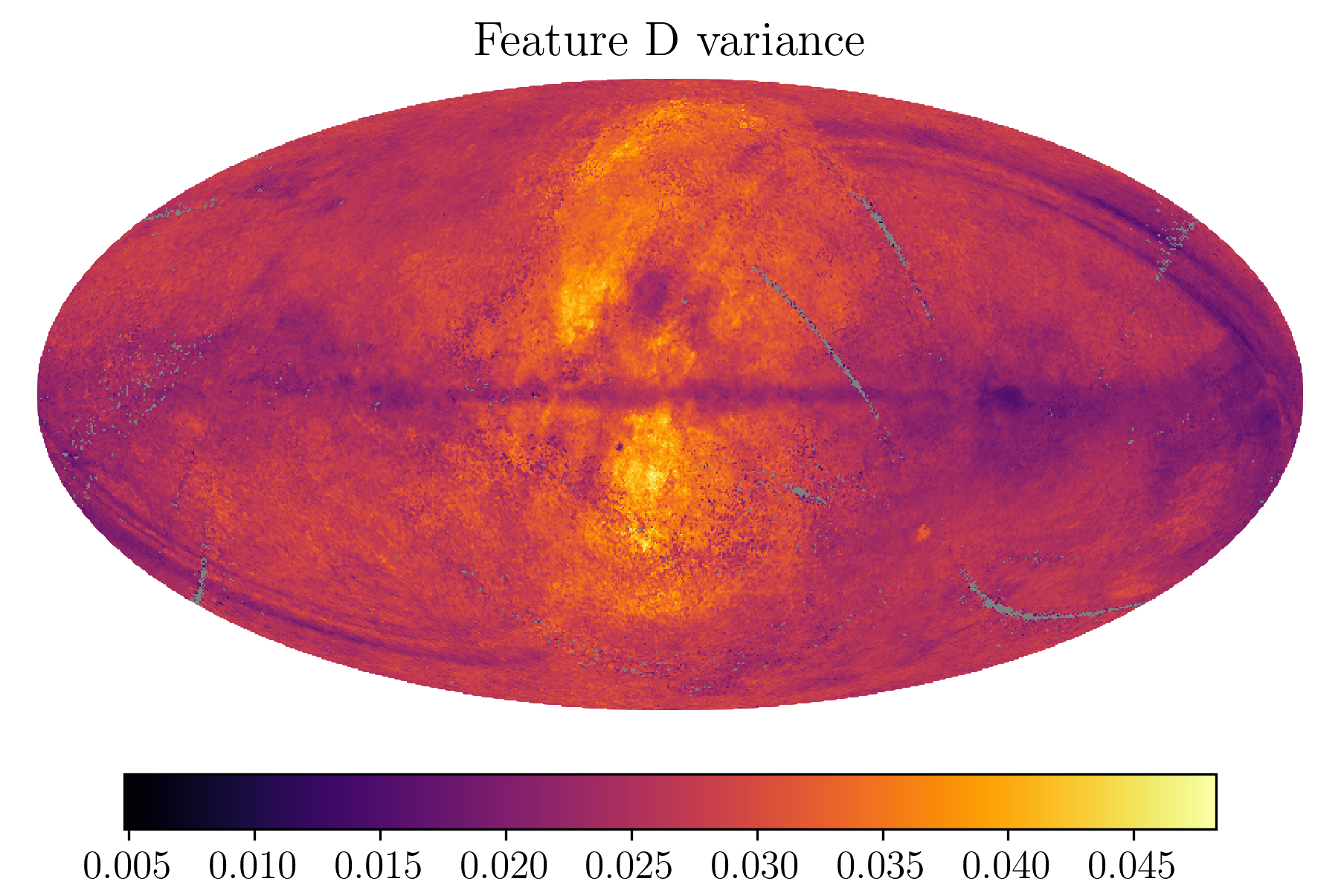}  
\end{subfigure}
\par\bigskip 
\begin{subfigure}{\textwidth}
\centering
  \includegraphics[width=0.45\linewidth]{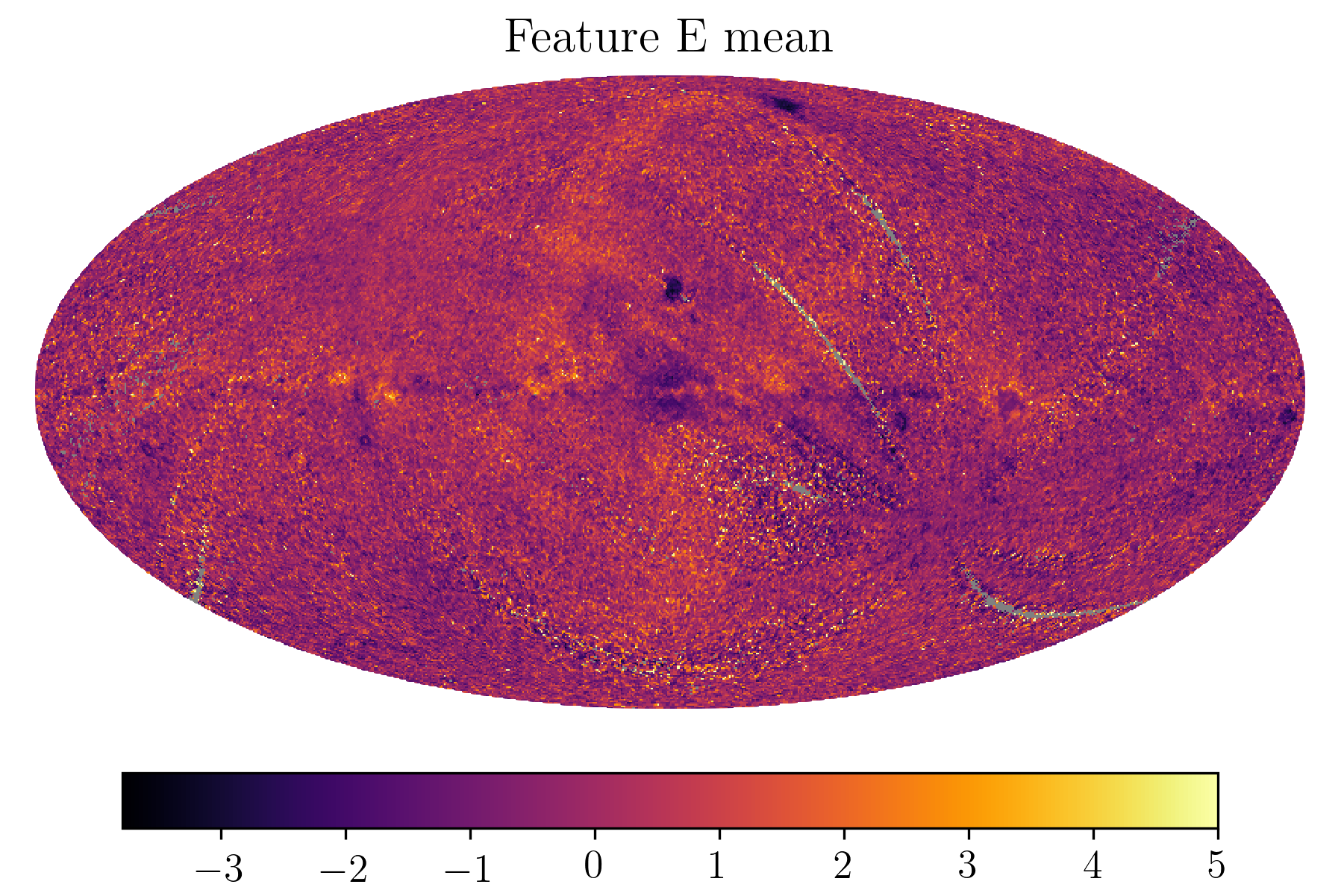}  
  \includegraphics[width=0.45\linewidth]{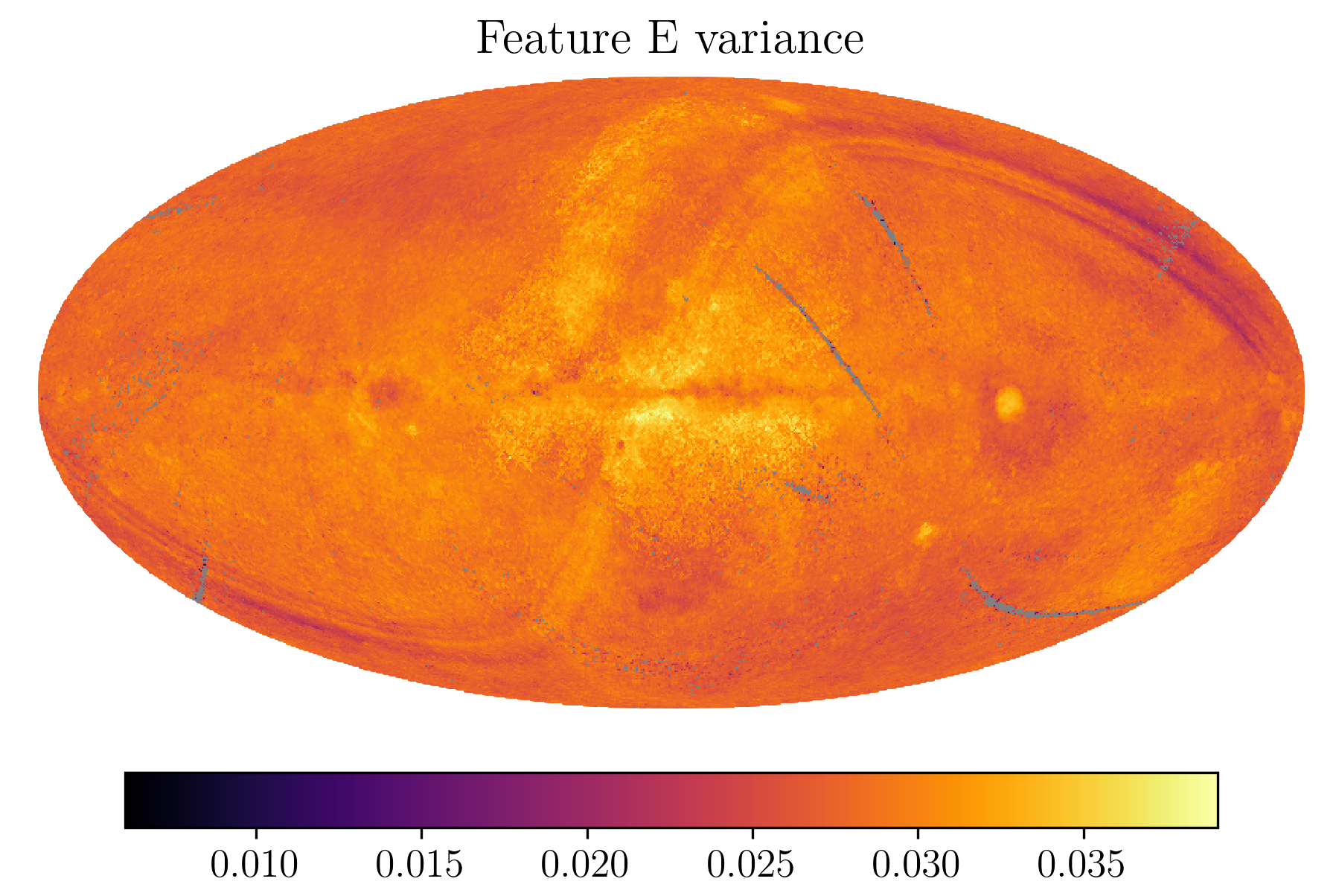}  
\end{subfigure}
\par\bigskip 
\begin{subfigure}{\textwidth}
\centering
  \includegraphics[width=0.45\linewidth]{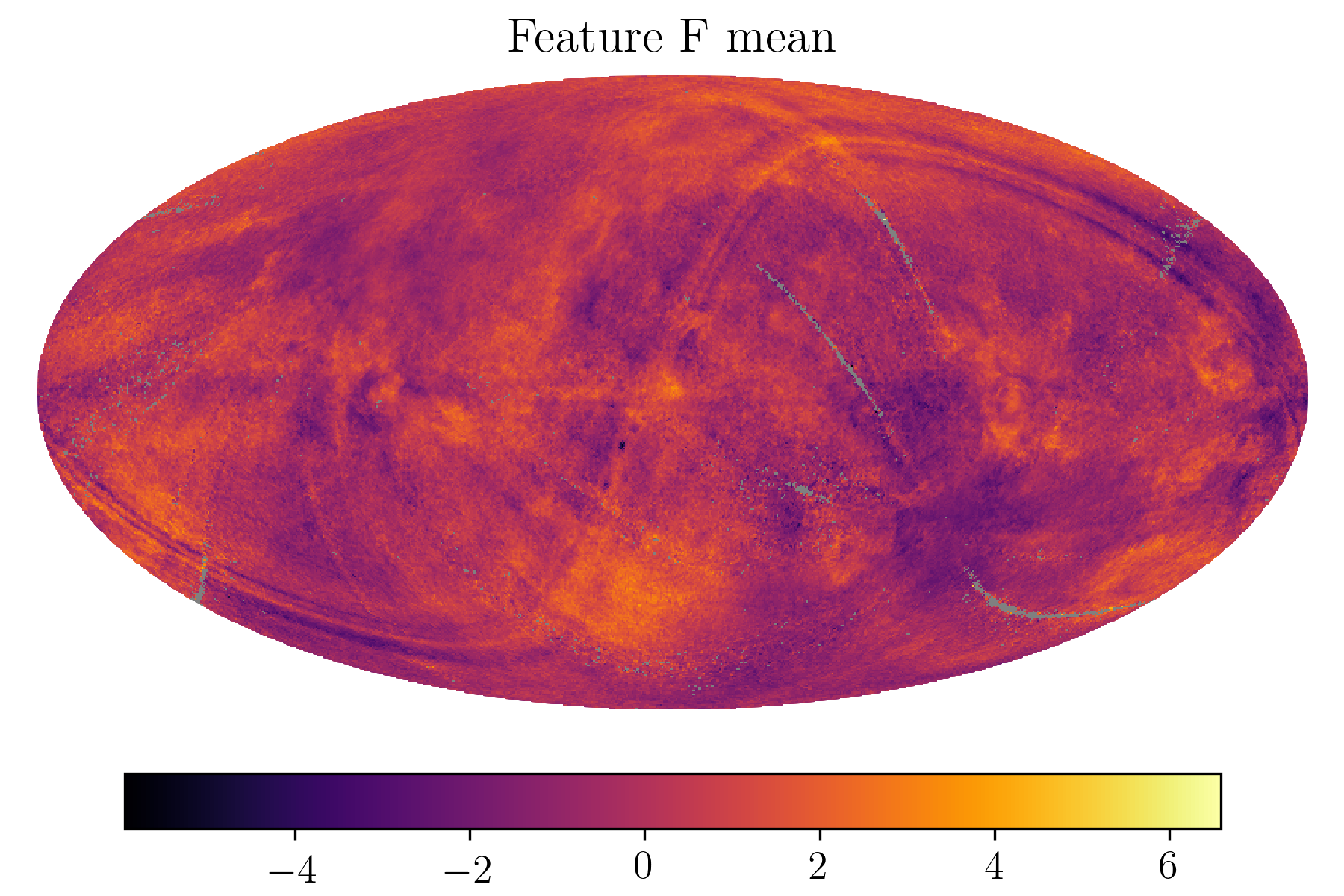}  
  \includegraphics[width=0.45\linewidth]{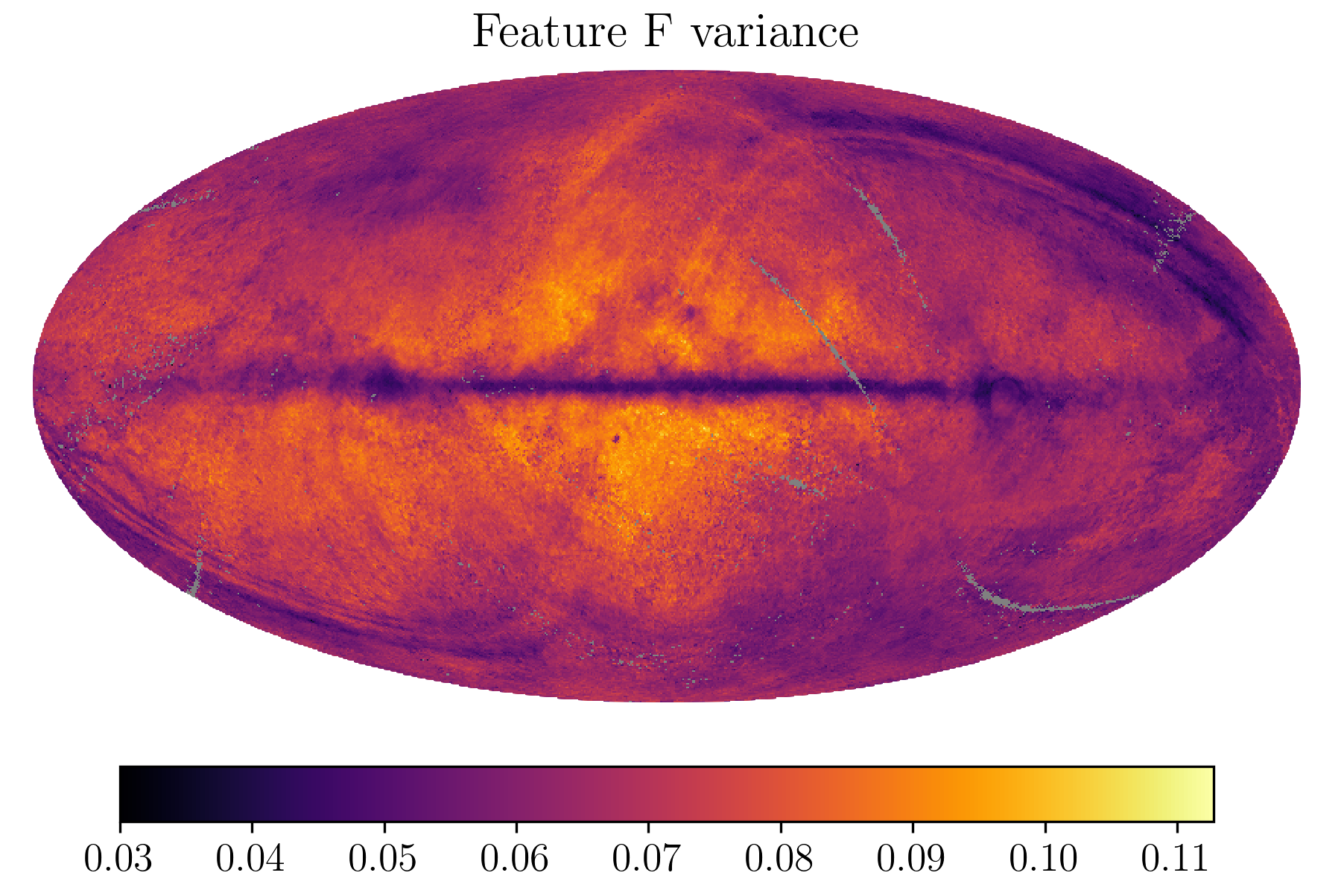}  
\end{subfigure}
\caption[Full latent space]{Full latent space representation with 10 neurons in descending order of significance. Significance values are: $S(\mathrm{feature \, D}) = 38.18$, $S(\mathrm{feature \, E}) = 34.07$, and $S(\mathrm{feature \, F}) = 13.77$.}
\label{fig:latspacefull}
\end{figure}

\begin{figure}\ContinuedFloat
\begin{subfigure}{\textwidth}
\centering
  \includegraphics[width=0.45\linewidth]{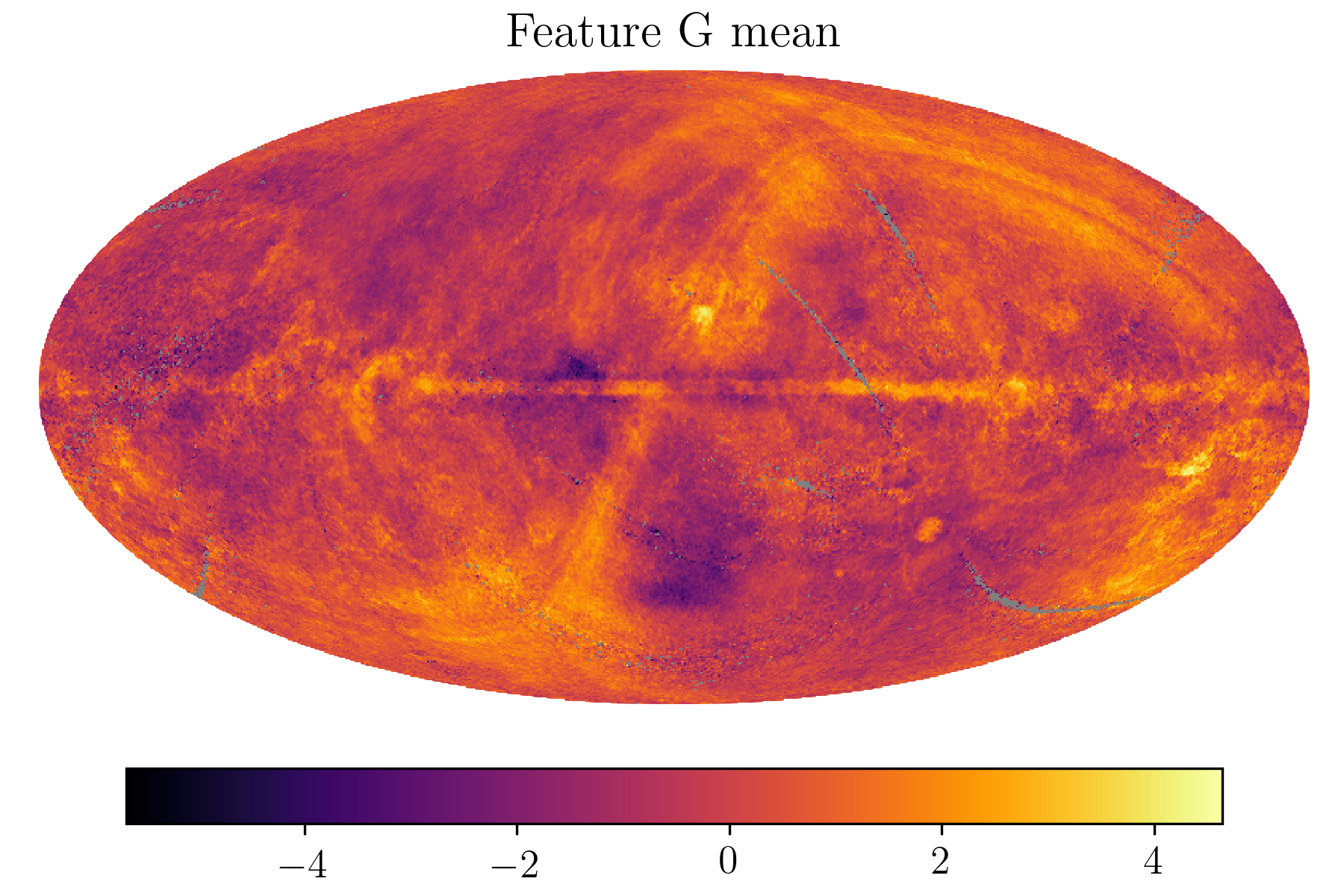}  
  \includegraphics[width=0.45\linewidth]{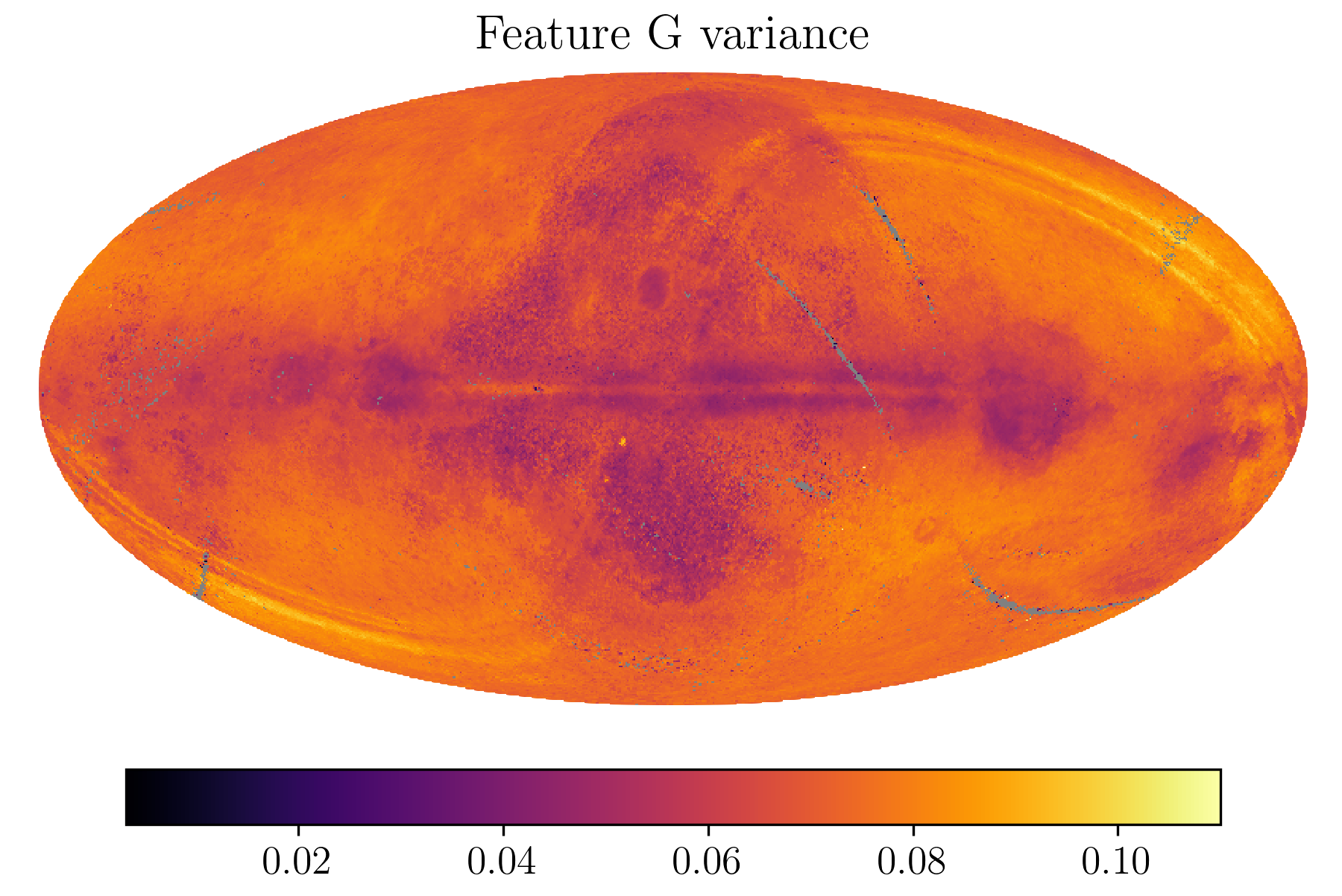}  
\end{subfigure}
\par\bigskip 
\begin{subfigure}{\textwidth}
\centering
  \includegraphics[width=0.45\linewidth]{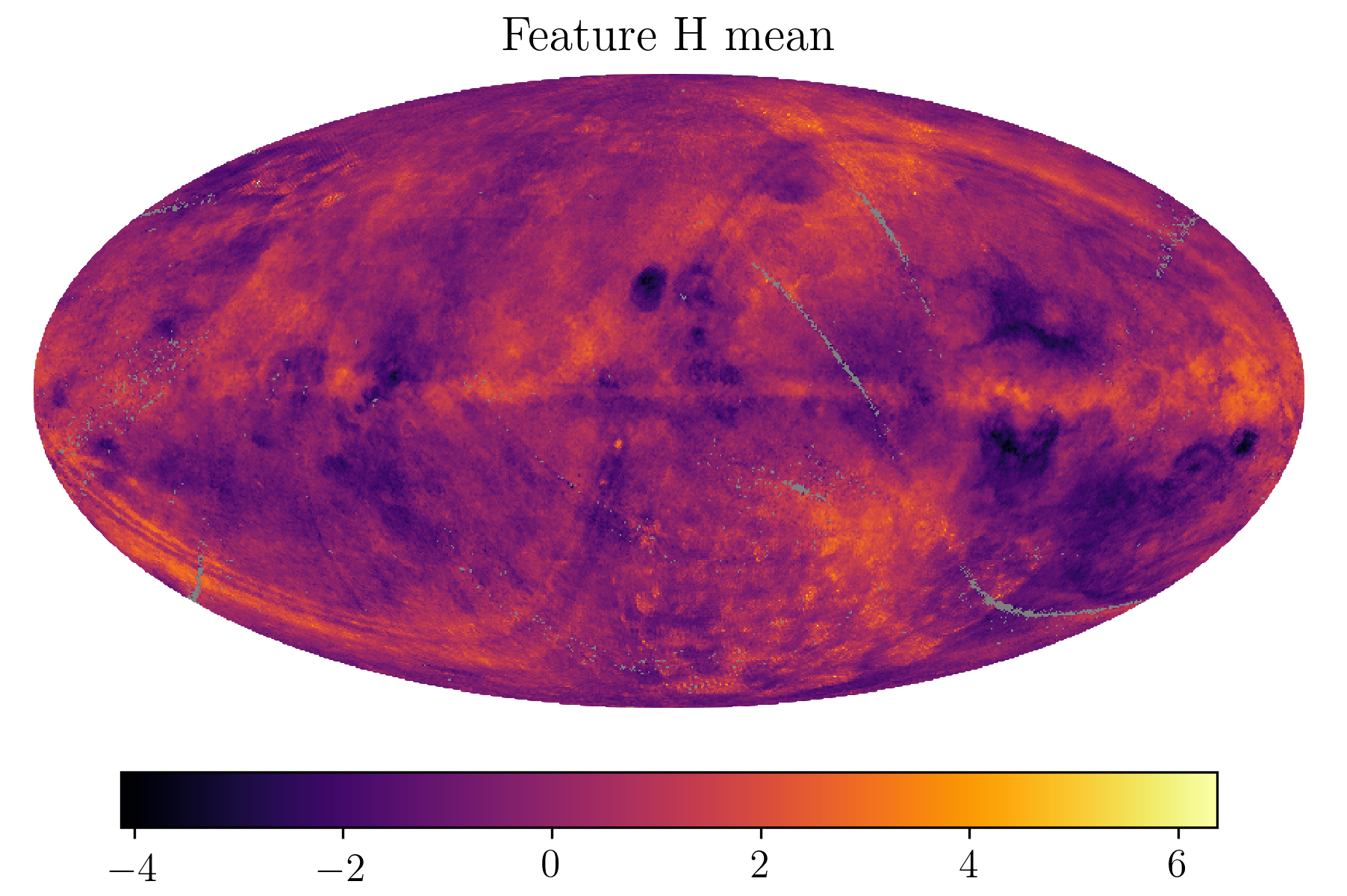}  
  \includegraphics[width=0.45\linewidth]{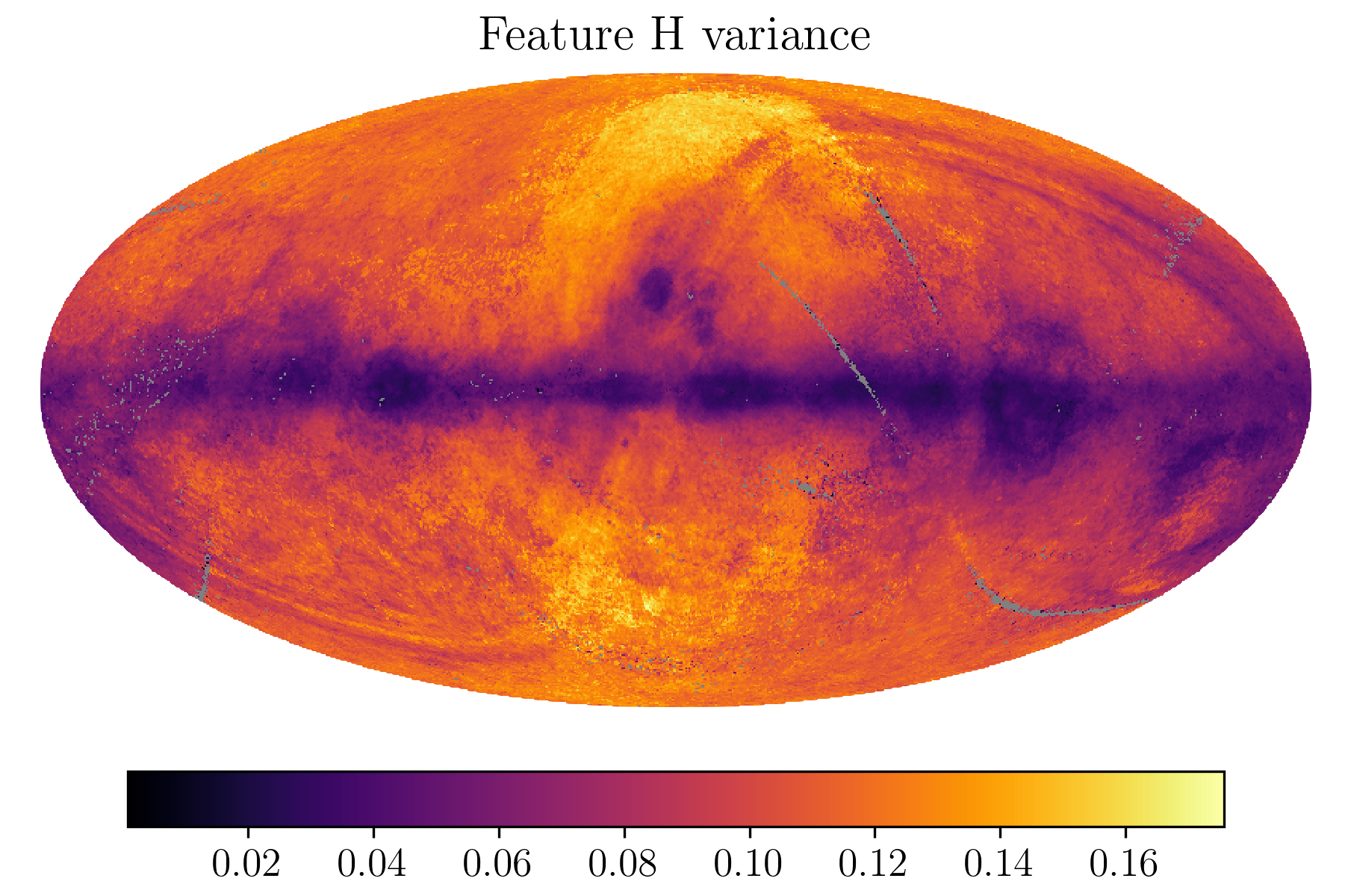}  
\end{subfigure}
\par\bigskip 
\begin{subfigure}{\textwidth}
\centering
  \includegraphics[width=0.45\linewidth]{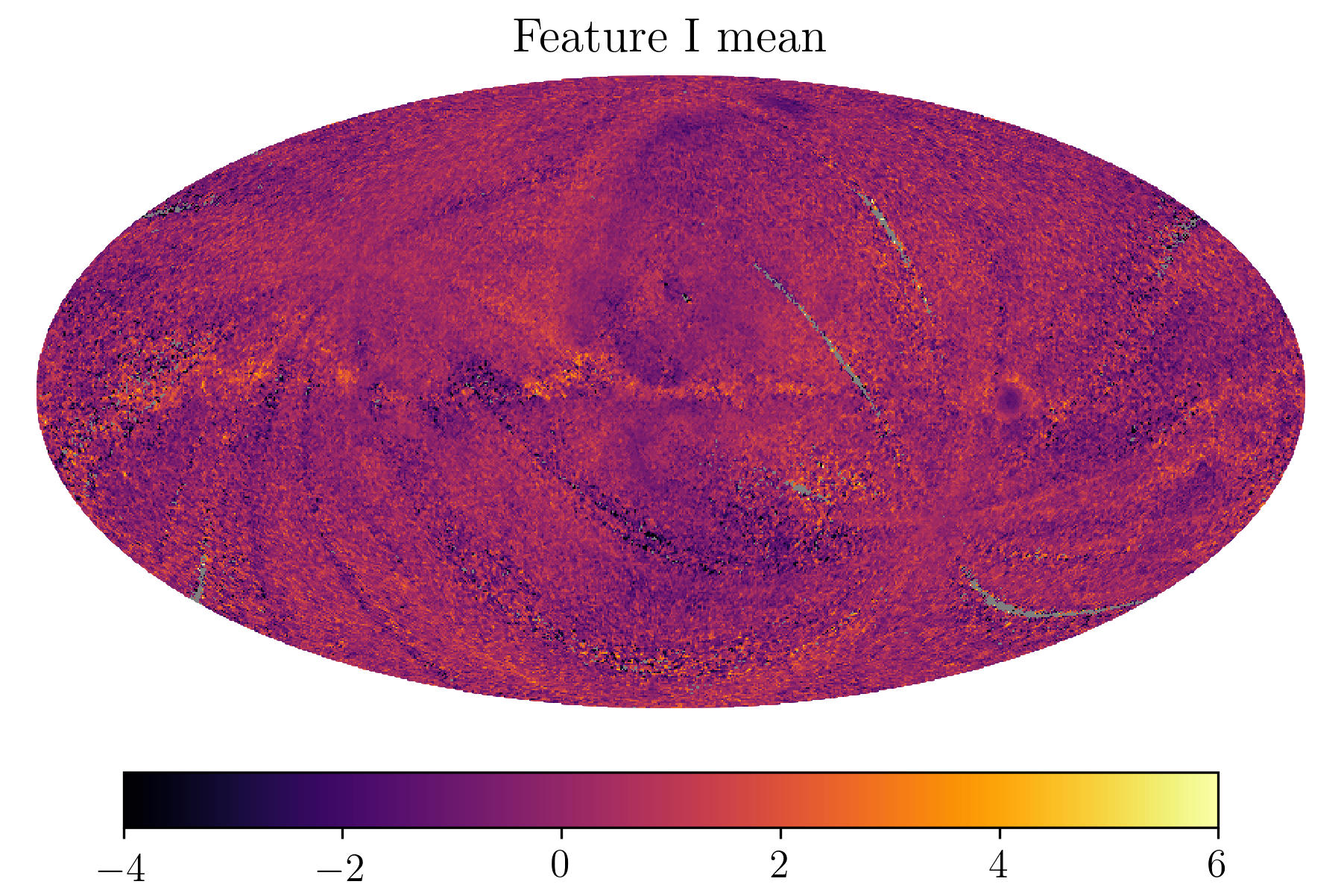}  
  \includegraphics[width=0.45\linewidth]{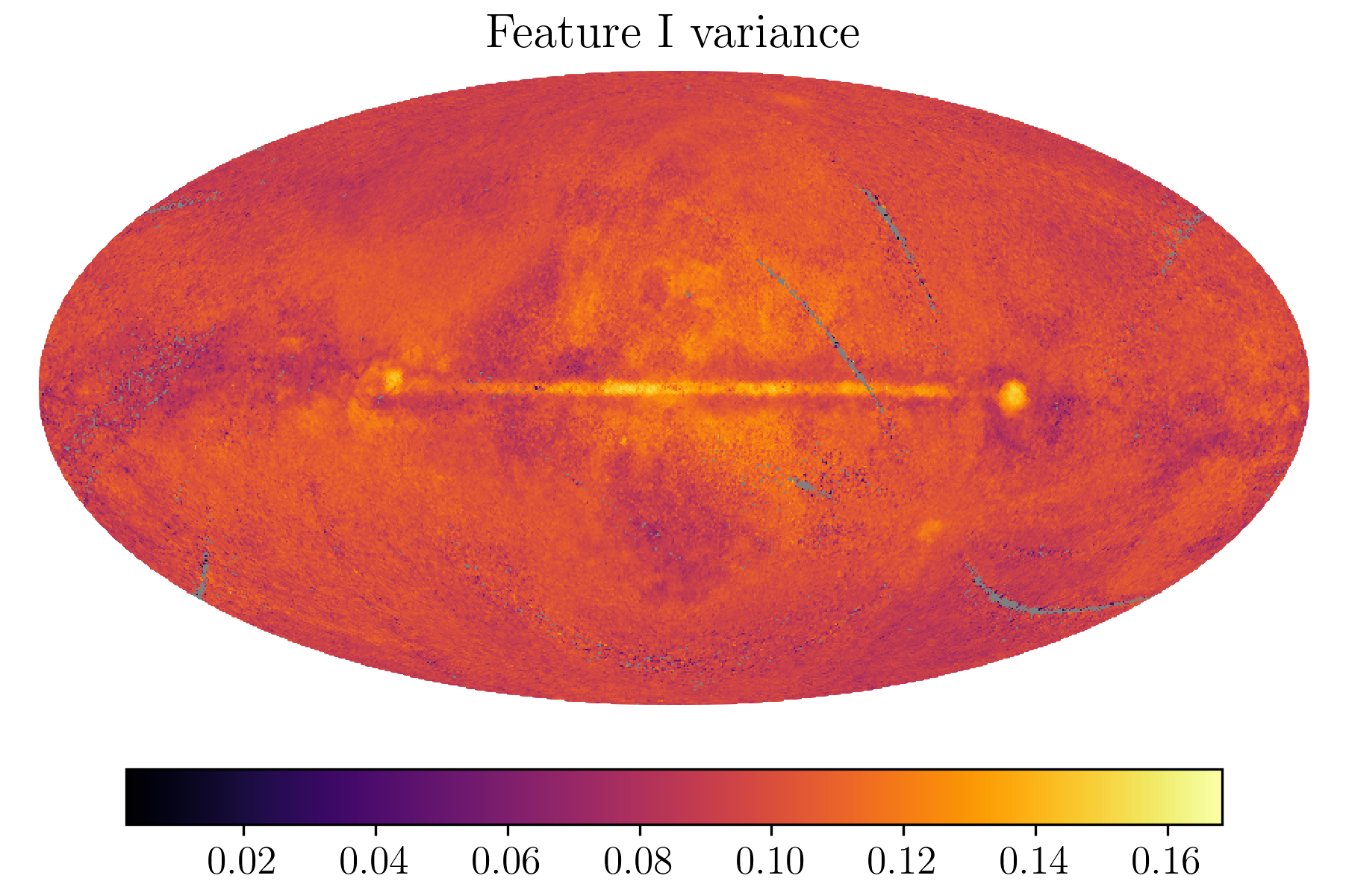}  
\end{subfigure}
\par\bigskip 
\begin{subfigure}{\textwidth}
\centering
  \includegraphics[width=0.45\linewidth]{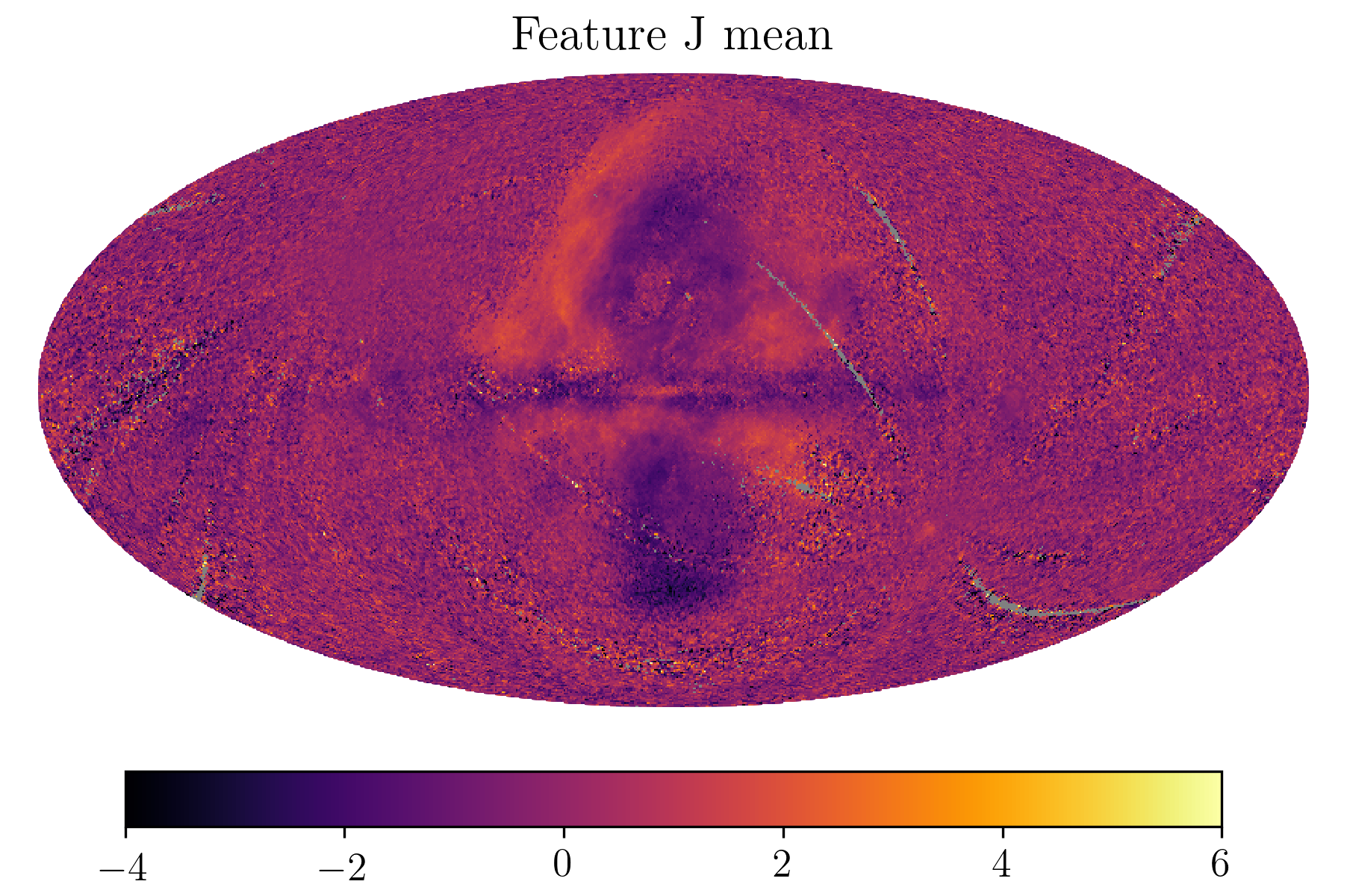}  
  \includegraphics[width=0.45\linewidth]{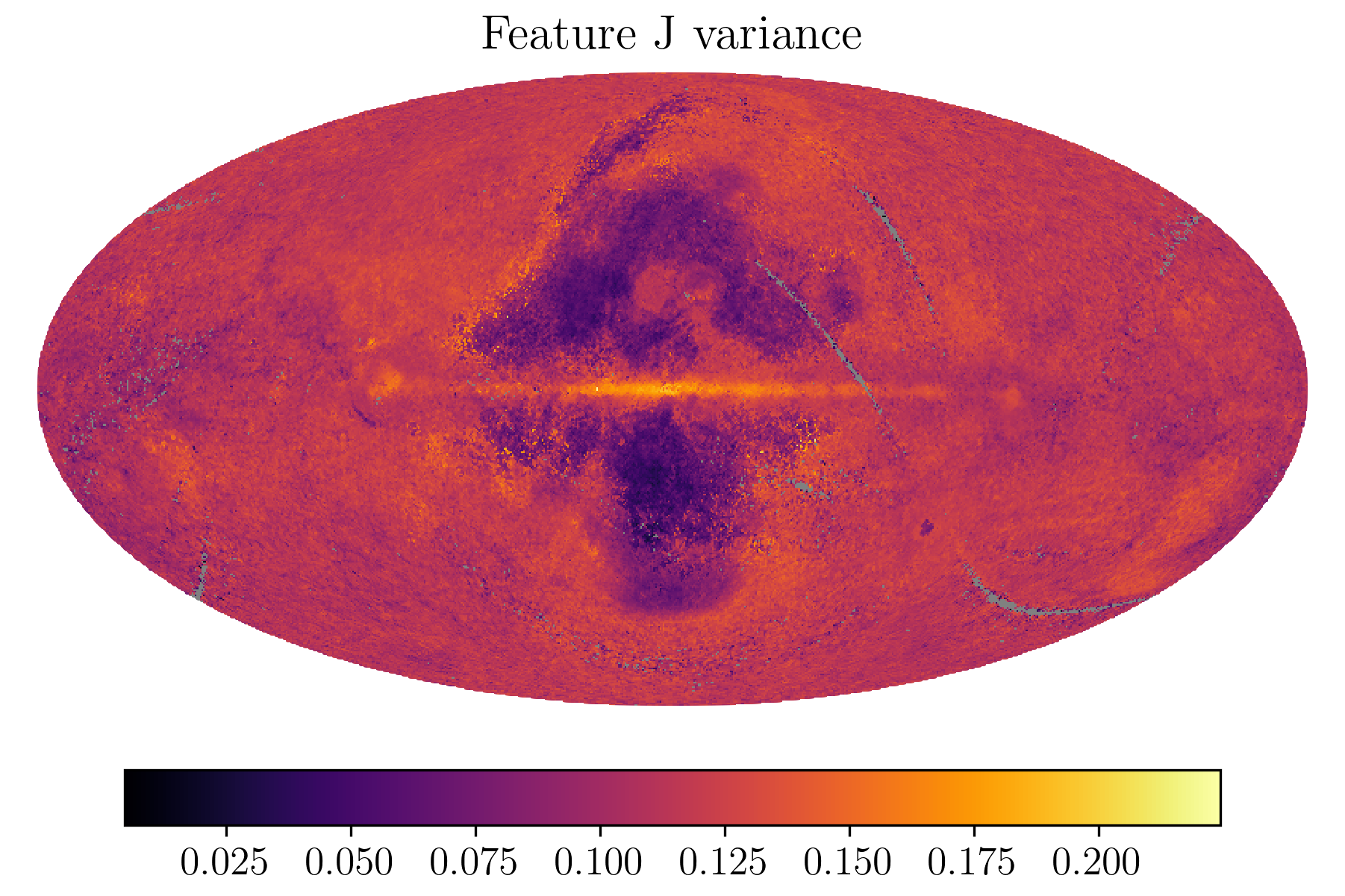}  
\end{subfigure}
\caption[Full latent space]{Least significant latent space feature maps. Significance values are: $S(\mathrm{feature \, G}) = 13.63$, $S(\mathrm{feature \, H}) = 11.85$, $S(\mathrm{feature \, I}) = 10.34$, and $S(\mathrm{feature \, J}) = 8.75$.}
\label{fig:latspacefull}
\end{figure}
\clearpage

\newpage
\section{Decoder Jacobian maps}
\label{app:gradientmaps}
The following panels show the gradients of reconstructed Galactic all-sky maps (that is the output of our NEAT-VAE algorithm) with respect to the latent space features A, B and C. The Galactic input data is displayed in the leftmost column, while the single features mark the top row. The resulting gradient values in each pixel are shown as a HEALPix map connecting a Galactic map and a feature map, with red colors indicating positive gradient values and blue colors denoting negative gradient values.
\begin{figure}[h]
\begin{subfigure}{\textwidth}
\centering
\includegraphics[width=\textwidth]{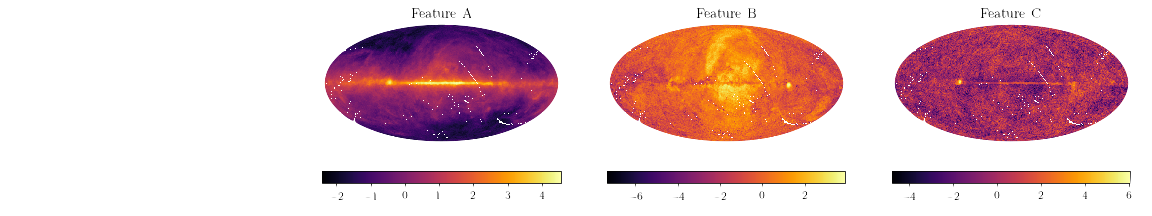}
\end{subfigure}
\begin{subfigure}{\textwidth}
\centering
\includegraphics[width=\textwidth]{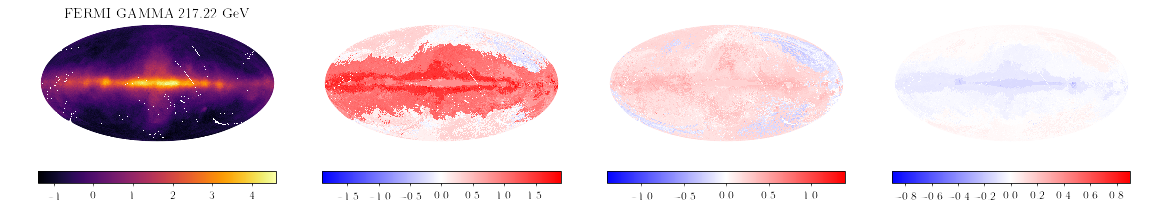}
\end{subfigure}
\begin{subfigure}{\textwidth}
\centering
\includegraphics[width=\textwidth]{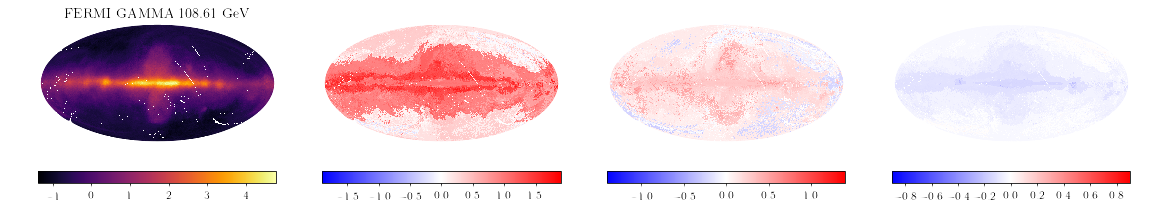}
\end{subfigure}
\begin{subfigure}{\textwidth}
\centering
\includegraphics[width=\textwidth]{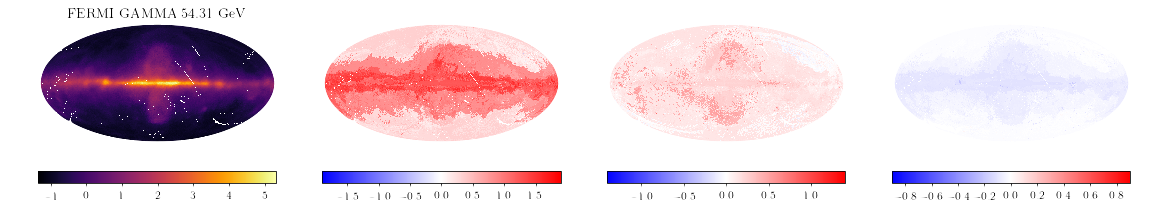}
\end{subfigure}
\begin{subfigure}{\textwidth}
\centering
\includegraphics[width=\textwidth]{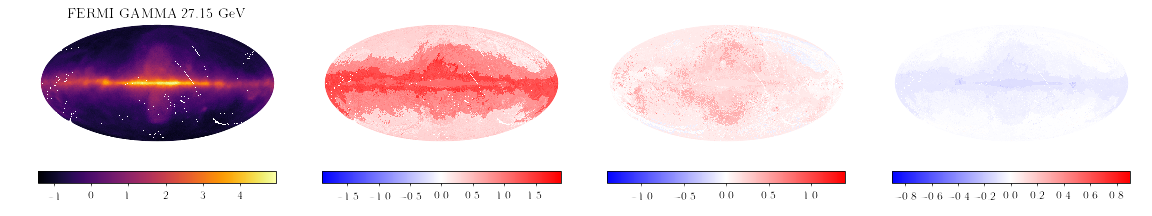}
\end{subfigure}
\begin{subfigure}{\textwidth}
\centering
\includegraphics[width=\textwidth]{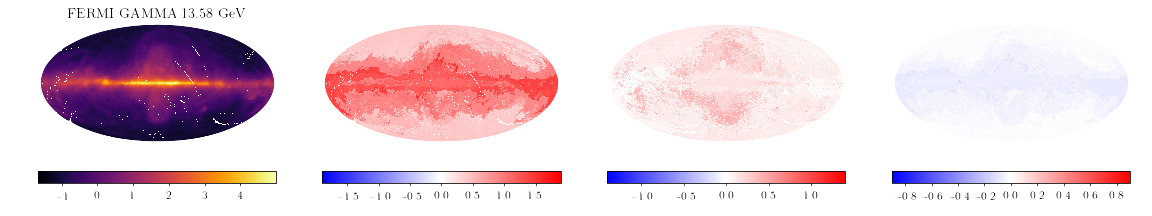}
\end{subfigure}
\caption[Gradient maps]{Decoder Jacobian maps. Derivatives of reconstructed Galactic all-sky maps with respect to latent feature maps A, B and C (top panel). }
\label{fig:grads}
\end{figure}

\begin{figure}\ContinuedFloat
\begin{subfigure}{\textwidth}
\centering
\includegraphics[width=\textwidth]{{Gradsfeaturemaps}.png}
\end{subfigure}
\begin{subfigure}{\textwidth}
\centering
\includegraphics[width=\textwidth]{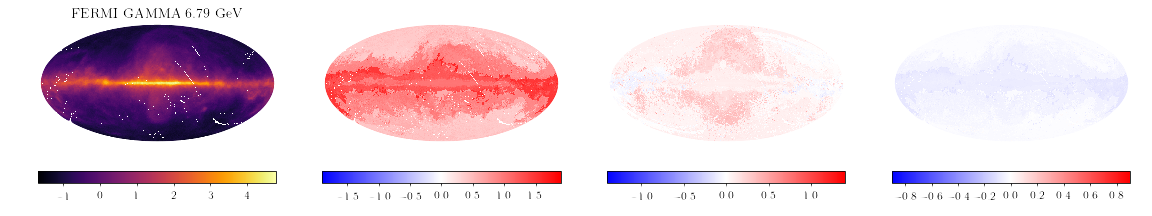}
\end{subfigure}
\begin{subfigure}{\textwidth}
\centering
\includegraphics[width=\textwidth]{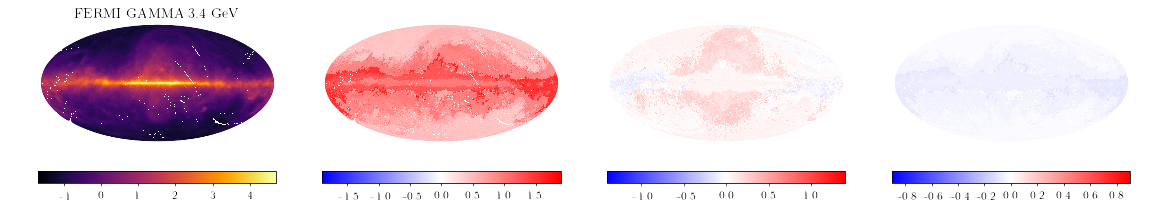}
\end{subfigure}
\begin{subfigure}{\textwidth}
\centering
\includegraphics[width=\textwidth]{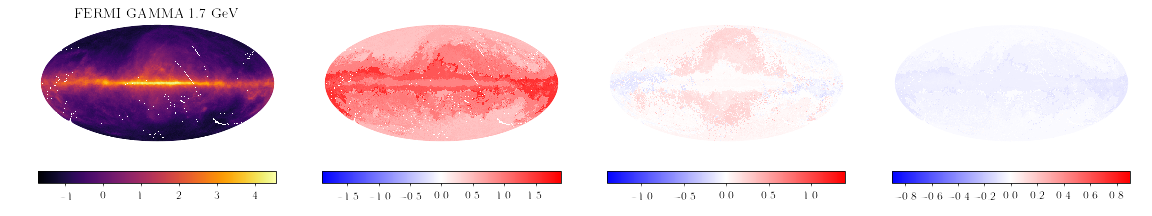}
\end{subfigure}
\begin{subfigure}{\textwidth}
\centering
\includegraphics[width=\textwidth]{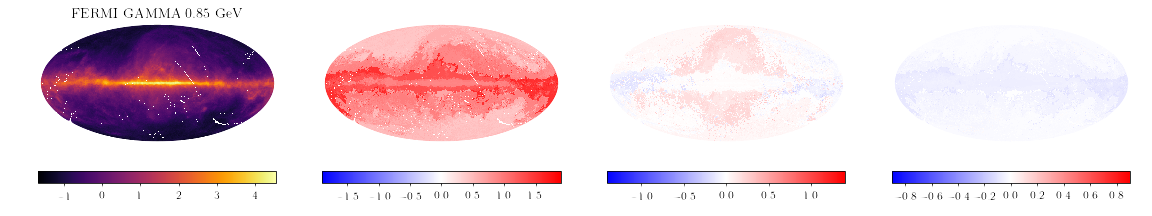}
\end{subfigure}
\begin{subfigure}{\textwidth}
\centering
\includegraphics[width=\textwidth]{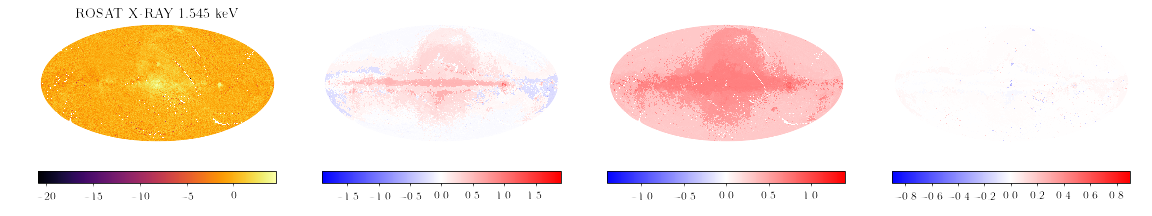}
\end{subfigure}
\begin{subfigure}{\textwidth}
\centering
\includegraphics[width=\textwidth]{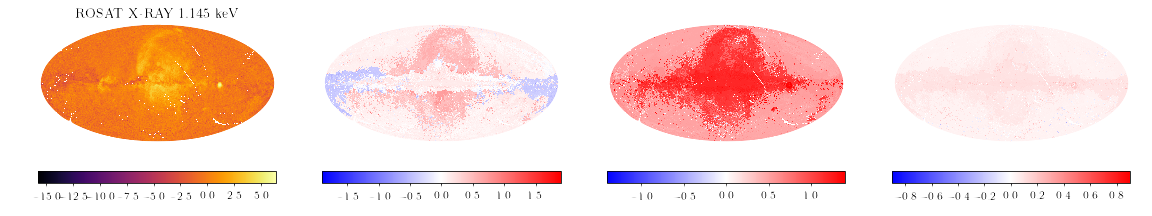}
\end{subfigure}
\caption[Gradient maps]{Decoder Jacobian maps continued.}
\label{fig:grads}
\end{figure}

\begin{figure}\ContinuedFloat
\begin{subfigure}{\textwidth}
\centering
\includegraphics[width=\textwidth]{{Gradsfeaturemaps}.png}
\end{subfigure}
\begin{subfigure}{\textwidth}
\centering
\includegraphics[width=\textwidth]{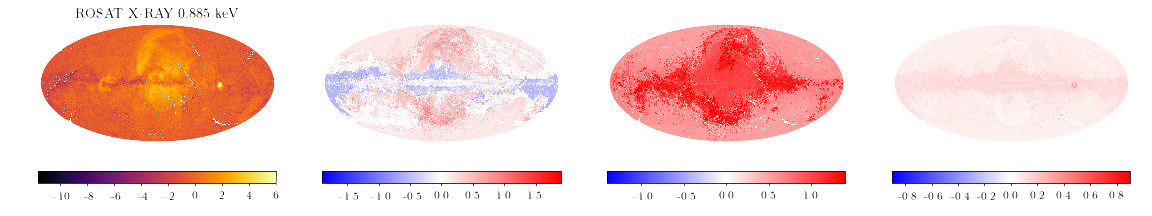}
\end{subfigure}
\begin{subfigure}{\textwidth}
\centering
\includegraphics[width=\textwidth]{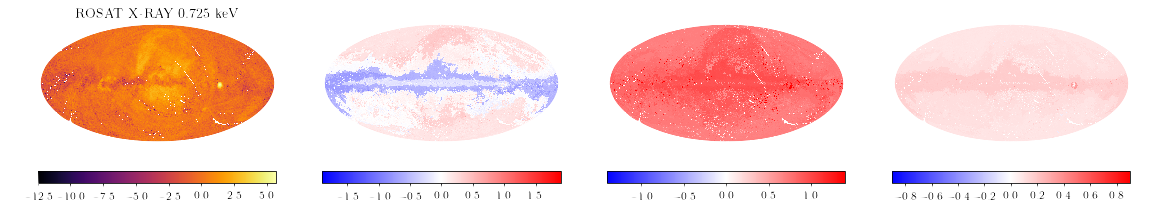}
\end{subfigure}
\begin{subfigure}{\textwidth}
\centering
\includegraphics[width=\textwidth]{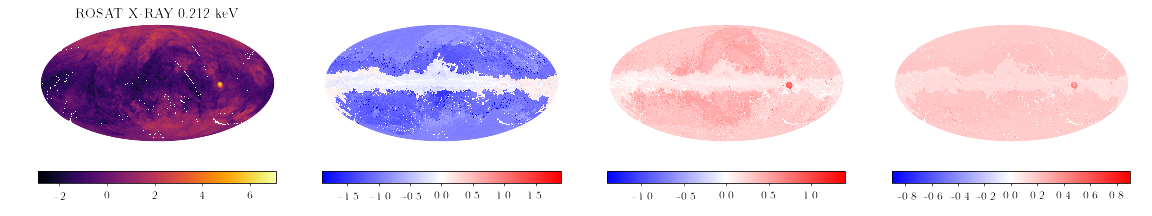}
\end{subfigure}
\begin{subfigure}{\textwidth}
\centering
\includegraphics[width=\textwidth]{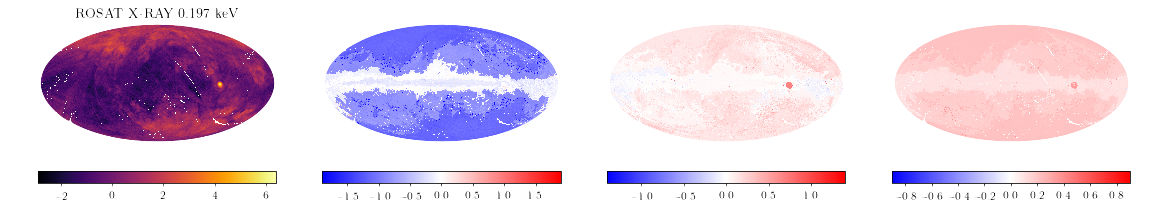}
\end{subfigure}
\begin{subfigure}{\textwidth}
\centering
\includegraphics[width=\textwidth]{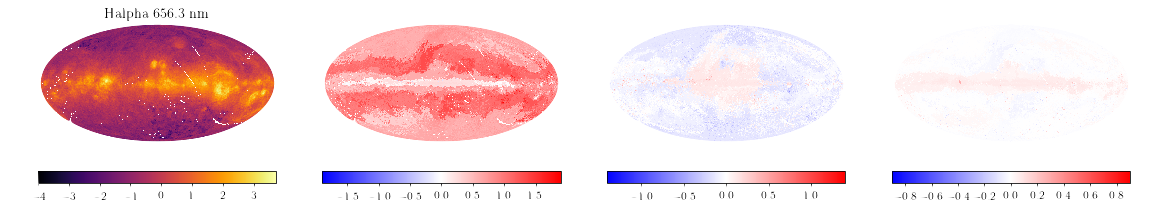}
\end{subfigure}
\begin{subfigure}{\textwidth}
\centering
\includegraphics[width=\textwidth]{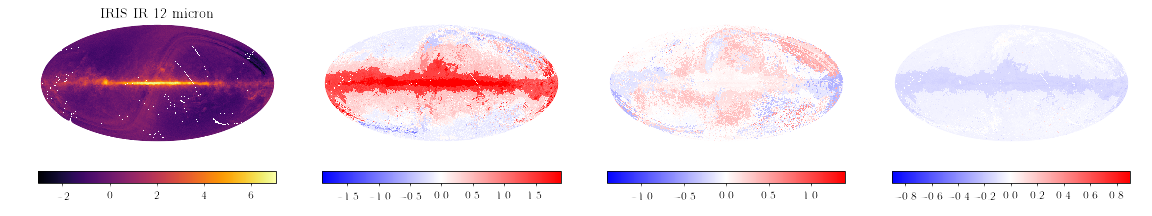}
\end{subfigure}

\caption[Gradient maps]{Decoder Jacobian maps continued.}
\label{fig:grads}
\end{figure}

\begin{figure}\ContinuedFloat
\begin{subfigure}{\textwidth}
\centering
\includegraphics[width=\textwidth]{{Gradsfeaturemaps}.png}
\end{subfigure}
\begin{subfigure}{\textwidth}
\centering
\includegraphics[width=\textwidth]{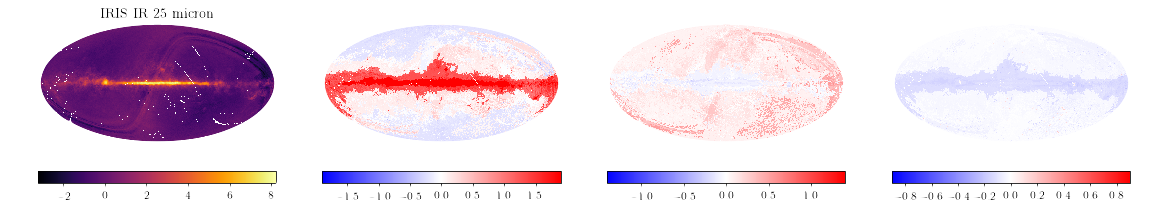}
\end{subfigure}
\begin{subfigure}{\textwidth}
\centering
\includegraphics[width=\textwidth]{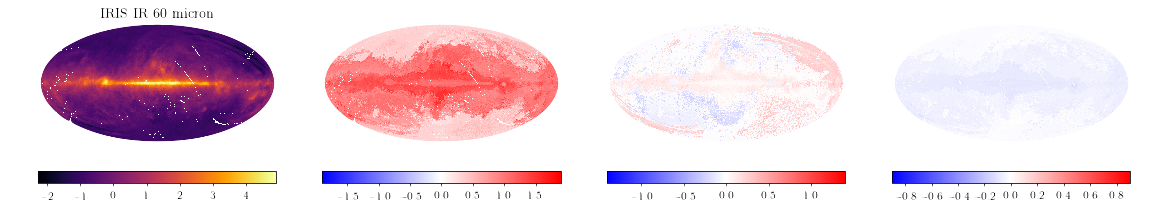}
\end{subfigure}
\begin{subfigure}{\textwidth}
\centering
\includegraphics[width=\textwidth]{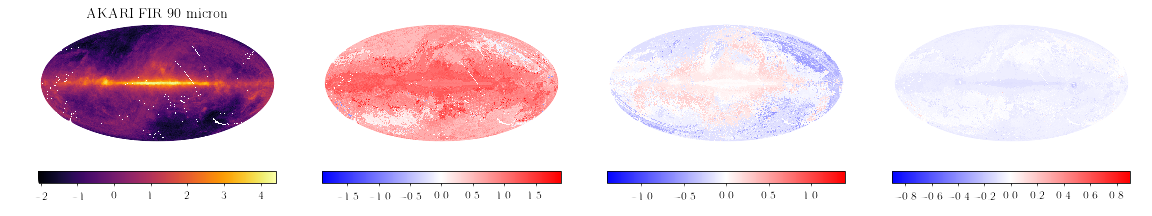}
\end{subfigure}
\begin{subfigure}{\textwidth}
\centering
\includegraphics[width=\textwidth]{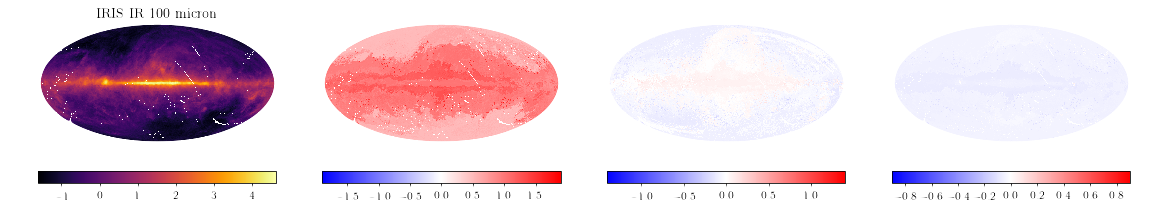}
\end{subfigure}
\begin{subfigure}{\textwidth}
\centering
\includegraphics[width=\textwidth]{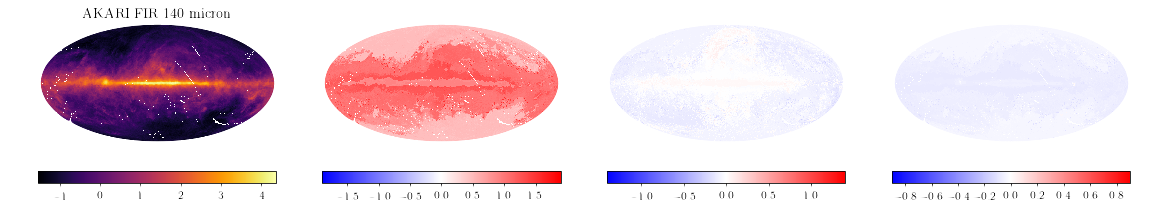}
\end{subfigure}
\begin{subfigure}{\textwidth}
\centering
\includegraphics[width=\textwidth]{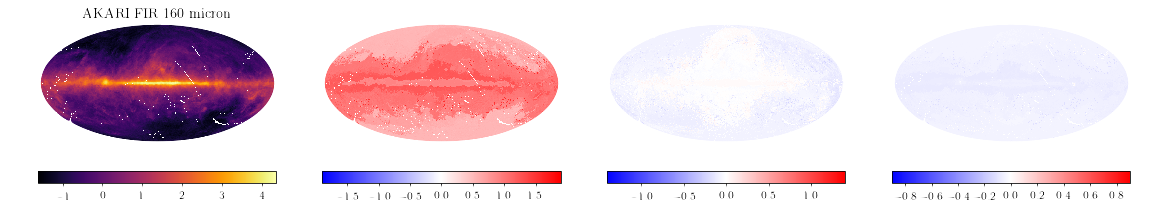}
\end{subfigure}

\caption[Gradient maps]{Decoder Jacobian maps continued.}
\label{fig:grads}
\end{figure}

\begin{figure}\ContinuedFloat
\begin{subfigure}{\textwidth}
\centering
\includegraphics[width=\textwidth]{{Gradsfeaturemaps}.png}
\end{subfigure}
\begin{subfigure}{\textwidth}
\centering
\includegraphics[width=\textwidth]{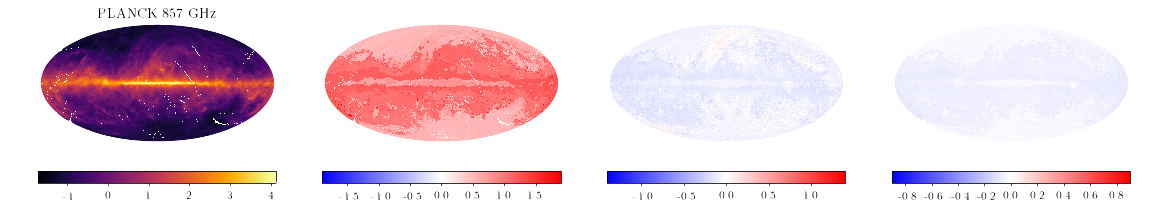}
\end{subfigure}
\begin{subfigure}{\textwidth}
\centering
\includegraphics[width=\textwidth]{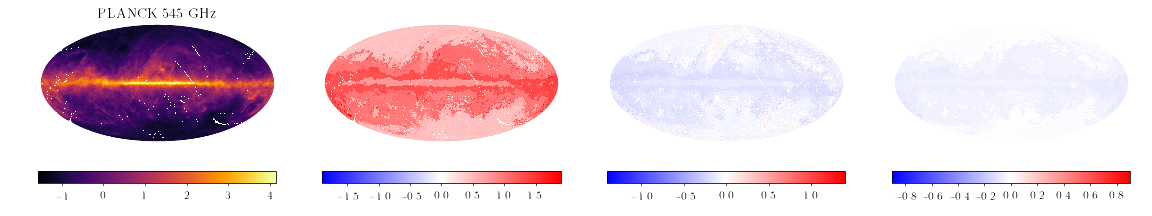}
\end{subfigure}
\begin{subfigure}{\textwidth}
\centering
\includegraphics[width=\textwidth]{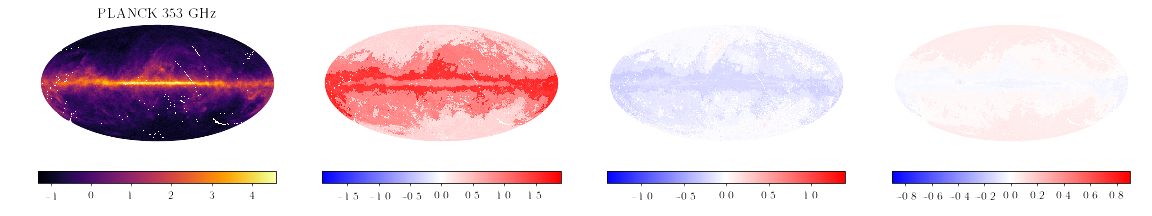}
\end{subfigure}
\begin{subfigure}{\textwidth}
\centering
\includegraphics[width=\textwidth]{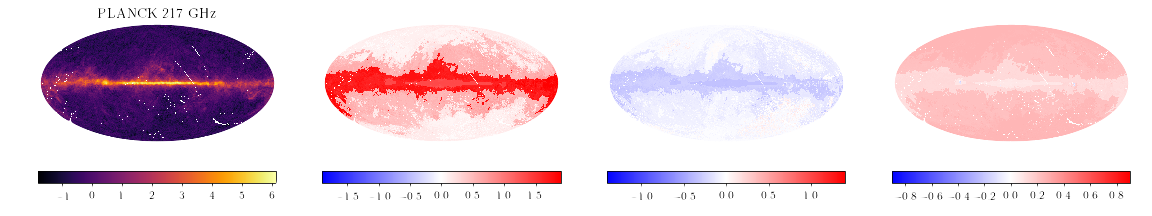}
\end{subfigure}
\begin{subfigure}{\textwidth}
\centering
\includegraphics[width=\textwidth]{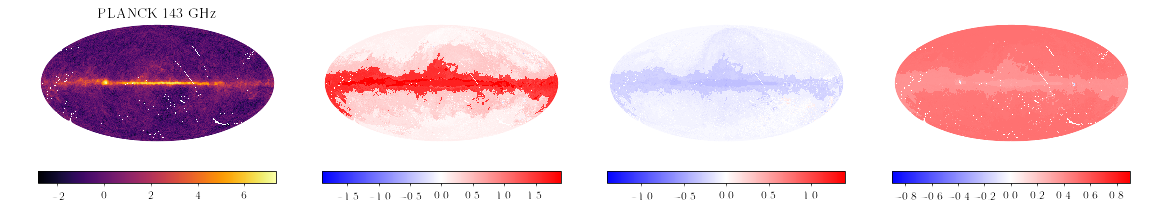}
\end{subfigure}
\begin{subfigure}{\textwidth}
\centering
\includegraphics[width=\textwidth]{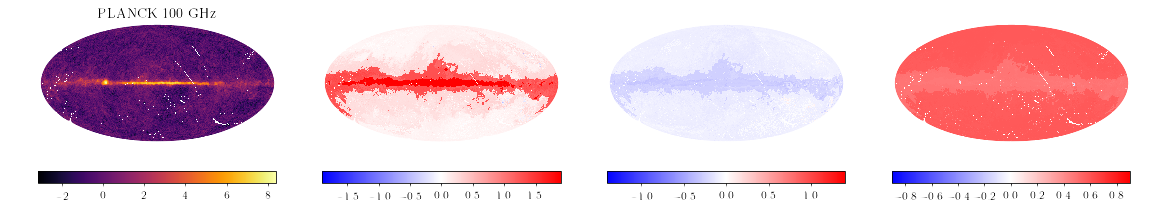}
\end{subfigure}

\caption[Gradient maps]{Decoder Jacobian maps continued.}
\label{fig:grads}
\end{figure}

\begin{figure}\ContinuedFloat
\begin{subfigure}{\textwidth}
\centering
\includegraphics[width=\textwidth]{{Gradsfeaturemaps}.png}
\end{subfigure}
\begin{subfigure}{\textwidth}
\centering
\includegraphics[width=\textwidth]{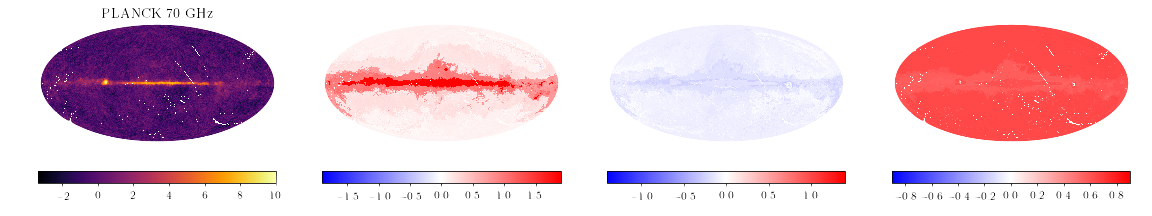}
\end{subfigure}
\begin{subfigure}{\textwidth}
\centering
\includegraphics[width=\textwidth]{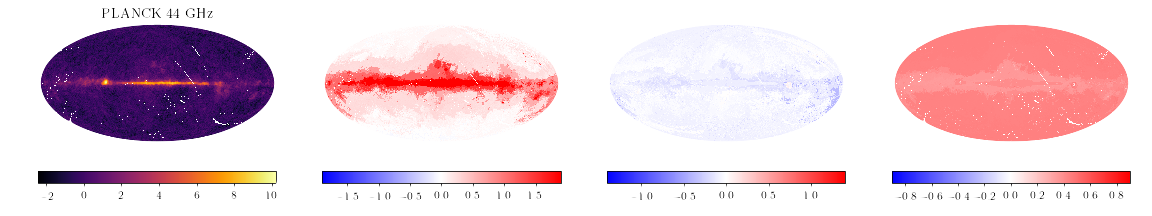}
\end{subfigure}
\begin{subfigure}{\textwidth}
\centering
\includegraphics[width=\textwidth]{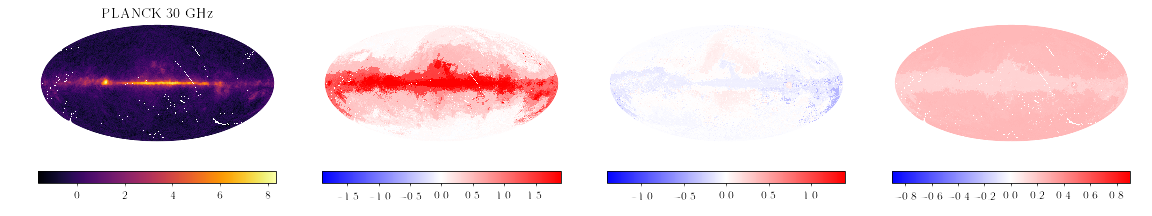}
\end{subfigure}
\begin{subfigure}{\textwidth}
\centering
\includegraphics[width=\textwidth]{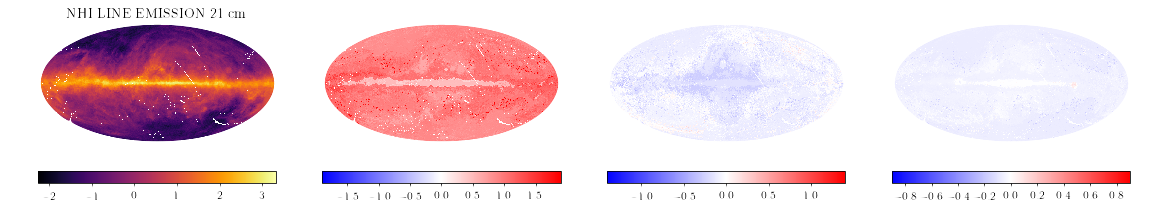}
\end{subfigure}
\begin{subfigure}{\textwidth}
\centering
\includegraphics[width=\textwidth]{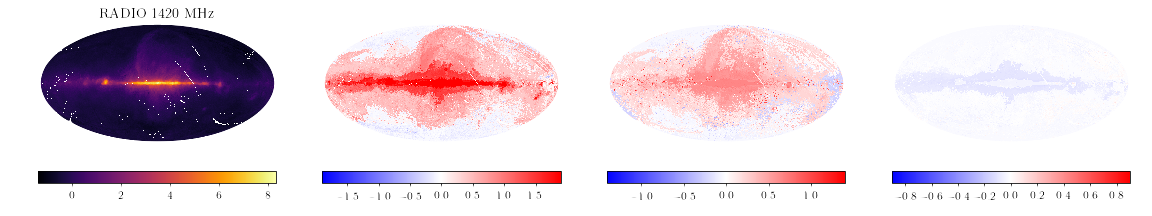}
\end{subfigure}
\begin{subfigure}{\textwidth}
\centering
\includegraphics[width=\textwidth]{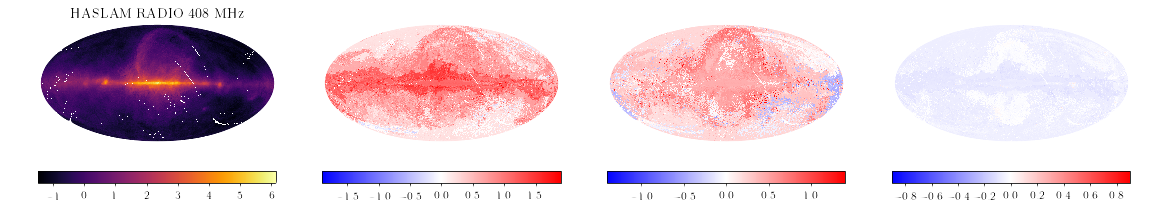}
\end{subfigure}
\caption[Gradient maps]{Decoder Jacobian maps.}
\label{fig:grads}
\end{figure}
\clearpage

\end{appendix}
\end{document}